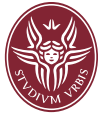 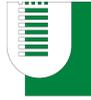 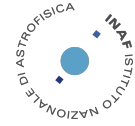

Ph.D. in Astronomy, Astrophysics,
and Space Science
Cycle XXXIV

# THE IMAGING X-RAY POLARIMETRY EXPLORER (IXPE): PROSPECTS FOR SPATIALLY-RESOLVED X-RAY POLARIMETRY OF EXTENDED SOURCES AND IN-ORBIT CALIBRATIONS

Riccardo Ferrazzoli

March 2022

**Supervisor:** Dr. Paolo Soffitta
**Co-supervisor:** Dr. Immacolata Donnarumma
**PhD Coordinators:** Prof. Nicola Vittorio,
                              Prof. Francesco Piacentini

Thesis defended on 28th March 2022
in front of a Board of Examiners composed by:

Prof. Marco Bersanelli (chairman)
Prof. Salvatore Capozziello
Prof. Paolo Natoli

---

**THE IMAGING X-RAY POLARIMETRY EXPLORER (IXPE): PROSPECTS FOR SPATIALLY-RESOLVED X-RAY POLARIMETRY OF EXTENDED SOURCES AND IN-ORBIT CALIBRATIONS**
Ph.D. thesis. Sapienza – University of Rome

I, Riccardo Ferrazzoli, declare that this thesis and the work presented in it are my own.
I confirm that:

- This work was done wholly or mainly while in candidature for a research degree at this University.

- Where any part of this thesis has previously been submitted for a degree or any other qualification at this University or any other institution, this has been clearly stated.

- Where I have consulted the published work of others, this is always clearly attributed.

- Where I have quoted from the work of others, the source is always given. With the exception of such quotations, this thesis is entirely my own work.

- I have acknowledged all main sources of help.

- Where the thesis is based on work done by myself jointly with others, I have made clear exactly what was done by others and what I have contributed myself.

Signed:

---

Date: April 14, 2022

---

This thesis has been typeset by LaTeX and the Sapthesis class.

Version: April 14, 2022

Author's email: riccardo.ferrazzoli@inaf.it

*To Valeria.*
*I love you.*

# Abstract

**Context:** X-ray polarimetry provides two missing observables in the high energy domain, namely the polarization degree and angle, making it possible to obtain information on the geometry and emission processes of high energy celestial sources. Unfortunately, the field has been dormant for decades because of technological limitations and competition with other experiments.

The launch of the NASA/ASI Imaging X-ray Polarimetry Explorer (IXPE) mission in December 2021 opens a new era for X-ray polarimetry. Thanks to the imaging and the polarimetric capabilities of the Gas Pixel Detectors, IXPE will investigate the polarimetric properties of complex fields and extended sources.

**Aims:** This thesis has been focused on two relevant tasks among those foreseen in the pre-launch phase of a new mission: the plan for in-orbit calibrations and for observations of faint extended sources. Since no X-ray celestial sources are available for in-orbit calibrations (the only known source in the X-rays, the Crab Nebula, cannot be employed because of its variability), IXPE will have on board a set of polarized and unpolarized calibration sources. Since the celestial sources will be observed by IXPE for long and segmented integration times, monitoring the detector performance during the mission lifetime will be of fundamental importance, because the characteristics of the GPD are expected to slightly evolve in time. I will present the acceptance tests of the Flight Models of the polarized and unpolarized calibration sources and their validation in thermal vacuum when combined with the Flight Models of the IXPE detectors.

While negligible for the observation of point sources, the depolarizing effect of unpolarized instrumental and diffuse sources of backgrounds will be another challenge for the X-ray polarimetric observation of faint, extended sources. I will describe the effect of the main sources of background (instrumental, diffuse Galactic plane emission, and cosmic X-ray background) on the X-ray polarimetric observations of faint, extended sources.

Finally, I will present a feasibility study of the IXPE observation of two extended sources: the Tycho supernova remnant, and the molecular clouds of the Sgr A complex.

**Methods:** The acceptance and validation tests of the Flight Models of the polarized and unpolarized calibration sources were performed with calibrated commercial detectors such as Charged Couple Devices and Silicon Drift Detectors, and with the IXPE Flight detector Units harboring the Gas Pixel Detectors in a thermal vacuum chamber.

The evaluation of the impact of the sources of background on the detectability of the polarization for a subset of faint extended sources in the IXPE observing plan, and the simulations of the IXPE observations were performed with the Monte Carlo software ixpeobssim.

**Results:** I obtained the counting rates, spectra, image and polarization information, from each Flight Model of the on board calibration sources, and determined the time necessary to achieve the needed precision. I demonstrated that the on-board calibration system will enable us to assess and verify the functionality of the GPD and validate its scientific results in orbit. In particular, the calibration sources illuminate the whole detector, or just a part of it, measuring properties such as the detector response to both polarized and unpolarized radiation, and gain variation in time and space. This information will be used to perform the on orbit calibration to check the performance in time of the GPD.

For the faintest extended sources, such as SN1006 and the molecular clouds of the Sgr A complex, background mitigation techniques will be necessary, while for other sources such as Cas A, Tycho, and the PSW MSH 15-52, the effects will be negligible. For the former, the impact of the instrumental background will require the application of rejection techniques based on the event properties in order to discriminate between real events and background produced by the interaction of cosmic rays with the detector.

The feasibility studies presented for the IXPE observation of Tycho and the molecular clouds of the Sgr A complex show that IXPE will be able to (a) distinguish between magnetic field geometries and detect polarization of synchrotron-emitting structures, (b) determine the recent past of our Galactic center, and (c) put constraints on the position of the clouds along the line of sight. The data analysis techniques presented here make it possible to reconstruct the intrinsic, undiluted polarization degree of these sources.

**Conclusions:** With the successful launch of IXPE, we can finally perform spatially resolved polarimetry in X-rays, we can add two observables (polarization angle and degree), and finally answer questions about source and magnetic field geometry and determine emission processes. The preliminary data acquired in the first weeks after the launch show that the the on-board calibration sources are performing as expected, that the residual instrumental background level is very close to the anticipated value, and that the imaging capabilities are fully compliant with the requirements.

The work done during this PhD project has been crucial for the IXPE mission, gathering information about the possibility of performing calibrations in space with the on board calibration sources, and observing extended sources (supernova remnants and faint molecular clouds) by means of realistic simulations, and careful data analysis.

# Acknowledgments

*This thesis faced many hurdles, and would have not see the light if not for the help of many people along the way. First, I want to thank Paolo Soffitta and Immacolata Donnarumma: they convinced me that I had "the right stuff" to do well in the challenging adventure that is the PhD. They supervised, advised, and mentored me even before the start of this formative experience, and I am grateful for the many opportunities they gave me. Physicists love objects in a vacuum, but Science does not thrive in it, and I had the luck and pleasure to work with many talented people that taught me when I asked for help, and scolded me when I deserved it. Fabio Muleri and Sergio Fabiani entrusted me with delicate and precious laboratory equipment, I am grateful for their trust. I am grateful for the history lessons of Enrico Costa, for the bountiful collaboration with Laura di Gesu, Fei Xie, and Fiamma Capitanio, and for the help provided by Alessandra De Rosa, Andrea Marinucci, and Simonetta Puccetti. I want to thank Steven Ehlert, Frèdèric Marin, and Pat Slane for acknowledging my contribution on the international stage. I am especially grateful to Frèdèric for inviting me to Strasbourg to show my work. I want also to acknowledge my STS master and office buddy, Raffaele Piazzolla, for the invaluable moral support. I thank the INAF-IAPS, ASI, and the Sapienza and Tor Vergata universities for providing the funding and amministrative support needed to sustain the PhD activities. As if getting a PhD in Astrophysics wasn't hard enough, I acknowledge COVID-19 for increasing its difficulty to the extreme. I thank Derrik Evans for keeping me sane and healthy through the long lock-down months. I want also to thank the referees for their evaluation and comments that improved this thesis. Finally, words cannot tell how grateful I am for my family: my parents, grandparents, and for the other half of my Universe, Valeria, for they believed in me when I did not.*

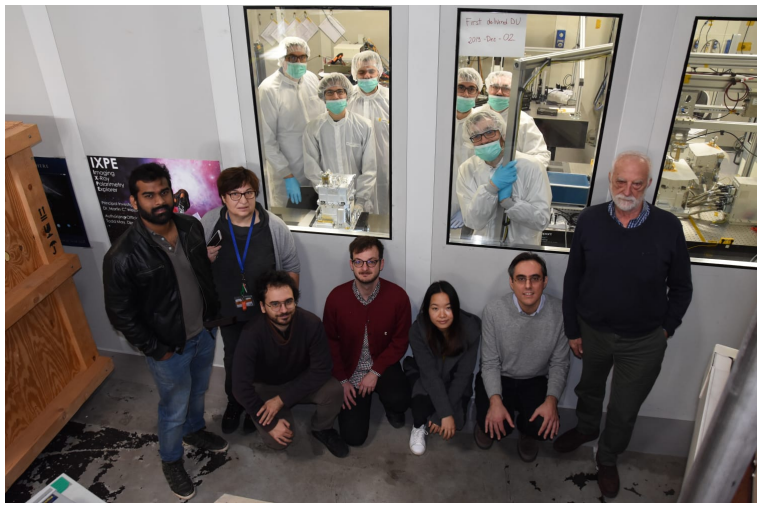

**Figure 0.1.** The INAF-IAPS IXPE team in occasion of the delivery of the first IXPE Detector Unit Flight Model in December 2019. Inside the Clean Room, from left to right, Paolo Soffitta, John Rankin, Fabio La Monaca, Sergio Fabiani, Fabio Muleri, and Alessandro Di Marco. Outside the Clean Room, from left to right, Ajay Rateesh, Alda Rubini, myself, Antonino Tobia, Fei Xie, Ettore del Monte, and Enrico Costa. Picture taken by Yuri Evangelista.

# Contents

# Chapter 1

# Introduction

## 1.1   X-ray polarimetry

In the current "multi-messenger era" of astronomy, most of the information from celestial sources still comes from electromagnetic radiation. However, even in the electromagnetic channel, we are missing information on the region of the parameter space probed by polarization in the X-ray band [154, for a review].

Polarimetry is an established measurement technique in astronomy that is successfully employed in the radio, infrared, and optical band. It is an invaluable tool for the study of cosmic sources, because it adds two observables to the parameter space, that is the polarization degree and the polarization angle in X-rays. The polarization of electromagnetic waves is a phenomenon that arises when there is a preferred direction in a system, so it may be induced, for example, by strong, ordered magnetic fields, or by the presence of geometrical asymmetries, such as jets and disk. Hence, by giving insights on the geometry and emission processes that occur in the celestial sources, X-ray polarimetry complements the information coming from other techniques such as spectroscopy, timing and imaging. In Fig. 1.1 a multi-frequency comparison of observations of the Crab Nebula in the radio, infrared, optical, and X-ray band with the respective polarization information is shown. The X-ray measurement, crucial for studying freshly accelerated electrons, is the only significant one available to date [165, 169]. It was achieved with the X-ray polarimeter on board the OSO-8 satellite, but it is not representative of the morphological complexity of the X-ray image, nor it is spatially resolved like in the other energy bands. For other sources only a handful of barely significant upper limits were estimated.

Given the processes involved in the polarization of electromagnetic waves, one would expect that the emission of all celestial sources to be polarized in X-rays at some extent. In fact, it was immediately recognized that polarization of the X-ray emission of celestial sources was connected to the emission processes themselves: cyclotron, synchrotron, non-thermal bremsstrahlung [59, 118, 175]; scattering on aspherical accreting plasmas, disks, blobs, columns [58, 95, 147]; vacuum polarization and birefringence through extreme magnetic fields [60, 94, 159]. X-ray polarimetry studies started right after the birth of X-ray astronomy in the '60s with the polarimeters based on Bragg diffraction and Thomson scattering flying on sounding rockets. However, because of technological limitations, few results were obtained, with the only significant measurement being the aforementioned polarization from the Crab Nebula. After this result, the field has remained dormant for decades. The difficulties came from the need of the satellites to spin in order to detect the signal. This was overcome by the introduction of the X-ray optics with the Einstein space telescope in the '80s that made the spinning observatories, and hence the available X-ray polarimeters, obsolete. Moreover, the previous instruments had large backgrounds and not enough sensitivity, and they had to compete with other experiments on board of the existing and planned missions. X-ray polarimeters based on the technology available at that time would have been limited to the brightest sources, and with a poor sensitivity if compared to those achieved with instruments dedicated to X-ray spectroscopy, timing, and imaging. This "dark age" ended finally with the development in Italy of photoelectric X-ray polarimeters based on the Gas Pixel Detector [GPD, 14, 36], that could be put at the focus of X-ray optics. The GPD exploits the photoelectric effect, that is the dominant interaction process in $2-8$ keV X-ray energy band, to resolve the photoelectron tracks in a gas cell. This makes it possible to reconstruct their initial emission direction which brings memory of the direction of the electric field of the absorbed photon, and hence its polarization properties. The GPD is not only capable of measuring linear polarization, but it also enables imaging, thus performing for the first time spatial-spectral-time resolved X-ray polarimetry. The first mission entirely dedicated to spatially-resolved X-ray polarimetry is the NASA/ASI Imaging X-ray Polarimetry Explorer [IXPE, 172, 173], that was successfully launched on December 9th 2021.

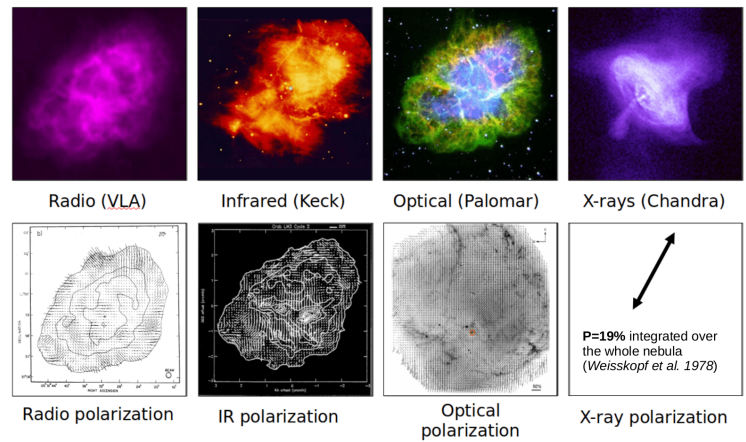

**Figure 1.1.** Multifrequency view in energy (top row) and polarization (bottom row) of the Crab Nebula [50]. The radio and infrared synchrotron emission comes from the pulsar wind confined by the magnetic field, with enhancements along filaments. The optical synchrotron emission (blue-green) is surrounded by emission lines from filaments (red). Finally the X-ray synchrotron emission comes from jets and wind downstream of the termination shock (i.e. the distance at which the pulsar wind ram pressure is balanced by the nebula internal pressure), marked by the bright inner ring. The polarization maps from the radio to optical show complex, spatially resolved fields, while in the X-rays only a spatially integrated measurement exists. Note that these pictures are not to scale, as the size of the synchrotron nebula decreases in size going from the radio to the X-ray band.

## 1.2 Scope of the thesis

This thesis was carried out at the Institute for Space Astrophysiscs and Planetology (IAPS) in Rome in collaboration with the Italian Space Agency (ASI), and took place during most of the development phase of the IXPE mission. I took part to the characterization and validation of the detector and the on-board calibration sources, to the preparation of the observation plan, and to the definition of the scientific objectives. From February 2018 to October 2021 my role inside the IXPE mission was that of "science participant", that is a junior member of the team, sponsored by a science mission collaborator, that has access to the mission scientific topical working groups (STWG). Starting from November 2021 I became a "science collaborator", tasked with scientific modeling of the sources, simulations of IXPE observations, providing input to the observing plans, and analysis and interpretation of the flight data. Moreover, I will be coauthor the scientific papers of the IXPE STWG and I could have other IXPE science participants under my guidance. This thesis aims to present in a coherent way my contributions to this space mission, and to describe how the results here presented will inform and help the data acquisition and scientific analysis.

The goal of the thesis is the calibrability of the instrument in orbit and the observability of challenging extended sources.

My contribution to the mission has been on both the laboratory/hardware and astrophysical side.

Since the Gas Pixel Detector is a new instrument and gas detectors are known to "evolve" in time, it will be fundamental to check the detector properties during the mission lifetime. As said earlier, the Crab Nebula is the only source known to be polarized in X-rays, but it cannot be used as a celestial calibration source because it is not known how its polarization properties change with time, and the X-ray emission itself is known to evolve in time [48, 65]. This means that for this mission our own calibration sources are required on board. To this end, each Detector Unit includes a Filter and Calibration Set (FCS) hosted in a Filter and Calibration Wheel (FCW) that allows to place in front of the GPD different calibration sources, thus monitoring the detector response to both polarized and unpolarized X-rays.

The calibration set consists of four sources powered by radioactive nuclides to produce X-rays and allow to study the detector response at different energies across all the detector surface. One source is polarized (CalA) and three are unpolarized (CalB, CalC, CalD). I performed the acceptance tests of the four Flight Models of the IXPE on-board polarized and unpolarized calibration sources. The goal was to study in the laboratory the performances of the four sets of calibration sources before installing them into the detector units and then test them in thermal-vacuum conditions. From these measurements I obtained the counting rates, spectra, images and polarization information for each source of the four sets, and I verified their capability to check the relevant detector properties during the mission lifetime.

On the Astrophysical side, I am involved in the activities of the Science Analysis and Simulations Working Group (SASWG), that is in charge of the simulations and analysis tools for the observations, and in those carried out in the STWG, actively participating in the definition of the IXPE observations of Supernova remnants (SNRs) and the Galactic center (GC) through simulations and theoretical modeling.

The aforementioned sources are all faint and extended, posing the challenge of correctly evaluating and mitigating the different sources of background, astrophysical or instrumental in nature, that will affect the X-ray polarimetric observations. These sources of background are unpolarized and dilute the polarization signal, which means that they have to be properly taken into account and then subtracted.

In this thesis I will describe the sources of background affecting the IXPE observations, such as the instrumental background, the Galactic plane diffuse background, and the Cosmic X-ray background. I contributed to estimate the IXPE instrumental background, and in this thesis I will evaluate its effects, together with those due to the astrophysical background, on the detectability of polarization in faint extended sources.

Because the IXPE data will be public as soon as the observations are complete, it is important for the collaboration to prepare in advance the observations through dedicated simulations.

I will present a simulated IXPE observation of the Tycho SNR, showing how IXPE will allow to distinguish between different magnetic field topologies and detect polarization of synchrotron-emitting structures. For SNRs, likely the site of acceleration of Galactic cosmic rays, spatially resolved X-ray polarimetry will be invaluable, because the polarization maps will allow to trace the magnetic field geometry and strength in the shocks, filaments, and magnetic structures that characterize them.

Finally, I will present a feasibility study that we performed for the IXPE observation of a fascinating and challenging target: the molecular clouds in the Galactic center. This measurement will allow to explore the past activity of Sgr A*, the Supermassive Black Hole (SMBH) at the Center of our Galaxy, possibly demonstrating that it underwent an active phase as early as $\sim$300 years ago. Since SgrA * is the closest SMBH available to us, the interest in studying its accretion history as a model for other SMBHs is obvious. Many attempts have been made to reconstruct the past light-curve of Sgr A* using the X-ray spectral and morphological variability of these reflection nebulae. However, the poor constrain of the distance along the line of sight of the clouds is a major source of uncertainty which makes inferring the time-delay of the X-ray emission difficult. An independent way to address this ambiguity is provided by X-ray polarimetry, as the reflected emission from a compact illuminating source is linearly polarized by scattering. The polarization direction is perpendicular to the scattering plane where the direction of external illuminating source lies. The detection of polarization from two or more sources would allow to pinpoint the position of the source that in the past illuminated the molecular clouds. The polarization degree, instead, depends on the scattering angle, and so on the position of the cloud along the line of sight. With X-ray polarimetric measurements, IXPE would not only identify the external source of illumination of the clouds, but also determine their distribution in the Galactic core. Because the molecular clouds are faint and diffuse, they will be subject to dilution by the diffuse emission that permeates the Galactic center and other sources of background. Moreover, they are typically distributed over angular regions that encompasses the whole IXPE field of view so that they will usually be observed off-axis. Hence, I will also present a follow up study in which I will focus on an IXPE observation of the so-called Sgr A complex. I will introduce data analysis techniques to reconstruct the intrinsic polarization degree of the clouds, and to determine how their detectability changes when observed off-axis.

The synergy among on-orbit calibrations, the knowledge of the detector, and the preparation in advance of the observations and data analysis thanks to realistic simulations described in this thesis, will contribute to properly interpret the scientific results of the IXPE mission.

The reopening after many decades of the observational window of X-ray polarimetry will finally allow us to answer many questions on the acceleration mechanism and geometry of celestial sources.

## 1.3   Summary of the thesis

The thesis is organized as follows. The first part is dedicated to the state of the Art of X-ray polarimetry up to the launch of IXPE. In Chapter 2, I will briefly present the fundamentals of polarized X-ray radiation, and in Chapter 3 I will trace the history of past X-ray polarimetry missions and instruments. In Chapter 4 I will describe the X-ray polarization processes that are relevant to the study of extended sources such as Supernova Remnants and the Molecular Clouds in the Galactic center. In Chapter 5 I will present the IXPE mission, I will describe the sources of background that are relevant to the IXPE observation of extended sources, and finally I will present the methods to simulate and analyze X-ray polarimetric data. The second part of the thesis aims to present my original contribution to the mission. I will present, in Chapter 6, the results of the test performed on the flight models of the IXPE on-board calibration sources; in Chapter 7 I will describe the impact of the sources of background on the IXPE observation of extended sources. In Chapter 8 I will present a simulated IXPE observation of the Tycho SNR, also discussing the scientific impact of those measurements. In Chapter 9 I will present the feasibility study of the IXPE observation of the molecular clouds in the Galactic center, and in Chapter 10 a follow-up study dedicated to data analysis techniques of the Sgr A * molecular cloud complex. Finally, in Chapter 11 I will show some preliminary results from the very first data acquired by IXPE in its first weeks after the launch. In Chapter 12 I will outline my conclusions and the future prospects of my research.

# Part I

# Part 1

# Chapter 2

# Polarimetry fundamentals

In this Chapter I will review the classical approach to the determination of the polarization properties of X-ray radiation (Section 2.1), and in Section 2.2 I will define the modulation factor. Then, in Section 2.3, I will present the most important figure of merit for the determination of the significance of an X-ray polarimetric measurement, that is the Minimum Detectable Polarization.

## 2.1   Wave polarization and the Stokes' parameters

The polarization plane of electromagnetic radiation is defined by the oscillation direction of the electric vector $\vec{E}$, with the polarization vector assumed to be coincident with the electric vector itself. In a Cartesian system with the $z$-axis parallel to the direction of propagation, the polarization vector can be decomposed along the two orthogonal axis $x$ and $y$ so that, given the three dimensional wave equation for an arbitrary vector field $\vec{E}$:

$$\nabla^2 \vec{E}(\vec{r}, t) = \frac{1}{c^2} \frac{\partial^2 \vec{E}(\vec{r}, t)}{\partial t^2} \quad . \tag{2.1}$$

Two general solutions in $x$ and $y$ direction can be found:

$$E_x(\vec{r}, t) = E_{0x} cos(\vec{k} \cdot \vec{r} - \omega t + \delta_x) \quad , \tag{2.2}$$

$$E_y(\vec{r}, t) = E_{0y} cos(\vec{k} \cdot \vec{r} - \omega t + \delta_y) \quad . \tag{2.3}$$

Where $E_{0x}$ and $E_{0y}$ are the maximum amplitudes, $\vec{k}$ is the wave vector and $\delta_x$ and $\delta_y$ are arbitrary phases. Because the linear combination of the two solutions $E_x$ and $E_y$ is still a solution, Eq. 2.2 and 2.3 can be re-written in the form of the equation of the polarization ellipse:

$$\frac{E_x^2}{E_{0x}^2} + \frac{E_y^2}{E_{0y}^2} - 2\frac{E_x}{E_{0x}} \frac{E_y}{E_{0y}} \cos \delta = \sin^2 \delta \quad , \tag{2.4}$$

where $\delta = \delta_x - \delta_y$ is the phase difference. The tip of the electric field vector $\vec{E}$ draws an ellipse for any time $t$, and the angle $\phi$ between the x axis and the semi-major axis of the ellipse is given by:

$$\phi = \frac{1}{2} tan^{-1} \left( \frac{2E_{0x}E_{0y} cos\delta}{E_{0x}^2 - E_{0y}^2} \right) \quad . \tag{2.5}$$

The polarization ellipse and its orientation with respect to the x-y plane are illustrated in Figure 2.1. Polarization is a vector quantity, that's why it cannot be measured as easily as regular total fluxes (scalar quantities) with simple photometers.

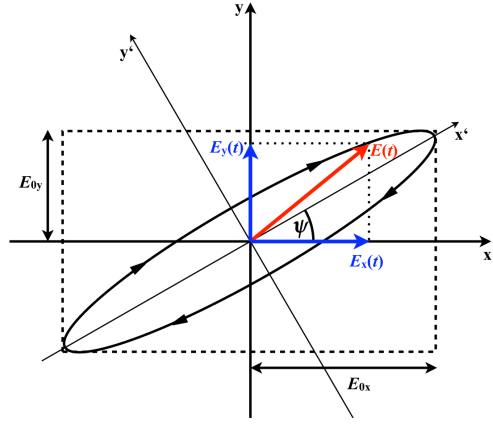

**Figure 2.1.** The polarization ellipse: the red vector represents the direction and intensity in time of the electric field vector $\vec{E}(t)$, the vector in blue are its projections on the reference frame.

In 1852 Sir George Gabriel Stokes introduced the four Stokes parameters [142], which completely fix the parameters of the polarization ellipse. The Stokes parameters allow to fully characterize the polarization state of the radiation in terms of intensities that, differently from the polarization degree and angle, can be summed. In the following we derive the Stokes parameters: in order to measure the time-dependent amplitudes $E_x(t)$ and $E_y(t)$ we average them over the observational time, so that Eq. 2.4 becomes

$$\frac{\langle E_x^2(t) \rangle}{E_{0x}^2} + \frac{\langle E_y^2(t) \rangle}{E_{0y}^2} - 2 \frac{\langle E_x(t)E_y(t) \rangle}{E_{0x}E_{0y}} \cos \delta = \sin^2 \delta \quad . \tag{2.6}$$

Using the expression

$$\langle E_i(t)E_j(t) \rangle = \lim_{T \to \infty} \frac{1}{T} \int_0^T E_i(t)E_j(t)dt \quad i,j = x,y \quad , \tag{2.7}$$

and a different form of Eq. 2.6 multiplied by $4E_{0x}^2 E_{0y}^2$:

$$4E_{0y}^2 \langle E_x^2(t) \rangle + 4E_{0x}^2 \langle E_y^2(t) \rangle - 8E_{0x}E_{0y}\langle E_x(t)E_y(t) \rangle \cos \delta = (2E_{0y}E_{0y}\sin\delta)^2 \quad , \tag{2.8}$$

we find:

$$\langle E_x^2(t) \rangle = \frac{1}{2}E_{0x}^2$$

$$, \langle E_y^2(t) \rangle = \frac{1}{2}E_{0y}^2 \tag{2.9}$$

$$, \langle E_x(t)E_y(t) \rangle = \frac{1}{2}E_{0x}E_{0y}\cos\delta \quad .$$

Substituting into Eq. 2.8, and expressing the result in terms of intensity $E_0$, we obtain

$$(E_{0x}^2 + E_{0y}^2)^2 - (E_{0x}^2 - E_{0y}^2)^2 - (2E_{0x}E_{0y}\cos\delta)^2 - 2E_{0x}E_{0y}\sin\delta = 0 \quad , \tag{2.10}$$

or

$$S_0^2 - S_1^2 - S_2^2 = S_3^2 \quad . \tag{2.11}$$

$S_0$, $S_1$, $S_2$ and $S_3$ are the Stokes parameters, often combined into a vector (Stokes vector) $\vec{S}$:

$$\vec{S} = \begin{pmatrix} S_0 \\ S_1 \\ S_2 \\ S_3 \end{pmatrix} = \begin{pmatrix} I \\ Q \\ U \\ V \end{pmatrix} = \begin{pmatrix} E_{0x}^2 + E_{0y}^2 \\ E_{0x}^2 - E_{0y}^2 \\ 2E_{0x}E_{0y}\cos\delta \\ 2E_{0x}E_{0y}\sin\delta \end{pmatrix} \tag{2.12}$$

The Stokes parameters are related to measurable quantities:

- **I**: total intensity of the photon flux;

- **Q**: difference between the vertical and horizontal polarization state;

- **U**: difference between the linear polarization oriented at $+45°$ and $-45°$ from the vertical polarization state;

- **V**: difference between the clockwise and anticlockwise rotational directions. This parameter is related to the measurement of circular polarization.

Therefore, by using the Stokes formalism, the polarization state of a radiation beam can be determined by a set of individual intensity measurements. Moreover, the Stokes parameters are additive quantities, so that any radiation beam can be split into an unpolarized component and a completely (elliptically) polarized component. The following relation is always valid:

$$I \geq \sqrt{Q^2 + U^2 + V^2} \geq 0 \quad , \tag{2.13}$$

the first equality is true for fully polarized radiation, while the second equality describes unpolarized radiation. Any other case is referred to as a partially polarized beam. From Eq. 2.13 it follows that for each radiation source its generic degree of polarization $P$ is given by:

$$P = \frac{\sqrt{Q^2 + U^2 + V^2}}{I} = \frac{I_p}{I} \quad , \tag{2.14}$$

In the case of X-ray polarimetry, techniques and instrumentation capable of measuring circular polarization do exist, such as the ones based on circular dichroism [157]. They find application in biology, chemistry, and material science [28] requiring, however, fluxes that can only be reached in particle accelerators [e.g. $>3{\times}10^{18}$ ergs/s/cm$^2$ of the FERMI free electron laser 126]. In astronomical applications of X-ray polarimetry (where fluxes ranges between $\sim 10^{-8} - 10^{-12}$ ergs/s/cm$^2$) only linear polarization can be sensitively measured, so that from now on we will assume $V = 0$. Therefore the Equation 2.14 becomes:

$$P = \frac{\sqrt{Q^2 + U^2}}{I} \quad . \tag{2.15}$$

Finally, the angle $\phi$ given by Equation 2.5 is called the polarization angle and it is fully determined by $Q$ and $U$ according to:

$$\phi = \frac{1}{2} tan^{-1} \left( \frac{U}{Q} \right) \quad , \tag{2.16}$$

counted in counter-clockwise direction with $\phi = 0°$ in $+Q$ direction.

## 2.2 The modulation factor

To measure polarization, we need a polarimeter. Such instrument can be obtained by coupling an analyzer of angular directions with a detector (see Figure 2.2).

The analyzer distinguishes the angular directions (the $\phi$ angle in Figure 2.2) of polarization and the detector detects the signal for each angular direction. Therefore the instrument response depends on the preferential direction of the polarized radiation. If the detected radiation is unpolarized there is no preferential angular direction and as a consequence the polarimeter response will be flat, being the same for each analyzed angular direction (see top of Figure 2.3). On the other hand, if the detected radiation is polarized the instrument will have a modulated response (see bottom of Figure 2.3) that can be fitted by a function with a $\cos^2 \phi$ term:

$$N(\phi) = A + B \cos^2(\phi - \phi_0) \quad , \tag{2.17}$$

where $A$ and $B$ are constants and the $\cos^2 \phi$ term depends on the physical process which the detector is based on. The radiation whose degree of polarization must be measured will produce a modulated pattern in the detector which will be fitted by the function of Equation 2.17. To calculate the unknown degree of polarization in a generic case, we introduce a normalization that we call "modulation factor".

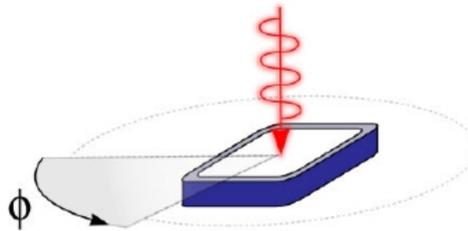

**Figure 2.2.** Scheme showing the concept of a polarimeter. The analyzer (in white) analyses the photons angular directions of polarization $\phi$, the detector (in blue) detect the signal for each angular direction [46].

The modulation factor is defined as the detector response to 100% polarized radiation:

$$\mu = \frac{N_{100\%}^{max} - N_{100\%}^{min}}{N_{100\%}^{max} + N_{100\%}^{min}} \quad , \tag{2.18}$$

where $N_{100\%}^{max/min}$ is the maximum and minimum detector response (as defined in Equation 2.17) for completely polarized radiation. In other words, the modulation factor is the semi-amplitude of the modulation curve normalized to its average value. It is also possible to express the modulation curve by using the Stokes parameters. By means of trigonometric transformations, Eq. 2.17 can be rewritten as

$$N(\phi) = A + \frac{B}{2}\Big(\cos 2\phi \cos 2\phi_0 + \sin 2\phi \sin 2\phi_0 + 1\Big) \quad . \tag{2.19}$$

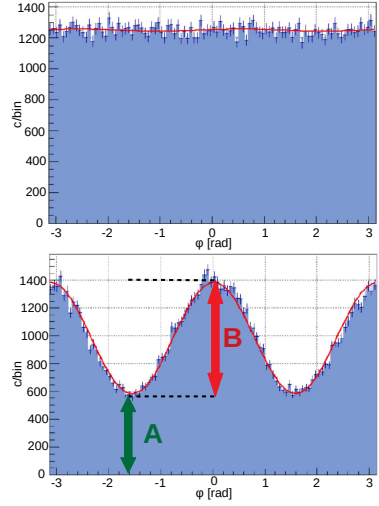

**Figure 2.3.** Real modulation curve derived from the measurement of the emission direction of the photoelectron in a Gas Pixel Detector (GPD) X-ray polarimeter (see later Section 3.5). In the top panel is shown the response to unpolarized radiation is flat because there is no a preferential direction of polarization. The bottom panel shows how the signal obtained from polarized radiation is modulated. The amplitudes $A$ and $B$ are the same as in Eq. 2.17 and 2.23.

Since the total intensity $I$ is the average of Equation 2.19 on the $\phi$ angular directions:

$$I = \frac{1}{2\pi}\int_0^{2\pi} N(\phi)d\phi = A + \frac{B}{2} \quad , \tag{2.20}$$

introducing the Stokes parameters as

$$\begin{cases} Q = \frac{1}{\mu}\frac{B}{2}\cos 2\phi_0 \\ U = \frac{1}{\mu}\frac{B}{2}\sin 2\phi_0 \quad , \end{cases} \tag{2.21}$$

equation 2.19 finally becomes:

$$N(\phi) = I + \mu(Q\cos 2\phi + U\sin 2\phi) \quad . \tag{2.22}$$

The polarization degree can be rewritten by substituting the Equations 2.21 and Equation 2.20:

$$P = \frac{\sqrt{Q^2+U^2}}{I} = \frac{1}{\mu}\frac{\sqrt{\frac{B^2}{4}\cos^2\phi_0 + \frac{B^2}{4}\sin^2\phi_0}}{A + \frac{B}{2}} = \frac{1}{\mu}\frac{B}{2A+B} \quad . \tag{2.23}$$

Comparing Equation 2.23 with Figure 2.3, we can see that $B$ represents the amplitude of the modulation ($B = 0$ for unpolarized radiation), $A$ is the response of the detector for unpolarized radiation and, in photoelectric and Bragg polarimeters, the maximum of the modulation curve corresponds to the polarization angle $\phi$ (see Fig. 2.3). We anticipate here that polarimeters based on photoelectric effect in gas that image the photo-electron track are characterized by a single analyzer/detector. These make this kind of device very suitable for X-ray polarimetry in space (see later Section 3.5).

## 2.3   Minimum Detectable Polarization

Concerning the statistics of X-ray polarimetry, one of the most important figure of merit for defining the sensitivity of a polarimeter is the Minimum Detectable Polarization (MDP, [171]).

The MDP is the minimum degree of polarization that can be detected within a certain confidence level $CL$, and is defined as the maximum amplitude of the polarization that can be due to statistical (i.e. Poissonian) fluctuations in a certain confidence interval. From this definition, Weisskopf et al. [171] calculate the MDP as

$$MDP_{CL} = \frac{\sqrt{-2\ln(1-CL)}}{\int \mu(E)A(E)\epsilon(E)F(E)} \times \sqrt{\frac{\int \mu(E)A(E)\epsilon(E)F(E) + \int B_{diff}(E) + B_{res}(E)}{T}} \quad,$$
(2.24)

where $\mu$ is the modulation factor, $A$ is the telescope effective area, $\epsilon$ is the detector efficiency, $F$ is the source intensity (in ph/s/cm$^2$/keV), $B_{diff}$ is the diffuse background counting rate, $B_{res}$ is the residual (instrumental) background, and $T$ is the observation time.

This translates into

$$MDP_{CL} = \sqrt{-2\ln(1-CL)} \times \frac{\sqrt{2(N_S + N_B)}}{\mu N_S} \quad,$$
(2.25)

where $N_S$ is the number of source counts and $N_B$ the total background counts.

With $R = N_S/T$ the detected source rate and $B = N_B/T$ the total background rate at the detector, at the 99% of confidence level the MDP is given by

$$MDP(99\%) = \frac{4.29}{\mu \cdot R}\sqrt{\frac{R+B}{T}} \quad.$$
(2.26)

It is important to underscore that the MDP does not represent the uncertainty in the polarization measurement, but rather the degree of polarization which has a certain probability (for Equation 2.26 it is the 1%) of being equaled or exceeded by chance.

If a polarization is detected with a value equal to $MDP_{99}$, then the measurement has a significance of only about 3 $\sigma$ [43]:

$$n_\sigma = \sqrt{-2\ln(1-CL)} = 3.03 \quad.$$
(2.27)

This means that an appropriate polarimeter must have an MDP much smaller than the degree of polarization to be measured from the source. The historical detection for the Crab Nebula from Weisskopf et al. [169], for instance, had a significance of 19 $\sigma$. For much fainter sources, such as the extended ones, we could claim detection for $\sigma > 3$.

If the observation is largely source dominated (that is, if $B \ll R$) the MDP formula simplifies in:

$$MDP_{99} \simeq \frac{4.29}{\mu\sqrt{RT}} \quad . \tag{2.28}$$

that in term of source detected events $N$ is:

$$MDP_{99} \simeq \frac{4.29}{\mu\sqrt{N}} \quad . \tag{2.29}$$

Therefore, to achieve, for example, an $MDP_{99}$ of 1%, with $\mu = 0.5$ about $7.36 \times 10^5$ events must be detected from the source.

X-ray polarimetry is indeed a "photon hungry" science that requires much more statistics than, for example, X-ray spectroscopy, where only ~hundreds of counts are enough to determine a spectral slope, of even ~tens of counts are sufficient to claim a source detection with imaging.

The large number of counts required to obtain low values of MDP is a key point for polarimeter measurements and is one of the reasons why X-ray polarimetry has been missing from the astronomical scene for so long.

To understand why the X-ray polarimetry window is finally reopening only now, in the next Chapter I will recall the history of the previous attempts at performing this kind of measurements.

# Chapter 3

# History and instruments of X-ray polarimetry

In this Chapter I will provide a historical *excursus* of the first instruments and mission that attempted at performing X-ray polarimetry of celestial sources based on the path traced by the father of X-ray astronomy himself, Riccardo Giacconi (Section 3.1: from the first X-ray polarimeters flown on rockets, Section 3.2, to spinning satellites 3.3). I will explain why at a certain point the X-ray polarimetry window closed (Section 3.4), and why it is opening again now thanks to a new generation of instruments based on the photoelectric effect (Section 3.5).

## 3.1 The "four seasons"

With the discovery of extra-solar X-Rays in 1962 [54], and hence the birth of observational X-ray astronomy, many sources were found to be characterized by non-thermal emission processes, by radiation transferred in highly asymmetric systems, or both. Hence, polarization of the X-ray emission was to be expected. It was immediately recognized that polarimetric capabilities in this band would have been a crucial asset, even more than at longer wavelengths (such as optical and radio), as X-rays probes the sources at smaller scales. Vitaly Ginzburg was one of the first to explain the role of synchrotron in the emission of X-rays [57] and, as a consequence, of polarimetry as a diagnostic. Ginzburg predicted, way ahead of his time, the role of polarimeters based on photoelectric process in gas [56], while most of scientists were developing instruments based on scattering. Among them, Herbert Schnopper was the first to understand that Bragg diffraction around 45° could be the basis of a technique to perform polarimetry [131].

The X-ray polarization coming from the emission processes themselves received much attention: Gnedin and Sunyaev [59], Rees [118], Westfold [175] predicted the X-ray polarization from cyclotron, synchrotron, and non-thermal bremsstrahlung; Gnedin and Silantev [58], Meszaros et al. [95], Sunyaev and Titarchuk [147] from scattering on aspherical accreting plasmas such as disks, blobs, and columns; Gnedin et al. [60], Meszaros and Ventura [94], Ventura [159] predicted polarized signal from vacuum polarization and birefringence through extreme magnetic fields.

While the theoretical ground-work of X-ray polarimetry was being laid down, Ric-cardo Giacconi outlined a path for the future of this new window on the high-energy sky [55]. He envisioned four "seasons" for the development of X-ray Astronomy, starting with instruments on sounding rockets, then moving to spinning satellites without optics, then satellites with focusing optics, and finally advanced space missions.

Remarkably, this path has been followed with great success: the rockets flown when Giacconi first proposed this path were followed by the UHURU and HEAO-2 (Einstein) missions, and finally the contemporary Chandra space telescope.

Initially, X-ray polarimetry seemed to follow this same pattern, with polarimeters first flown on rockets and then on the OSO-8 satellite. However, the field failed to reach the last two steps.

How and when did this happen?

## 3.2 The first season: rockets

The first attempt at measuring X-ray polarimetry from a celestial source was made by Angel et al. [5] in 1968. Using a sounding rocket carrying a scattering polarimeter sensitive to X-rays in the energy range from 6 to 18 keV, they targeted Sco X-1 (the brightest X-ray source known at that time), but were only able to set an upper limit for the polarization. The group of Columbia University headed by Robert Novick (Fig. 3.1) then spearheaded the field of X-ray polarimetry at the end of the 60's: in April 1969 a sounding rocket carrying two Lithium-block Thomson-scattering polarimeters was launched to search for polarization in the Crab Nebula. It obtained, in the $5.5-22$ keV energy range, only an upper limit of $P < 27$ % at 99% confidence [176]. A second launch in July of the same year targeted again Sco X-1 [108]. The rockets, spinning with rotation axis pointed towards the target in the sky, collected data for a few minutes while above 80 km. Each Lithium block was 5 cm × 5 cm and 12.7 cm in height and surrounded by proportional counters (see 3.2 (a)). In 1972 a more advanced sounding rocket carrying Lithium scattering polarimeters and Bragg crystal polarimeters was launched targeting again the Crab Nebula [164]. The detector consisted in four panels of mosaic graphite with a 45° inclination with respect to the spin axis, that Bragg-diffracted X-ray photons on proportional counters. The energy range of the instrument was of $2.0-3.3$ keV for first order reflection, and $4.0-6.6$ keV for second order reflection. A drawing of the detector is shown in Fig. 3.2 (b). Since the efficiency of the Bragg diffraction is modulated with a $\cos^2$ law around the plane of polarization of the radiation, the modulation of the counts with the azimuth angles provided a measurement of the polarization of the beam. While loss of telemetry reduced the significance of the flight data, by over imposing the data of all flights a high amount of polarization was found with the needed significance [109]. This result encouraged the building of new instrumentation based on Bragg diffraction, to be put aboard satellites. In the following sections I will briefly describe the working principles of these first detectors.

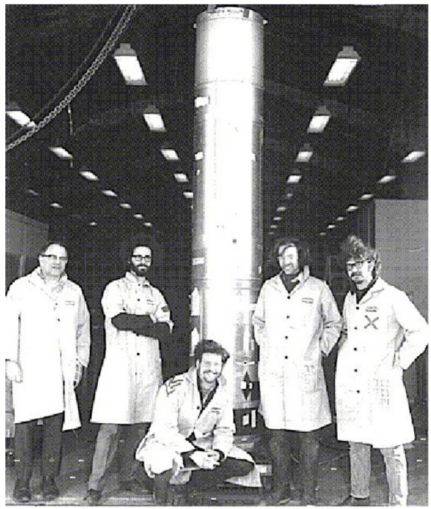

**Figure 3.1.** The Columbia University X-ray polarimetry team of Robert Novick in front of a sounding rocket [170].

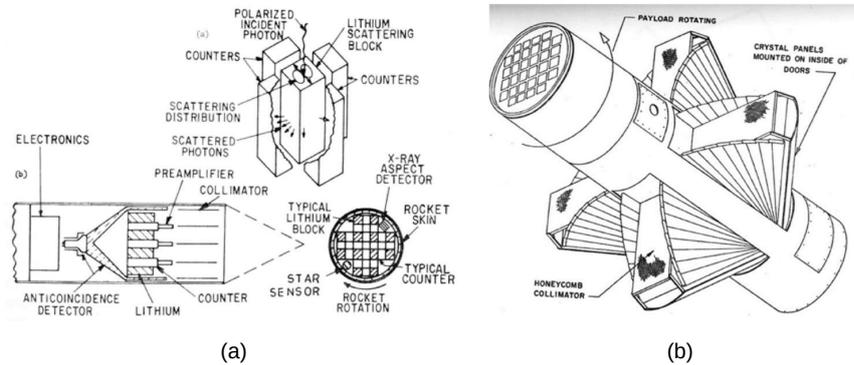

**Figure 3.2.** (a) Lithium scatterer X-ray polarimeter, payload of the first sounding-rocket experiments [5, 108, 176]. (b) Aerobee-350 X-ray polarimeter payload shown in tight configuration after target acquisition[164]: the honeycomb-shaped collimator and flaps on the sides were used as a radio-frequency shield and to prevent direct illumination of the proportional counter by the diffuse X-ray background.

### 3.2.1 Bragg diffraction polarimeters

At energies below a few tens of keV, X-rays interact more strongly via the photoelectric process than via scattering. However, superposition of coherent scatterings off a periodic medium, such as an atomic crystal, can produce efficient diffraction. This process is known as Bragg diffraction and occurs when the difference in path length for scattering from two adjacent crystal planes ($2d\sin\theta$, where $d$ is the crystal plane spacing and $\theta$ is the angle between the incident ray and the scattering planes is an integer multiple, $n$, of the photon wavelength, $\lambda$, see Figure 3.3). This condition, known as "Bragg's law", can be expressed as:

$$n\lambda = 2d\sin\theta \quad \text{or} \quad \mathrm{E} = \frac{2\mathrm{d}}{\mathrm{nhc}}\sin\theta \quad , \tag{3.1}$$

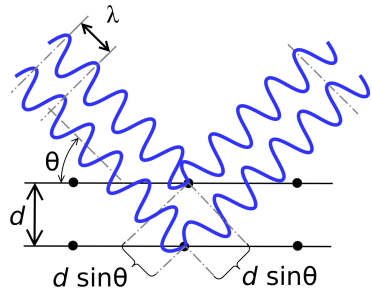

**Figure 3.3.** In Bragg diffraction the two outgoing waves are in phase if the difference in path length for scattering from two adjacent crystal planes, $2d \sin \theta$, where $d$ is the crystal plane spacing and $\theta$ is the angle of incidence, is an integer multiple of the photon wavelength, $\theta$ (adapted from Kaaret [77]).

where $E$ is the photon energy for which the condition is satisfied.

Because the reflectivity for radiation that is polarized parallel to the incidence plane is ∼zero for angles $\theta$ close to the Brewster angle (which is 45° for X-rays and a continuum X-ray source), Bragg diffraction can be used for polarization analysis. The modulation factors for Bragg polarimeters are typically very high and can exceed 99%. In order to produce a modulation curve, Bragg diffraction polarimeters must either rotate, or at least three crystals must be put at 45° offsets to measure instantaneously the Stokes parameters. In the early stages of X-Ray Astronomy, instruments both on rockets and satellites used slat collimators and spun around the pointing axis scanning the sky. Equipping them with Bragg polarimeters was then a straightforward and robust choice.

However, Bragg diffraction has a problem: because of the condition given by Eq. 3.1, very few photons are diffracted in a very narrow energy band and, as explained in Sec. 2.3, one of the major problems of X-ray polarimetry is the starvation of photons. Efficient reflection can be obtained for X-rays exactly satisfying the Bragg condition, but the efficiency drops off rapidly as the photon wavelength or incidence angle changes. The efficiency of a Bragg polarimeter for an astrophysical source with a broad spectrum and for perfect atomic crystals is defined by its effective width $\Delta E(\theta)$, that typically is of a few eV. The effective width is the integral of reflectivity $R(E, \theta)$ over all energies at fixed angle:

$$\Delta E(\theta) = \int R(E, \theta) dE \quad . \tag{3.2}$$

Many solutions were then tried in order to increase the efficiency of the polarimeters based on this technique and/or decrease the background. Increasing the area of the crystal allows to gather more photons, however also increasing the background in the process. Employing mosaic crystals, i.e. imperfect crystals that are a mosaic of small crystal domains with random orientations with angles of the order of 0.2°, 0.5°, 1.0°. The crystal domains are thin compared with the X-ray absorption length, so an X-ray may pass through multiple domains until it finds one oriented to finally satisfy the Bragg condition. This allows to increase the energy bandwidth by some tens of eV. Finally, the use of bent crystals provide a focusing effect and reduce background, with some trade-off on the modulation.

### 3.2.2 Thomson/Compton polarimeters

At energies above a few tens of keV, the dominant interaction process of X-rays with matter is Thomson/Compton scattering. In Compton scattering, when the X-ray energy is an appreciable fraction of the rest mass energy of an electron, the electron will recoil during the interaction, taking energy from the photon. The cross section of the process is [93]

$$\frac{d\sigma}{d\Omega} = \frac{r_e^2}{2} \Big(\frac{E'}{E}\Big)^2 \Big(\frac{E'}{E} + \frac{E}{E'} - 2\sin^2\theta\cos^2\phi\Big) \quad , \tag{3.3}$$

where $r_e$ is the classical electron radius, $E$ is the initial photon energy, $E'$ is final photon energy, and we have averaged over the polarization of the final photon.

The relationship between the photon energies and the scattering angle $\theta$, is given by

$$E' = E\Big[1 + (1 + \cos\theta)\frac{E}{m_e c^2}\Big]^{-1} \quad . \tag{3.4}$$

For scattering angles near 90°, the azimuthal distribution of the scattered photon is strongly dependent on the X-ray polarization, thus Compton scattering is effective for polarization analysis as it depends on the angle $\phi$ as seen in Eq. 3.3.

At low X-ray energies, the electron recoil becomes negligible. In this limit, known as Thomson scattering, modulation reaches 100% for 90° scattering. The basic principle of all Compton/Thomson polarimeters is shown in Fig. 3.4. The typical scattering X-ray polarimeter (shown in Fig. 3.4) is made of two detector elements to determine the energies of the scattered photon and of the electron. The first detector element, or "target", provides the physical medium for the Compton interaction to take place [35]. The target is usually made of a low atomic number material, such as Lithium. This increases the path length traveled by the scattered photons, minimize the photoelectric interaction cross section, and maximize the scattering one, thus increasing the polarization sensitivity. The second detector element is called "absorber" or "calorimeter" (since it absorbs the majority of the photon energy). Its role is to absorb the scattered X-rays, and for this reason is often composed by a high atomic number material, such as CsI/YAP (Caesium Iodide/Yttrium Aluminium Perovskite).

The target/absorber geometry is typically arranged to maximize scatterings through polar angles of 90° and the detector records the azimuthal distribution $\theta$ of scattered photons [35]. If the incoming radiation is polarized, the azimuthal distribution of the scattered photons will be modulated. At X-ray low energies, in the Thomson limit, only the scattered photon is detected. This is referred to as "passive scattering". For sufficiently energetic X-rays, in the Compton limit, a recoil electron is produced, allowing to detect both the initial interaction point and the scattered photon. This is instead referred to as "active scattering" and has many advantages, as the instrumental background can be reduced by imposing a coincidence requirement. It is worth noting that, in principle, Compton polarimeters do not require a distinction between target and absorber. Hence, Compton polarimetry is also possible in uniform detector arrays. The advantage of scattering polarimeters is their efficiency over a large energy bandwidth, allowing to perform energy-resolved polarimetry. The principal disadvantage is a modulation factor lower than 100%, as it will approach this value only for a 90° scattering. To increase the efficiency of such detectors, integration over a range of scattering angles is needed. In this case, realistic modulation factors are of the order of 50%. Moreover a large detector area to encircle the scattering element is necessary, resulting in a high background rate. Historically, instrumental systematics of the scattering X-ray polarimeters have been handled through calibrating variations in efficiency between the detector elements, or more commonly by rotating the instrument along the pointing axis.

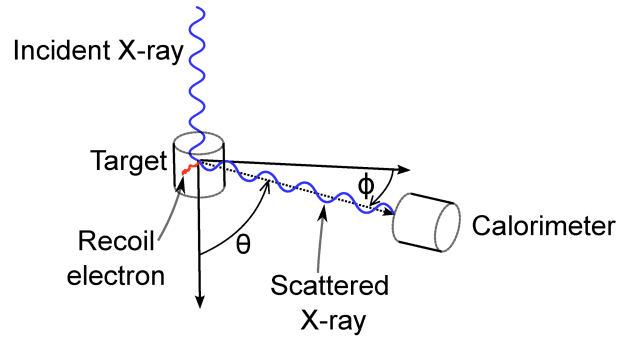

**Figure 3.4.** Compton/Thomson polarimeter scheme: the incident X-ray photon in scattered on a target and detected by the absorbing element. Adapted from Kaaret [77].

## 3.3 The second season: spinning satellites

In 1970 the Soviet Intercosmos-1 was the first satellite to measure the X-ray polarization from solar flares [151]. Then, the British satellite ARIEL-5 hosted a Bragg crystal spectrometer-polarimeter (shown in Fig. 3.5) but, because of low performance of the instrument and competition of other experiments on the satellite, it only produced an upper limit of 7.7% at three-sigma confidence for Sco X-1 [63] after a 10 days observation. Meanwhile in the USA, the Eight Orbiting Solar Observatory

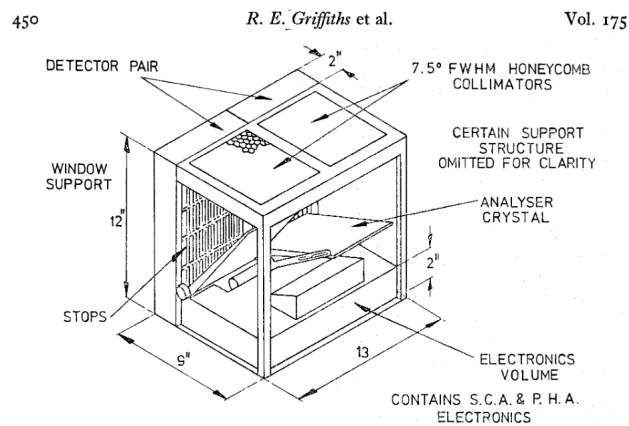

**Figure 3.5.** The X-ray Bragg polarimeter on board the Ariel 5 satellite [63].

(OSO-8) hosted a Bragg polarimeter, conceived and built by the Novick team [166]. In order to maximize the throughput, slightly curved mosaic crystals of pyrolitic graphite with a mosaic spread of 0.8° and an effective width of 40 eV were adopted. A broader set of angles around 45° were thus accepted, resulting in a partially reduced modulation, but the detectors had a surface much lower than that of the crystal, so reducing significantly the background. The instrument contained two orthogonal polarimeters and rotated at a rate of 6 rpm. The OSO-8 polarimeter is show in Fig. 3.6 (a). OSO-8 was more successful than its predecessors and made a first survey searching for polarization of sources, the only firm result being again the Crab Nebula, showing that its polarization at 2.6 keV is 19.2% ± 1.0% at a position angle of 156.4° ± 1.3°, detected with a single campaign and with high significance (19 $\sigma$) [165, 169]. This historical result is shown in Fig. 3.6 (b) and (c). An upper limit on the Crab pulsar was also derived [169] exploiting a Lunar occultation of the nebula. A tight upper limit on Sco X-1 was derived [168], while a marginal detection at a few percent level could be claimed for Cyg X-1 [167]. The comparison of Ariel-5 and OSO-8 suggest one consideration: while both satellites included a Bragg diffraction polarimeter, the ARIEL 5 one was conceived to be a spectrometer. Its crystal was flat and the angle would select the energy to analyze so that at 45° the crystal would work as a polarimeter. However, Ariel-5 ended up having much worse background and performance than OSO-8. The moral of the story is that X-ray polarimetry has such needs and difficulties compared to other measurements that it requires dedicated missions.

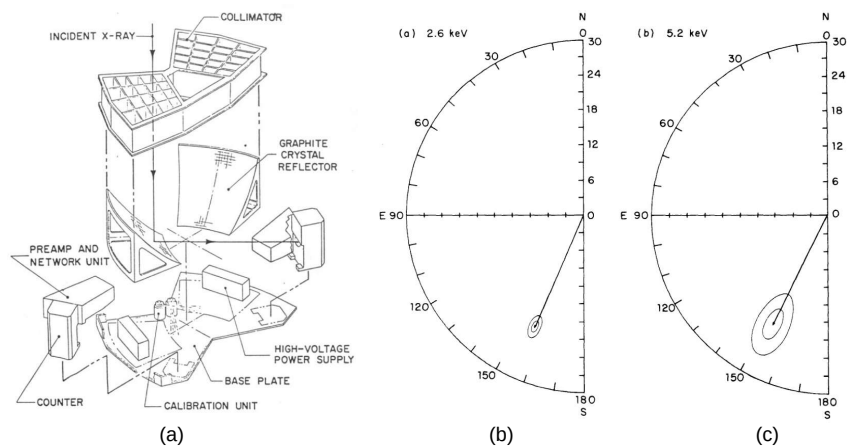

(a)                          (b)                          (c)

**Figure 3.6.** (a) The X-ray Bragg polarimeter on board the OSO-8 satellite [166]. The Crab Nebula polarization measurement performed by the polarimeter on board OSO-8 [169] at 2.6 keV (b) and 5.2 keV (c). The polarization degree and angle vector is shown with confidence contours at 67% and 99%.

## 3.4   The third season: satellites with optics, or when things went "wrong"

In the end, only Bragg based experiments were successful, but the X-ray sky was found to be less polarized than anticipated and, as a consequence, the measurement of the polarization more difficult than expected. Nonetheless, after the OSO-8 breakthrough, several missions and instruments were envisioned to follow up these first discoveries. However, only two years after the result on Crab and when OSO-8 was still active, the HEAO-2 satellite (better known as Einstein) was launched [18]. The passage from instruments based on collimators to ones based on optics triggered a paradigm-shift, changing the concept itself of an X-ray source: for sounding rockets, or for UHURU, a source was an excess of counts, following the triangular profile of the area exposed to a direction of the sky by a slat collimator off-set with respect to the rotation axis. With Einstein, a source become a cluster of counts in the sky imaged from a position-sensitive detector in the focus of the optics. Rotation was mandatory in the first case, not needed or even detrimental in the second one, because adding rotating elements in spacecrafts is more technologically challenging and costly than non-moving parts. Moreover, the classical detectors where "non local", that is it was not possible to determine the position of the interaction inside them as with detectors placed in the focus of X-ray optics. Thus, the classical X-ray polarimeters based on diffraction and scattering could no longer be included in Einstein and then into the modern X-ray space telescopes. X-ray polarimetry ultimately didn't manage to make the transit into the "third season" of the Giacconi path.

### 3.4.1   Intermission: the Stellar X-ray Polarimeter (SXRP)

A last attempt was however performed with the Stellar X-Ray Polarimeter (SXRP) on board of the Soviet Spectrum-X-Gamma (SRG) mission, an ambitious project with an important international contribution to the scientific instrumentation. Indeed SXRP was provided by NASA and Italy, and included both a Bragg polarimeter and a Thomson passive polarimeter [76] designed to work in the $2-15$ keV band. SRG had two large area X-ray telescopes with sliding devices in the focus, allowing to swap different instruments, the X-ray polarimeters among them. The Bragg polarimeter, made of a round flat pyrolitic graphite crystal at $45°$ diffracting photons at 2.6 and 5.2 keV toward a thin window detector, would produce a real image on a secondary focus, whose brightness was modulated through the rotation of the whole equipment around the optical axis of the telescope. From a technical point of view, SXRP was the best that could be done with conventional techniques in the focus of a large telescope. SXRP was designed [76], built, passed all the acceptance tests, and calibrated [136, 152]. However, with the fall of Soviet Union, SRG was never launched. Attempts to employ detectors non calibrated for polarization serving as Compton polarimeters at soft-$\gamma$ energies have been done, such as INTEGRAL/IBIS [29] that reported, for a few bright sources (Crab Nebula [52], Cygnus X-1 [74], and $\gamma$-ray bursts [61, 62]) linear polarization detection. However, none of these results are completely unambiguous. In the end, no X-ray polarimeter was to fly in the '80, '90 and '00...

## 3.5 A new Spring: photoelectric polarimetry with the Gas Pixel Detector

Finally, at the beginning of the 2000's, this "dark age" ended, with the development of photoelectric polarimeters such as the Gas Pixel Detector (GPD) in Italy [13, 14, 15, 36] and the Time Projection Chamber (TPC) in the USA [16].

Because it is employed in the mission that is at the center of this thesis, I focus my attention on the GPD: the GPD is the result of a collaboration between INFN-Pisa and INAF-IAPS in Rome that exploits the photoelectric effect, that is the dominant interaction process in $2-8$ keV "classical" X-ray energy band. In the following sections the physical principles of the functioning of the GPD will be briefly described.

### 3.5.1 Photoelectric polarimetry

During a photoelectric interaction, a photon is absorbed by an atomic electron that is then ejected from an inner shell and called "photoelectron".

For the process to take place, the energy $E$ of the absorbed photon must be greater than the binding energy $I$ needed to extract the electron that is emitted with a kinetic energy $E_e = E - I$. For the K-shell the photoelectric cross section is [66]:

$$\sigma_{ph}^K = 4\sqrt{2}\frac{8\pi r_0^2}{3}\alpha\Big(\frac{E}{mc^2}\Big)^{-7/2} \tag{3.5}$$

with $\alpha$ fine structure constant and $r_0$ classical electron radius. This formula in general holds for s-shell photoelectrons, so for sure for the K-shell, but also for other shells when the photoelectron is in this state. The emission directions of p-shell photoelectrons are also modulated by photon polarization but at a lower extent.

Photoelectric absorption strongly depends on the atomic number of the element, but even for low $Z$ photoelectric absorption remains the most probable interaction between X-ray radiation and matter up to $\sim$10 keV. This is shown in Fig. 3.7 that illustrates the mass attenuation coefficient of Neon (Z=10) as a function of energy and hence the most probable interaction process as its function: at energies >10 keV, the scattering processes start to compete with the photoelectric effect. The total cross section of the process is obtained by summing the contributions of each electron shell, but when the energy is sufficient to extract electrons from inner shells, their contribution become predominant with respect to that of outer shells. Indeed, above the binding energy, the photoabsorption cross section of each orbital quickly decreases with energy, $\sigma_{ph} \sim E^{-7/2}$ (see Equation 3.5) to the point that in the X-rays only the inner shells of the mirror of the X-ray optics are involved, because they reflect the highest energy photons. The emission of the photoelectron leaves a vacancy in an inner shell of the atom which is filled by an electron from an outer shell. Because the binding energy of the outer shell is almost negligible with respect to that of the inner one, the energy of the transition is almost equal to the binding energy of the emitted photoelectron. The energy release results in the production of an X-ray photon (fluorescence emission) or with the emission of a further electron, called Auger electron in at least half the cases for $Z \leq 30$. The photoelectron has a higher probability of being parallel to the electric field of the absorbed photon. Hence, the initial emission direction of the photoelectron brings memory of the linear polarization of the incident radiation. For a linearly polarized photon, in the non-relativistic case, the photoelectron angular distribution is given by [66]:

$$\frac{d\sigma_{ph}^K}{d\Omega} = r_0^2 \alpha^4 Z^5 \left(\frac{E}{m_e c^2}\right)^{-7/2} \cdot \frac{4\sqrt{2} \sin^2\theta \cos^2\phi}{(1 + \beta \cos\theta)^4} \tag{3.6}$$

where $\beta$ is the photoelectron velocity in unit of $c$, $\theta$ and $\phi$ are the angles that identify the photoelectron direction of emission with respect to the absorbed photon and its electric field (see Fig. 3.7 (a)). If incident photons are linearly polarized, the distribution of the photoelectrons emitted per azimuthal angle is modulated as a $cos(2\phi)$ function as in Fig. 2.3. The amplitude of the modulation ($B$ in Fig. 2.3) is proportional to the polarization degree, while the peak of the distribution corresponds to the polarization direction (that is the polarization angle). In the ideal case of 100% polarized photons, the directions of emission are completely modulated with a $\cos^2$ function of the azimuth angle and then photoelectric absorption from K-shell can be considered a perfect polarization analyzer.

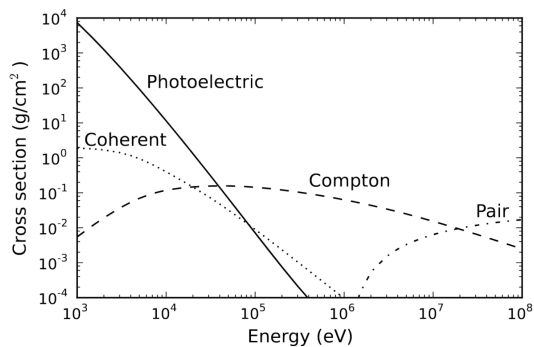

**Figure 3.7.** Dominant photon interaction as a function energy for Neon (adapted from Kaaret [77]).

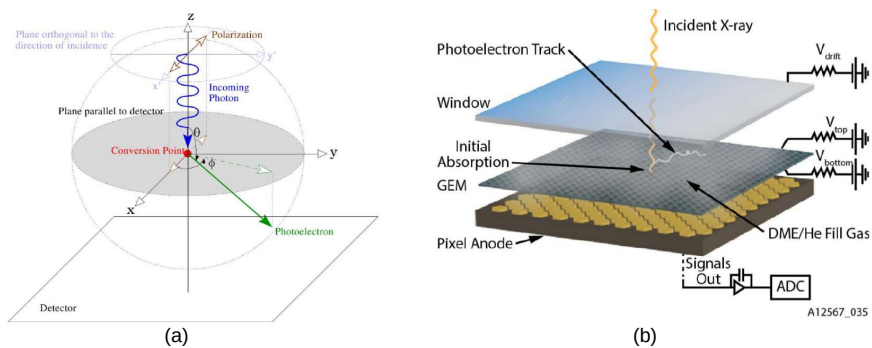

**Figure 3.8.** (a) Definition of the angles of emission of the photoelectron. (b) Exploded view of a Gas Pixel Detector. The volume of the gas cell is divided into two parts: the upper absorption gap, between the drift plane (which is also the entrance window) and the GEM top, and the lower transfer gap, between the GEM bottom and the readout ASIC [36].

### 3.5.2 The GPD

The GPD (a cut-out view of which is shown in Fig. 3.8 (b)) exploits the effect described in the previous section: a photon crosses a beryllium window, enters an active gas volume, is absorbed in the gas, and causes the emission of a photoelectron with more probability in the direction of its electric field. As the photoelectron moves in the gas looses its energy ionizing the gas and produces a track. The active gas cell in which the photoelectric interaction takes place is a 1 cm gap filled with pure dimethyl-ether (DME, $(CH_3)_2O$) at $\sim$800 bar pressure, that allows for a small diffusion (and hence limited track blurring), and a trade-off between quantum efficiency and modulation factor in the $2-8$ keV energy range. Under the action of an electric field parallel to the optical axis, the primary ionization electrons generated by the photoelectron drift toward the Gas Electon Multiplier (GEM), which is a dielectric foil with 9 $\mu$m copper metallization on both sides, and perforated by microscopic holes (30 $\mu$m diameter, 50 $\mu$m pitch). The differential voltage applied on the GEM induces electron multiplication. The GEM hence provides the gain while preserving the track shape. Indeed, the charge generated in the avalanche is collected by the pixellated anode plane that is the upper layer of an ASIC (Application Specific Integrated Circuit) CMOS (Complementary Metal-Oxide Semiconductor) chip (15×15 mm$^2$, 105600 hexagonal pixels with a 50 $\mu$m pitch). Each pixel is connected to an underlying electronics chain that includes a signal pre-processing function for the automatic localization of the event coordinates [14]. The polarization information is derived directly on a statistical basis from the angular distribution of the emission direction of the tracks produced by the photoelectrons, reconstructed by imaging the track projections onto the readout plane. The direction of the incoming photons is parallel to the drift direction and the impact point is derived with high precision as shown in Fig. 3.9. The GPD preserves the azimuthal symmetry in the response, allowing finally for a non-rotating device that can be put at the focus of X-ray telescope optics. Indeed, the GPD in addition to being a polarimeter has very good imaging capabilities, limited only by the Point Spread Function (PSF) of the optics [45, 139].

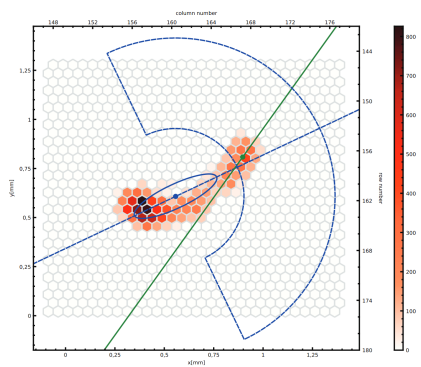

**Figure 3.9.** Real photoelectron track at 5.9 keV with reconstructed direction of emission (green solid line) and absorption point (green dot) [140]. The blue dashed line is the first-step direction estimation based on the barycenter of the track (blue dot) from which the actual absorption point and emission direction is evaluated inside the blue half-circled area.

### 3.5.3    The new X-ray polarimetry missions

The development of the GPD brought a renown interest in X-ray Polarimetry, finally allowing for the transit to the "third season" and the development of new missions. In the last decade and a half, 9 final-design GPDs have been assembled and tested, collecting more than ∼25 years of data, bringing the technology readiness level to flight standards [8]. Indeed in the last 15 years the GPD has been proposed as a detector for many X-ray astronomy missions. A mission called POLARIX was selected by the Italian Space Agency, but the whole program of small scientific missions was unfortunately canceled. Also the X-ray Evolving Universe Spectroscopy (XEUS) was developed until its scope was reduced into the polarimeter-less International X-ray Observatory (IXO, then rechristened ATHENA). The Gravity and Extreme Magnetism (GEMS) Small Explorer mission based on the TPC detector [149] was selected by NASA in 2009. Differently from the GPD, the TPC is a non-imaging detector with large quantum efficiency, that however need rotation like the missions of old due to spurious effects, intrinsic to the readout methodology. GEMS was canceled in 2012 because of cost overruns and schedule problems. GEMS acted again as a show-stopper for any other projects of X-ray Polarimetry until 2015, when an ESA mission, the X-ray Imaging Polarimetry Explorer [XIPE 138] was selected for competitive phase A study. Eventually, XIPE was not approved for flight, because in the meantime, in January 2017 the Imaging X-ray Polarimetry Explorer [IXPE, 172, 173], already proposed in 2008, was selected for development as a NASA SMEX mission. IXPE being the first mission of its kind, and performing spatially resolved X-ray polarimetry in the 2−8 keV band with the GPD, will be for X-ray polarimetry what Einstein has been for X-ray astronomy. It's success will enable larger "fourth season-class" missions, that indeed are already being developed.

The Chinese-European enhanced X-ray Timing and Polarimetry mission [eXTP 181] will have on board four GPDs and is expected to launch in 2028. Even more ambitious missions are being proposed, such as the X-ray Polarimetry Probe [XPP 73], and the Next Generation X-ray Polarimeter [NGXP 141]. Also new Compton-based balloon borne hard X-ray polarimeters, such as X-Calibur [12] and XL-Calibur [2] have been developed following the renewed interest in X-ray polarimetry. A GPD, was actually flown on board a CubeSat [Polarlight 47] and is currently collecting scientific data from the Crab Nebula, possibly detecting a time variation of polarization of the pulsar [48], and other bright sources.

IXPE successfully launched on December 9th 2021, and the window of X-ray polarimetry is being opened again, becoming a mature field and a fundamental part of the X-ray astronomy.

# Chapter 4

# X-rays polarization processes in extended sources

IXPE will be the first imaging X-ray polarimetry to be flown, ever. As such, a handful, but potentially very rewarding, observations will be dedicated to extended source characterized by an emission from spatial structures larger than the typical angular resolution of its X-ray telescopes. This thesis deals with such structures. This Chapter serves as an introduction to the sources whose IXPE observation I simulate in the following Chapters. I focus on the processes that are relevant for the X-ray polarimetry of Supernova Remnants (SNRs, Section 4.1), and molecular clouds (MC) in the Galactic center (GC) (Section 4.2). Parts of Section 4.2 are based on my works published in Di Gesu, L. et al. [38], Ferrazzoli, R. et al. [51].

## 4.1 X-ray polarization processes in Supernova Remnants

Supernova remnants (SNRs) are the result of the violent death of a star. Young ($\lesssim$500 years old) SNRs are relativistic particle accelerators, and are the most likely sources of Galactic cosmic rays (CRs). The first observational evidence of this theory was provided by the discovery that their radio emission was due to synchrotron radiation [133]. The acceleration mechanism is generally accepted to be diffusive shock acceleration [DSA, 88, for a review on the subject]: according to the DSA collision-less shocks, produced by supersonic ejecta, expand into the interstellar medium, simultaneously accelerating charged particles by repeatedly reflecting them on magnetic mirrors, hence amplifying the local turbulent magnetic field.

However, while the radio synchrotron emission is produced by electrons accelerated to "only" a few GeV energies, the Galactic CR spectrum extends up to TeV energies, up to the so called "knee" at 3 TeV. Magnetic fields much stronger and turbulent than the average Galactic magnetic field ($\sim$5 $\mu$G) are needed to produce the electrons responsible for Galactic CRs. Where can we find magnetic fields of the needed strength? Chandra X-ray observations found thin rims, filaments and other magnetic structures in the outer shells of many young SNRs such as SN1006 [87], Cassiopeia A [161], and Tycho [71]. Their continuum is considered to be mostly non-thermal and due to synchrotron emission [71, 87, 161]. The X-ray spectrum of Tycho exhibits in the soft ($2-4$ keV) band many line emission from Mg, Si, S, Ar, Ca, due to a multi-temperature plasma, and a Fe line at 6.4 keV. In the $4-6$ keV band, a line-less continuum, mostly non-thermal in origin, traces the synchrotron emission. The Tycho $4-6$ keV continuum emission is shown in Fig. 4.1 to highlight the non-thermal structures. These X-ray synchrotron emitting regions can be very narrow, with their flux sharply rising and falling on spatial scales of a few arcseconds. Their width is a measure of the average strength of the magnetic field: the width is determined by the combination of synchrotron energy losses and transport of electrons downstream to the shock, and the timescale of these energy losses is inversely proportional to the magnetic field strength [see e.g. 160]. Typical SNR magnetic field values range between $100-500$ $\mu$G [see 125, for recent estimates]. The structures might be due to the geometric projection of the thin regions accelerating the TeV electrons, with alternative scenario arising from rapid decay of amplified magnetic fields in the vicinity of the shock wave [112]. In any case, these observations support the model [122] that in SNRs, electrons can be accelerated by fast (supersonic, $> 3000 km/s$) shocks up to TeV energies and can efficiently emit synchrotron radiation up to X-ray energies in the amplified SNR magnetic fields. Much of their kinetic energy hence is deposited in the CRs. At such energies, the time scale of radiation loss is very short, implying that X-ray synchrotron emission can only occur in regions of active particle acceleration, such as the SNR shock fronts, where magnetic fields must be strong and turbulent. Indeed, according to the DSA theory, the CR themselves, as they diffuse ahead of the shock front, are responsible for the creation of the magnetic field turbulence. While this process has received substantial theoretical attention, the observational data are still poor. X-ray polarimetry would enable us to probe the magnetic field structure and provide constraints on models of diffusive shock acceleration with efficient magnetic field amplification. According to radio polarimetric observations, young SNRs are usually characterized by radial magnetic fields [e.g. 4, 86], while older ($\geq$2000 years) exhibit a tangential magnetic field structures [e.g. 123, 124]. The tangential magnetic fields of old SNR can be explained by the shock wave compression of the magnetic field component perpendicular to the shock normal. On the other hand, the origin of the radial magnetic field of young SNRs is less understood. Zirakashvili and Ptuskin [184] argue for magnetic field amplification in the CR precursors of SNR, while West et al. [174] suggest a selection effect due to the location where relativistic electrons pile-up downstream of the shock. This is where X-ray polarization comes into play. Spatially-resolved X-ray polarimetry would allow for measurement close to the shock, establishing whether the radial structure of the magnetic field is already present there, or if it is caused by processes further away from the shock [162]. It could also separate the thermalized plasma from non-thermal (synchrotron) components, locating the regions of ordered magnetic field on the site of shock acceleration, and thus explore the turbulence level of the magnetic field.

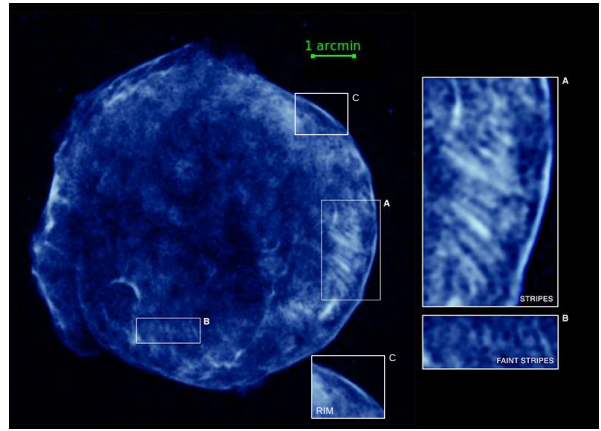

**Figure 4.1.** Chandra image of the Tycho SNR in the 4.1-6 keV energy band, adapted from Eriksen et al. [44]. Highlighted are some of the most prominent synchrotron structures: the more bright (A) and faint (B) "stripes", and the rim.

What levels of X-ray polarization can we expect in SNR? In the radio (e.g. at 2.8 cm), because of depolarization due to the combined effects of random magnetic field orientations along the line of sight and Faraday rotation [e.g. 86, 144], polarization of young SNRs is observed to be lower than old SNRs [40], potentially making the detection of X-ray polarization challenging. However, in X-rays the rapid energy loss of TeV electrons means they are confined to a thin shell behind the remnant shock front. In this case, contrary to the radio observations, the Faraday de-polarization effects are negligible in X-rays. Moreover, the turbulent magnetic fields that reduce the average polarization can actually result in highly polarized patchy structures potentially observable in high resolution images at X-rays [23, 25], offering the opportunity to study magnetic field orientations and turbulence closer to the sites where particle acceleration takes place with respect to radio observations. So, the X-ray polarization fraction may be actually higher in X-ray than in radio. The maximum polarization fraction depends on the spectral index of the synchrotron emission. Synchrotron radiation is emitted when charged particles are accelerated in a magnetic field due to the Lorentz force and is intrinsically highly polarized [127] perpendicularly to the projection of the magnetic field on the sky. For a power law distribution of emitting particles in a perfectly ordered magnetic field the maximum degree of polarization $P_{max}$ depends on the spectral photon number index, $p$, and is given by the expression:

$$P_{max} = \frac{p+1}{p+\frac{7}{3}} \tag{4.1}$$

So, the steeper the spectral index, the higher the polarization fraction. The X-ray synchrotron spectral index is usually steeper than in the radio, since it is associated to photons with energies near the spectral cut-off. In the radio young SNR have typically $p \sim 0.6$, corresponding to $P_{max} \approx 54.5\%$, whereas in X-rays $p \sim 3$, corresponding to $P_{max} \approx 75\%$. On the other hand, there are effects that conspire to reduce the expected X-ray polarization degree. In the particular case of the SNR Cas A and Tycho, an important fraction of their X-ray emission comes from an unpolarized thermal emission from a multi-temperature out-of-equilibrium plasma, as probed by the prominent emission lines at energies <4 keV [71, 161]. This has the effect of diluting the polarization degree in the energy bands where the emission lines are more prominent, e.g. in the 2−4 keV band, and at the 6.4 keV Fe K$_\alpha$ line. This suggests that for the aforementioned SNRs, the search for polarization in the line-less continuum in the $4-6$ keV energy band between the Calcium and Iron emission lines may be preferable.

Finally, the proximity of X-ray synchrotron radiation regions to the shock front may preferentially select regions with high magnetic-field turbulence, resulting in lower polarization fractions. The magnitude of the magnetic-field fluctuations and their size distributions have been described quantitatively in Baring [10], Bykov et al. [23], Bykov and Uvarov [24], Bykov et al. [25], who show that X-ray polarization is fundamental for the study the magnetic field fluctuations, and then for a proper understanding of DSA in young SNRs.

Bykov et al. [23], Bykov and Uvarov [24] and [25] presented a study of the flux and polarization degree variations in turbulent field models of synchrotron-emitting remnants, outlining how the imaging capabilities provided by IXPE enables to sample various angular scales and provide information on field turbulence.

In Fig. 4.2 is shown the result from simulations performed by Bykov and Uvarov [24] for a DSA model that accounts for magnetic field amplification from a CR current driven instability. The resulting image (left panel) is strikingly similar to the stripe-like structures observed in Tycho (see Fig. 4.1) and the expected polarization degree (right panel) could be as high as ∼50%, mainly because of the peaked structure of the spatial spectrum, and because of the steep distribution of the synchrotron emitting electrons. Baring [10] presented a model for X-ray polarization signatures from energetic electrons moving in simulated Magneto-Hydro-Dynamic (MHD) turbulence of different levels of magnetic field randomization.

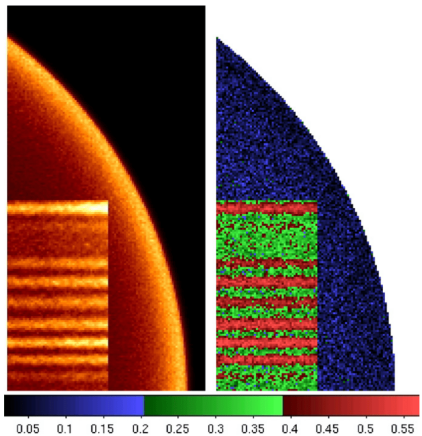

**Figure 4.2.** From Bykov and Uvarov [24], SNR synchrotron image simulated in the DSA model: left panel: synchrotron X-ray intensity at 5 keV; right panel: degree of polarization of the X-ray emission with values corresponding to the color bar.

I summarize here his findings, as they will be useful for the simulations I present in Chapter 8.

Considering a MHD turbulence with Kolmogorov [81] power spectrum of the fluctuations, the wave variance (that is, the average of the amplitude of the magnetic field fluctuations $\delta B$ over the unperturbed magnetic field $B_0$, squared) is

$$\sigma^2 = \langle \left(\frac{\delta B}{B_0}\right)^2 \rangle \tag{4.2}$$

For a uniform, non-turbulent field, $\sigma^2 = 0$. For a non-zero turbulence, with respect to the maximum theoretical value of the synchrotron polarization degree given by Eq. 4.1 the polarization degree $P(\sigma)$ is reduced to the value

$$P(\sigma) \simeq \frac{3}{4} - \frac{3}{2}\sigma^2 \quad . \tag{4.3}$$

In Chapter 8 I will present a realistic and up-to-date simulation of a 1 Ms IXPE observation of Tycho that will show that it is possible to determine the polarization degree produced by different turbulence levels (in Table 4.1 are shown some expected variance $\sigma^2$ values and average turbulent magnetic field fraction corresponding to different maximum synchrotron polarization obtained using Eq. 4.3), and magnetic field topology models.

**Table 4.1.** Expected variance $\sigma^2$ and average turbulent magnetic field fraction values corresponding to different maximum synchrotron polarization.

| $P_{max}$ (%) | $\sigma^2$ | $\langle \delta B/B_0 \rangle$ |
|---|---|---|
| 50 | 0.16 | 0.4 |
| 25 | 0.33 | 0.57 |
| 10 | 0.43 | 0.66 |

## 4.2 X-ray polarization processes in the Galactic center molecular clouds

Scattering of light by particles (e.g. electrons, atoms, molecules, dust grains, etc.) is one of the most efficient way to create or alter the polarization state of a photon. Indeed X-ray polarization induced by scattering of radiation in non-spherical geometries, such as disks, blobs and columns able to polarize radiation has been expected for quite a while [95, 130, 146].

In Chapter 2 I described the scattering processes from the point of view of the detector technology. Here I focus on X-ray polarization induced by Thomson scattering, considering a free, non-relativistic scattering volume (the simplest case being just an electron) at rest and an incoming electromagnetic wave with $h\nu \ll m_e c^2$. The geometry of the interaction is shown in Fig. 4.3. Let's assume that the incident wave is completely unpolarized: in this case, the incoming radiation can be considered to be the sum of two orthogonal completely linearly polarized waves, and then the associated scattering patterns can be summed. Since the choice of the orientation of these polarization is arbitrary, it is convenient to chose one of them as the one defined by the incident and scattered directions, and the other one perpendicular to this plane. The scattering can be then considered as the sum of two independent scattering processes, one with emission angle $\Theta$, the other with $\pi/2$. The scattering angle (i.e. the angle between the scattered wave and the incident

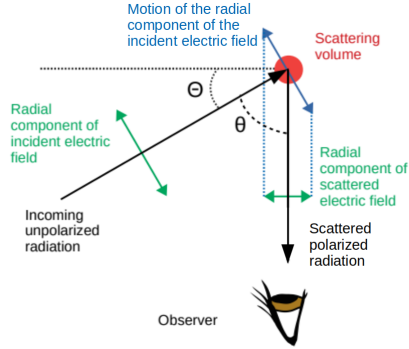

**Figure 4.3.** Geometry of a generic scattering process polarizing a beam of unpolarized photons interacting with a scattering volume. $\theta$ is the scattering angle, $\Theta$ the emission angle.

wave) is $\theta = \pi/2 - \Theta$, so we have

$$
\begin{aligned}
\left(\frac{d\sigma}{d\Omega}\right)_{unpol} &= \frac{1}{2}\Big[\Big(\frac{d\sigma(\Theta)}{d\Omega}\Big)_{pol} + \Big(\frac{d\sigma(\pi/2)}{d\Omega}\Big)_{pol}\Big] \\
&= \frac{1}{2}r_0^2(1+\sin^2\Theta) \\
&= \frac{1}{2}r_0^2(1+\cos^2\theta) \quad.
\end{aligned}
\tag{4.4}
$$

The cross-section of the interaction depends only on the scattering angle $\theta$, with $r_0$ being the classical electron radius. The difference between the two terms of the right-hand-side of Eq. 4.4 is then associated to the introduction of polarization by the scattering process, which is

$$
P = \frac{1-\cos^2\theta}{1+\cos^2\theta} \quad.
\tag{4.5}
$$

The induced polarization can range from 100% if the scattering angle is 90°, to zero for 0° and 180°.

An interesting application of scattering-induced polarization can be found in our Galactic Center (GC) region.

The GC, here intended as the central degree of the Milky Way, is a rich region of our Galaxy that hosts giant molecular and atomic clouds, star clusters, active binary systems with compact objects, and dynamic structures [see 114, for a review]. Its most important resident is, however, the supermassive black hole (SMBH) Sgr A*. Sgr A* is about $\sim 8$ kpc from the Earth and is the closest known SMBH, with a mass $\sim 4$ million times the solar one [64]. For a SMBH of its mass, Sgr A* appears to be dimmer than expected, with a luminosity of $\sim 10^{33}$ erg s$^{-1}$ (like the luminosity of our Sun!) and accreting at a quiescent level of $\sim 10^{-11}$ the Eddington luminosity [7]. However, past and more luminous phases of Sgr A* [see, again 114, for a review on the topic] could explain some observed phenomena in the GC region. Determining the luminosity history of Sgr A* would be of great interest for our understanding of the duty cycle of mass accretion onto SMBH, which is thought to drive the co-evolution of SMBH and galaxies, as they appear to share scaling relations and coupled growth [39]. For instance the gamma and X-ray bubbles that Fermi-LAT and eROSITA observed 10 kpc both below and above the Galactic plane are indicative of an Active Galactic Nucleus (AGN) phase of Sgr A* some million years ago [115, 145, 185].

However, there are hints of an even more recent activity. Within $\sim 100$ pc from Sgr A*, in the so called Central Molecular Zone [CMZ, 100], several molecular clouds (MC) are observed, for instance, in the thermal far infrared images obtained with the Herschel satellite [97]. Interestingly, the physical conditions in the CMZ inferred from infrared observations (i.e., the geometrical size, column density and gas dynamics) are reminiscent of an AGN torus [116]. Historically established MC complexes are the Sgr A, Sgr B, and Sgr C ones, shown in Fig. 4.4. These MC are

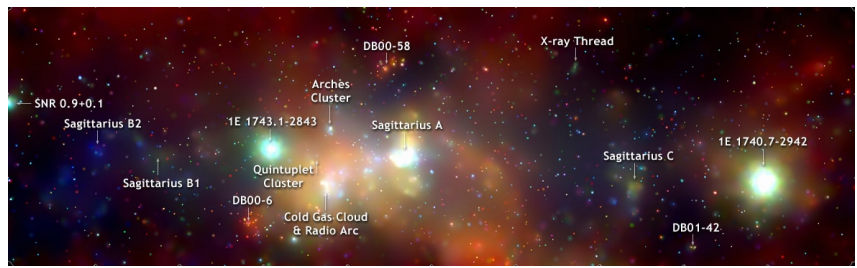

**Figure 4.4.** RGB Chandra image of the central 100pc of the Milky Way. In red $1 - 3$ keV, in green $3 - 5$ keV, in blue $5 - 8$ keV. Labeled are the X-ray reflection nebulae Sgr A, B and C and other prominent X-ray sources.

also found to be X-ray emitters, their spectra being characteristic of X-ray reflection by cold gas illuminated by an external source. Indeed, the clouds display X-ray reflection spectral features like a steep continuum plus a Fe $K_\alpha$, emission line whose brightness and morphology are variable in time (see Fig. 4.5). Because, no possible

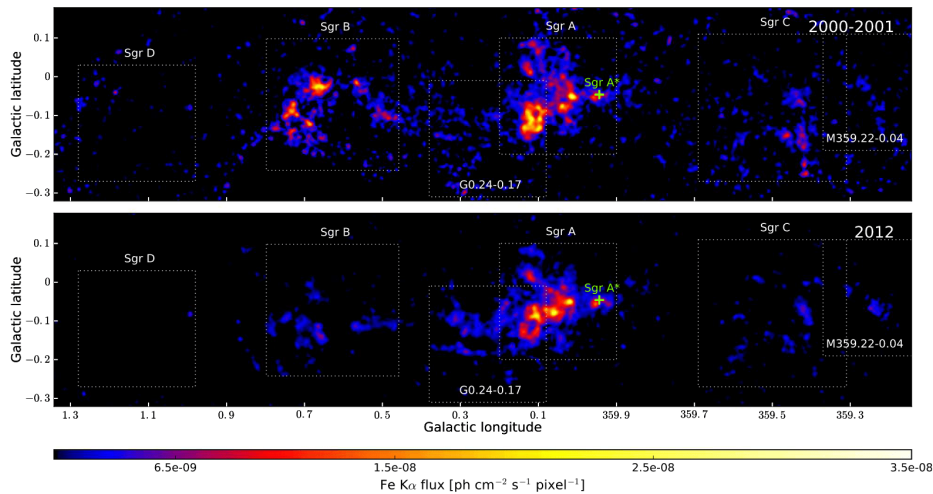

**Figure 4.5.** From Terrier et al. [150]: Fe $K_\alpha$ intensity maps (background and continuum subtracted) of the inner GC region measured by XMM-Newton at in 2000–2001 (top) and 2012 (bottom). It can be appreciated how on the timescale of years some molecular cloud complexes, such as the Sgr B one, almost disappeared.

X-ray illuminating source bright enough to explain their emission is present nearby, Sunyaev et al. [148] first suggested that the observed X-ray reflection spectrum from the MC is the echo of a past outburst of Sgr A*, delayed by the light travel time across the CMZ. Murakami et al. [106] estimated a peak luminosity of the external source of $\sim 10^{39}$ $(D/100pc)^2$ erg s$^{-1}$, where $D$ is the cloud-source distance. If true, this means that in the recent past (on the timescale of just $\sim$300 years!) our GC reached a $2-10$ keV luminosity of $5 \times 10^{39}$ erg s$^{-1}$, about $10^6$ times the current value and similar to that of a low luminosity AGN, and we can reconstruct the history of its energy release using the MC. In the last 30 years, X-ray spectral [e.g. 27, 30, 84, 105, 129, 148, 163] and timing [e.g. 34, 72, 104, 113, 150] studies of the MC have provided evidence of past single [113] or multiple [34, 150] outburst of Sgr A*.

There are alternative theories for other sources of illumination, such as X-ray transients that power different molecular complexes [91], or the interaction of low energy cosmic ray electrons with the molecular gas [41, 156, 180]. While the former mechanism cannot be excluded, it requires a population of sufficiently powerful transients to be located close to each molecular complex. As for the latter hypothesis, because the MC fluxes vary on time scales of $5-10$ years [e.g. 150], although it is unlikely that cosmic ray electrons could be responsible for the bulk of the variable part of the MC emission, they are not conclusively ruled out by current data [99, 182]. Still, it is possible that both alternative mechanisms could contribute to some extent to the observed X-ray reflection signature.

Despite many observational efforts, it is still difficult to unambiguously derive the past light curve of Sgr A* from the X-ray variability of the MC. This is mainly because the distance $\vec{d}_{los}$ of the clouds along the line-of-sight is loosely constrained [e.g., 27, 30, 119, 163], which, in turn, makes it challenging to accurately infer the light travel time $t_{light}$. The geometry of the MC-SgrA* system is sketched in Fig. 4.6, where $\vec{d}_{proj}$ is the the Sgr A*-cloud distance projected on the plane of the sky, $c$ is the speed of light and $\theta$ is the scattering angle.

A possible way to overcome the difficulty in determining $\vec{d}_{los}$ is provided by X-ray polarimetry. If the MC were illuminated by an external compact source like Sgr A*, the reflected X-ray radiation would be highly linearly polarized by scattering. The expected polarization degree $P$ depends on the scattering angle $\theta$ as in Eq. 4.5. In turn, the scattering angle is related to $\vec{d}_{los}$ by:

$$\vec{d}_{los} = \vec{d}_{proj} \cot \theta \quad . \qquad (4.6)$$

In this scenario, the polarization degree is 100% for a cloud located in the Sgr A* plane ($\vec{d}_{los} = 0$ pc, $\theta = 90°$), while the direction to the external illuminating source is perpendicular to the polarization direction [155]. Therefore, detecting the polarization degree of the molecular clouds would identify the external illuminating source and produce a map of the molecular gas in the GC region in three dimensions. From Eq. 4.5 and 4.6 there is an ambiguity on the sign of $\vec{d}_{los}$, that could be however solved by infrared (reddening) observations [e.g. 186], or by spectroscopic means studying the shape of the reflection continuum [31].

Thanks to the launch of IXPE, it is now possible, for the first time, to employ spatially resolved X-ray polarimetry to address the Sgr A* past outburst hypothesis

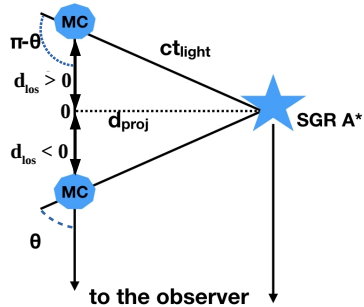

**Figure 4.6.** Scattering geometry for a MC located in front or behind the Sgr A* plane (Adapted from Ferrazzoli, R. et al. [51]). The two positions depicted have scattering angles $\theta$ and $\pi - \theta$ which results in the same polarization degree. In addition, $\vec{d}_{proj}$ is the cloud-Sgr A* distance projected in the plane of the sky, $\vec{d}_{los}$ is the line-of-sight distance of the cloud with respect to the Sgr A* plane, $c$ is the speed of light, and $t_{light}$ is light travel time between Sgr A* and the cloud. The vector $\vec{d}_{los}$ assumes negative values if the cloud is in front of the Sgr A* plane and positive if behind.

in an independent way.

In the years preceding the launch of IXPE, many X-ray polarimeters where being proposed, and the prospect of having soon an X-ray polarimeter was concrete. This led to a renewed interest in the modeling of the X-ray polarization properties in the GC region [31, 32, 33, 78] and in the evaluating the detectability of candidate molecular cloud targets [89, 90].

Churazov et al. [31] were the first to argue that X-ray polarimetry of the MC in the GC would have been the most convincing test of the origin of their illumination. The continuum emission from the scattering process, is indeed expected to be almost completely polarized, with a slow decrease with energy due to the increasing depolarization contribution from multiple scatterings: at higher energies the probability of photoelectric absorption decreases, while the scattering cross section remains almost constant in the $2-8$ keV energy range, so that the ratio between absorption and scattering decreases.

Marin et al. [90] simulated the expected polarization for the major MC complexes in the CMZ and produced the polarization map (shown in Fig. 4.7) that IXPE would obtain (at the time IXPE was still in the proposal stage). The model by Marin et al. [90] shows that a variety of polarization signatures, ranging from nearly unpolarized to highly polarized (up to ∼77%) could be expected.

More recently, Khabibullin et al. [78] suggested the possibility that an X-ray polarimetry observation of the MC, in addition to determine if Sgr A* was indeed the illuminating source, could help to infer the eventual intrinsic polarization properties of the flare emission.

A physical limit of the proposed X-ray polarimetry observations in the CMZ region is the fact that the MC are embedded in the diffuse, unpolarized emission of the GC [82, 134]. Besides the X-ray reflection from the molecular clouds, the $2-8$ keV emission in the GC region comprises the contribution of two diffuse emission compo-nents [see 114, and references therein] that hereafter we label as "soft-plasma" and "hard-plasma". The soft-plasma is traced by e.g. SiXII, SiXIII, SXV, and ArXVII lines. These are ascribed to a ∼1 keV collisionally-ionized plasma that pervades the GC and can be sustained by the supernova activity in the region. Conversely, the hard-plasma is traced by a FeXXV-Heα line emission at ∼ 6.7 keV that is morphologically peaked in the central degree. This component is often modeled as ∼6.5 keV thermal plasma. At least a part of it may be ascribed to unresolved point sources like accreting white dwarf and coronally active stars [120, 179]. The remaining emission might be associated with truly diffuse hot gas, possibly originating from supernova remnants.

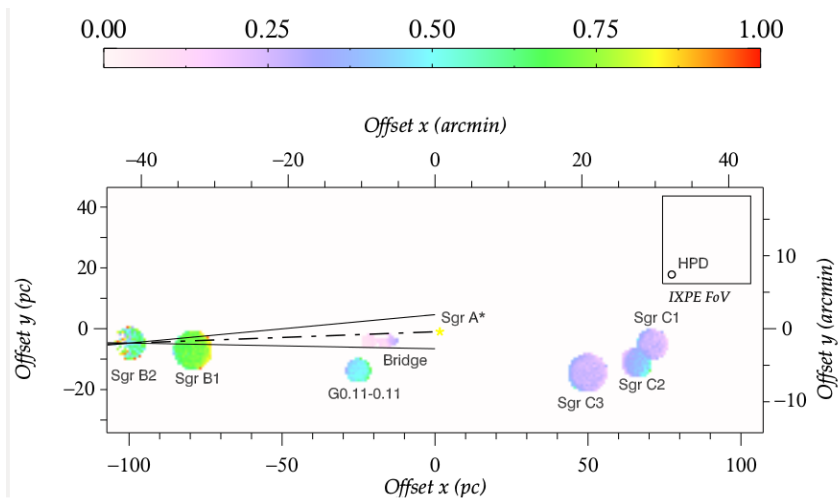

**Figure 4.7.** Polarization map of the GC showing how the angle of polarization would constrain the position of the illuminating source [adapted from 90]. The colorbar shows the polarized fraction. The IXPE field of view (FoV) is indicated with a box, and a yellow star indicates the position of Sgr A*.

The effect of this diffuse, thermal, unpolarized emission, is the heavy dilution of the (theoretically) high polarization degree of the MC.

Because of the complexity of the diffuse emission in the GC region, the synergy between polarimetric and imaging capabilities is a crucial asset for this study because it allows to resolve the faint MC from the diffuse emission in the background.

For this reason, in Di Gesu, L. et al. [38] we set up a method to perform realistic simulations of IXPE observations of the MC and of their environment considering the polarimetric, spatial, and spectral properties of all the components that contribute to the X-ray emission in the GC region. This work is described in detail in Chapter 9. Finally in Ferrazzoli, R. et al. [51] we presented data analysis techniques for the polarimetric mapping of the Sgr A molecular cloud complex and derived the effect of observing the clouds off-axis in the IXPE field of view.

I will present our results in details in Chapter 10.

# Chapter 5

# The Imaging X-ray Polarimetry Explorer

In this Chapter I will describe the IXPE mission: the first part of the Chapter (Section 5.1) is partially based on the work I published in Ferrazzoli and IXPE Collaboration [49] and will present the mission characteristics and my role in it.
In Section 5.2, I will briefly describe the sources of background IXPE will be subjected to. In Section 5.3, I will outline the techniques developed to analyze the IXPE data. Finally in Section 5.4, I will present the IXPE observation simulator "ixpeobssim".

## 5.1  Mission characteristics

As explained in Chapter 2, the technological maturity reached by the Gas Pixel Detector polarimeters has allowed the development of the space mission entirely dedicated to X-ray polarimetry: the Imaging X-ray Polarimeter Explorer [IXPE, 172]. IXPE will have a sensitivity two orders of magnitude higher than OSO-8 and, at the same time, will add imaging capabilities in combination with simultaneous spectral and temporal measurements. Hence, IXPE is posed to fill the gap with the polarimetric information available in other wavelengths. IXPE successfully launched on December 9th on top of a SpaceX Falcon 9 rocket. IXPE is born as a collaboration between the NASA Marshall Space Flight Center (MSFC) and the Italian Space Agency (ASI), with Martin Weisskopf as PI of the mission. NASA manages the Mirror Unit design and fabrication, the Science Operation Center (SOC), including data analysis, and mirror and telescope calibrations. Italian responsibilities are divided between ASI, INAF-IAPS,INAF-OAC, INFN and OHB-I and include the instrument management, Detector Units and Detector Service Units manufacturing and calibration. Italy also provided the track reconstruction algorithm and the Malindi Ground Station as the primary contact point for command, telemetry and data downlink. Ball aerospace built the spacecraft, that is shown after the integration in Fig. 5.1. A network of international scientific collaborators, institutions and Universities (such as Università Roma Tre, Stanford University, and the Massachusetts Institute of Technology) provides the backbone of the Scientific Topical Working Groups (STWG). The STWG are tasked with the selection of the sources to be put in the IXPE observing program, as well as the development of the

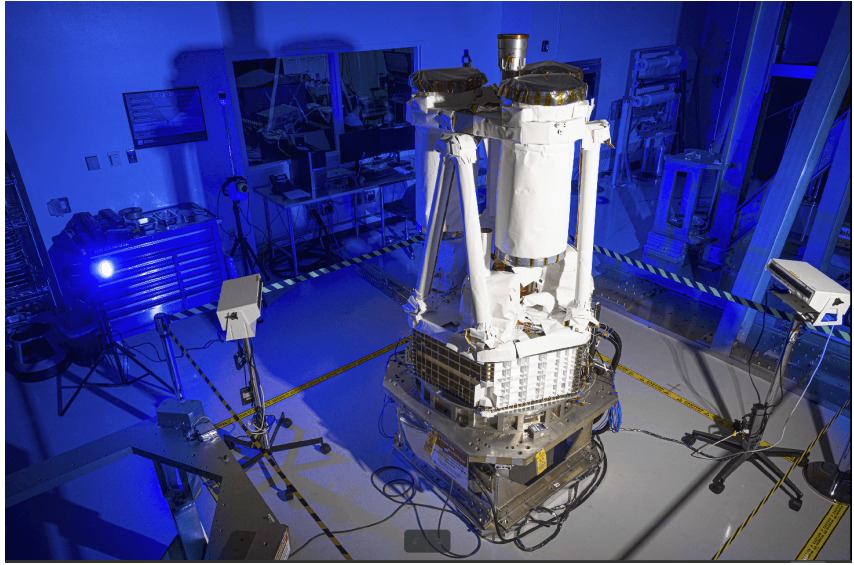

**Figure 5.1.** The IXPE spacecraft in stowed configuration after integration at Ball Aerospace.

theoretical predictions, analysis and interpretation of the data obtained during the observations. The institutions, Universities, and organization involved in the IXPE project are shown in Fig. 5.2. All scientific data will be publicly available through

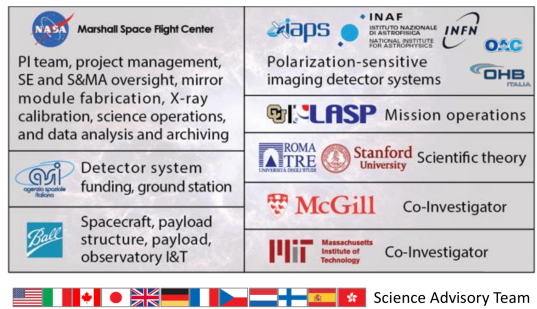

**Figure 5.2.** Institutions, Universities, and organization partners of the IXPE mission.

NASA's High-Energy Astrophysics Science Archive Research Center (HEASARC) one week after 90% of the data from an observation campaign has been collected. HEASARC will also manage the scientific data distribution and archiving. The Laboratory for Astronomy & Space Physics of Boulder will handle the Mission Operations. The baseline IXPE mission calls for two years of science operations, following a 1-month commissioning phase, but further extensions are possible since there are no consumables on board. IXPE was initially designed to be launched with a Pegasus XL airborne launcher (as seen on the left side of Fig. 5.3). Cost advantages allowed to move the launch to a SpaceX Falcon9 rocket, of which IXPE was the comfy sole payload (as can be appreciated from the right side of Fig. 5.3). IXPE was placed in a 600 km circular, equatorial orbit in order to minimize the passage over the South Atlantic Magnetic Anomaly and maximize the number of passages per day over the Malindi Ground Station.

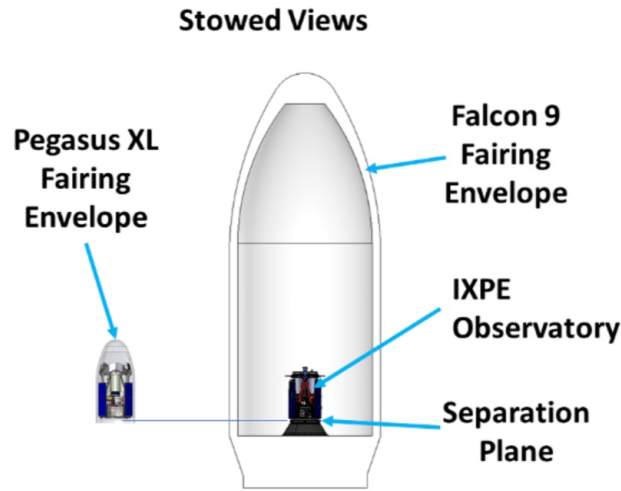

**Figure 5.3.** IXPE in stowed configuration inside the original launcher, Pegasus XL (left), and inside the payload faring of the Falcon 9.

In Fig. 5.4 are shown the Falcon 9 with IXPE on top (b) and the author on this thesis (b) the day of the launch at Cape Canaveral. A sketch of IXPE in its deployed configuration is shown in Fig. 5.5: it carries three grazing-incidence X-ray telescopes with identical Mirror Module Assemblies (MMA, [117]) at the end of a 4 meter long extensible boom, and three identical detector units with a GPD each. The detectors are clocked at a 120° angle with respect to each other. This disposition, once the images from the three detectors are rotated and summed, allows to reduce spurious effects and check for real astrophysical polarization. The polarization sensitivity is such to reach a Minimum Detectable Polarization at 99% confidence $< 5.5\%$ for a 0.5 mCrab source with a 10 day observation in $2-8$ keV energy band (1 mCrab = 2 $\times 10^{-11}$ erg s$^{-1}$cm$^2$ in $2-8$ keV). The spurious modulation for an unpolarized source is expected to be $< 0.3\%$ after calibration. The angular resolution is $\sim 25-30''$ over a $\sim 12.8' \times 12.8'$ overlapping field of view for the three detectors' polarization sensitive areas, enough to cover the vast majority of the extended sources that IXPE will observe. A characteristic of IXPE is the dithering during observations, i.e. the pointing is not "stared" but moves along a path that covers in time part of the detector: this has been introduced to reduce spurious effects and to avoid having to calibrate pixel by pixel, as it is done for example with Chandra and other observatories with imaging detectors. The energy band is the "classical" X-ray one of $2-8$ keV with an energy resolution of $\sim 20\%$ at 5.9 keV. This spectral resolution roughly scales as $\sim 1/\sqrt{energy}$, enabling energy-resolved polarimetry when statistics is large enough. A timing accuracy of $1-2$ $\mu$s can be reached using GPS-pulse-per-second signal and on-board clocks. This allows to perform time-resolved polarimetry in classes of sources such as accreting pulsars and binary systems.

A summary of IXPE scientific requirements is reported in Table 5.1.

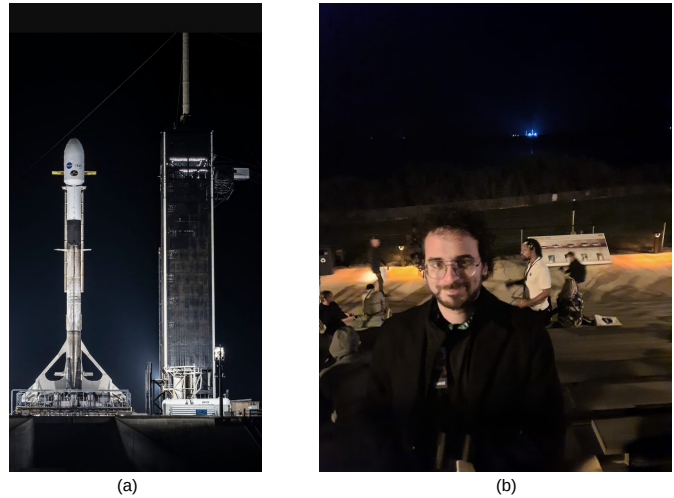

**Figure 5.4.** (a) The Falcon 9 booster B1061.5 carrying IXPE in its payload faring (credits: SpaceX); (b) the author of this thesis at the Banana Creek launch-viewing area minutes before the lift-off of IXPE: the pad is visible lightened-up in the background (credits: Valeria Cascone).

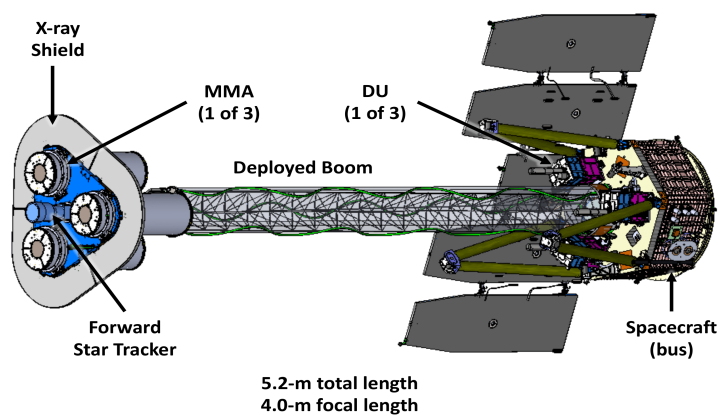

**Figure 5.5.** IXPE in deployed configuration.

**Table 5.1.** Summary of IXPE scientific requirements.

| Parameter | Observable | Property | Value |
|---|---|---|---|
| Launch date | - | - | November 2021 |
| Operational life | - | - | 2 years + 1 |
| Orbit | - | Equatorial, circular | 600 km |
| Polarization sensitivity | Degree $\Pi$, angle $\phi$ | $MDP_{99}$ for 0.5 mCrab in 10 days | $\leq 5.5\%$ |
| | | Systematic error in modulation | $\leq 0.3\%$ |
| | | Systematic error in polarization angle | $\leq 1°$ |
| Energy dependence | $F(E)$, $\Pi(E)$, $\phi(E)$ | Energy band $E_{min} - E_{max}$ | $2 - 8$ keV |
| | | Energy resolution $\Delta E \propto \sqrt{E}$ | $\leq 1.5$ keV |
| Spatial dependence | $F(x,y)$, $\Pi(x,y)$, $\phi(x,y)$ | Angular resolution | $\leq 30''$ |
| | | Field of view | $\geq 9'$ |
| Time dependence | $F(t)$, $\Pi(t)$, $\phi(t)$ | Time accuracy | $\leq 25$ $\mu s$ |
| Areal background rate | $R_B/A_{det}$ | $R_B/A_{det} \ll R_S/A_S$ per DU | $<$4E-3 s$^{-1}$ cm$^{-2}$ |

### 5.1.1 Observing plan

The main scientific cases that will be tackled by IXPE are: (1) Particle acceleration processes; (2) Radiation propagation in highly magnetized compact objects; (3) Scattering induced polarization; (4) Fundamental physics. IXPE will measure polarization in X-rays from neutron stars, from stellar-mass black holes and from Active Galactic Nuclei (AGN). For the brightest extended sources like Pulsar Wind Nebulae (PWNe), Super Nova Remnants (SNRs), large scale jets in AGN, and molecular cloud complexes in the Galactic center, IXPE will perform spatially resolved polarimetry. In some cases, IXPE will definitively answer important questions about source geometry and emission mechanisms or even exotic effects in extreme environments. In others, it will provide useful constraints including the possibility that no currently proposed model explains the data. In Table 5.2 the tentative first year IXPE observing plan is shown. The observing plan has been prepared by the STWG based on the theoretical expectations and scientific returns. More than 30 sources will be observed. In addition to them, transient sources will be observed as targets of opportunity, depending on trigger conditions, often in coordination with other multi-wavelength observatories in orbit and on ground. A challenge for assembling the observing plan is the long exposure times needed in case of sources with modest flux, so that in many cases the observation campaign has to be split in two or more sets of observations. Moreover, some highly polarized sources are often variable, and trade-offs are necessary in mission planning. Indeed the observing plan can, and will, be updated as the year progresses because of the addition of target of opportunities (i.e. transient sources such as Black Hole Binaries), or due to source-specific conditions (e.g. the observation of GRS 1915+105, initially scheduled for April 2022, has been moved to late 2022 because it entered a low luminosity state).

In the SNR STWG, I will be the lead author of the Tycho SNR discovery paper and I am part of the teams that will analyze the data of the IXPE observations of CasA and SN1006. In the Radio Quiet AGN & Sgr A* STWG I will be co-author of the Discovery paper of the IXPE observation of the molecular clouds in the Galactic center. I am also part of the Centaurus A team and I am a collaborator for the first Black Hole Binary in hard state transient to be observed. Finally I am also a member of the Science Analysis and Simulation WG, whose task is to develop the tools for the simulation and analysis of IXPE data. I worked on the implementation of the background sources in the observation simulator and participated to the Data-challenge launched to train the community to analyze the IXPE data.

### 5.1.2 The on-board FCW

IXPE is a so called "discovery mission", because employs a novel technology to perform observations that have never been done before. The observing plan is rich and the challenges many, in particular regarding the observation of faint extended sources. Hence the need to assess the calibrability and the performances of the instruments while in orbit, during the mission life-time, through a dedicated Filter and Calibration set, and the observability of the sources through realistic and detailed simulations.

**Table 5.2.** IXPE 1st year observing plan.

| Source type | Name of source | Observation time [ks] | Notes |
|---|---|---|---|
| PWN | Crab Nebula + Pulsar | 100 | Imaging |
| | MSH 15-52 + PSR B150 | 1500 | Imaging |
| | Vela | 1000 | Imaging |
| SNR | Cas A | 1000 | Imaging |
| | Tycho | 1000 | Imaging |
| | SN1006 | 1000 | Imaging |
| Magnetars | RXS J170849.0-400910 | 1000 | |
| | 4U 0142+61 | 1000 | |
| Persistent accreting NS | Cyg X2 | 100 | |
| | X Persei | 250 | |
| | Her X1 | 400 | |
| | GX9+9 | 100 | |
| | Vela X1 | 300 | |
| | GX 301-2 | 300 | |
| | 4U 1626-67 | 200 | |
| | GS 1826-238 | 100 | |
| Persistent accreting BH | Cyg X1 | 300 | |
| | GRS 1915+105 | 250 | |
| RQ AGN | MCG-5-23-16 | 500 | |
| | Circinus | 800 | |
| | IC 4329A | 500 | |
| Molecular Clouds in the GC | Sgr A complex | 1000 | Imaging |
| Blazars | Mrk 421 | 300 | |
| | Cen A | 200 | Imaging |
| | Mrk 501 | 300 | |
| | J0211+1051 | 400 | |
| | S5 0716+714 | 400 | |
| | 3C 454.3 | 200 | |
| | 3C 279 | 200 | |
| | 3C 273 | 200 | |
| | 1ES 1959+650 | 150 | |
| | BL Lac | 400 | |
| | 1ES 0229+200 | 400 | |
| Total observation time | | 15.85 Ms | |
| NS/BH transients | 21 sources | 12.55 Ms | |

As explained in Section 3.5, the prototype GPD has been extensively tested for more than fifteen years, but its complexity requires an accurate monitoring of the performance during the lifetime of the mission. More in general, detectors based on gas multiplication can evolve on the time scale of months, thus requiring a detailed plan for in-orbit calibrations. The major interest in this mission is its ground-breaking scientific expectations, but also operating and calibrating the GPD in orbit will pose a serious challenge. As of today, only one source is known to be polarized in X-rays, the Crab Nebula [169], but it cannot be used as a calibration source because it is found to vary and it is not known how its polarization properties change with time. Moreover, one of the scientific drivers of IXPE is the possibility of detecting polarization degrees down to a $\sim 1\%$ level, and since most X-ray sources are expected to be polarized, celestial calibration sources with a polarized signal well below the requirement of 1% are needed.

This ultimately poses the stringent requirement of calibrating with sources installed in the on-board instrumentation and powered by radioactive $^{55}$Fe sources. To this end, during the mission lifetime, the GPD response will be monitored through a Filter and Calibration Set (FCS) hosted in a Filter and Calibration Wheel (FCW) included in each Detector Unit (DU) [50].

Each FCS consists of four calibration sources and filters for special observations, they are:

- **CalA:** polarized calibration source;

- **CalB:** unpolarized calibration point-like source;

- **CalC:** unpolarized calibration source illuminating the whole detector;

- **CalD:** unpolarized calibration source illuminating the whole detector at a different energy from CalD;

- **Gray filter:** to attenuate the flux of very bright sources;

- **Closed position:** to perform internal background measurements;

- **Open position:** for regular observations.

I performed the acceptance tests of the Flight Models of the on-board calibration sources, and I analyzed the thermal vacuum measuremnt taken with the calibration sources integrated with the Flight Models of the DUs. The full description and role of each item of the FCS will be given in Chapter 6.

## 5.2   Sources of Background

In the analysis of extended sources, the contribution of both the astrophysical and instrumental background has to be properly considered. The astrophysical background can be extragalactic, due to unresolved point sources such as AGNs, or Galactic, coming from the diffuse emission from the Galactic plane.

The flux of the former in general will be uniform and present in every observation, the latter will instead depend on the distance of the source from the Galactic plane. Even if each of the unresolved sources that make up these backgrounds may have its own polarization properties, their average will be an unpolarized, diffuse X-ray source. For what concerns the extragalactic background, it must be noted that, according to **?** ] the optical polarization of quasars aligns to large-scale structures. However, the CXB represents only a fraction of the IXPE background, and we will not be able to achieve the sufficient significance to detect it's eventual X-ray polarization. For this reason in the following we will assume the astrophysical background to be unpolarized. From our simulations [178], we expect the instrumental background to be also unpolarized.

Although unpolarized, these background sources will have the effect of diluting the intrinsic polarization degree of a source according to the formula [92]:

$$P_{dil} = \frac{P_0}{1 + \frac{R_B}{R_S}} \quad .$$

(5.1)

Where $P_{dil}$ is the diluted polarization degree, $P_0$ the intrinsic source polarization, $R_B$ the unpolarized background counting rate, and $R_S$ the counting rate of the polarized source.

Moreover, as seen in Section 2.3, the presence of a background affects the Minimum Detectable polarization, and hence the significance of the measurement.

The presence of background makes an already challenging X-ray polarimetric observation even more complex in case of faint extended sources. So, it is fundamental to understand, mitigate, and subtract when possible these sources of noise. In Chapter 7 I will show the expected impact of this background component on the IXPE observation of a set of extended sources such as CasA, Tycho, the PWN MSH 15-52, and the G0.11-0.11 molecular cloud.

### 5.2.1 Instrumental background

The instrumental background is caused by the interaction of cosmic particles and photons with the spacecraft and the detector itself. This produces spurious tracks in the GPD that has to be distinguished from real photoelectron tracks. In Xie et al. [178], a paper that I co-authored, we presented for the first time the expected background of the GPD by Monte Carlo simulations and its impact on real observations of point and extended X-ray sources, presenting the Tycho SNR as an example.
We studied different background rejection techniques based on the analysis of the tracks collected by the Gas Pixel Detectors on board IXPE.

### 5.2.2 Diffuse Galactic plane background

The Galactic plane itself is permeated by a diffuse X-ray emission [see 83, for a recent review]. This Galactic plane diffuse emission (GPDE) exhibits emission lines from highly ionized heavy elements such as Si, S, and Fe, suggesting an origin from a hot, thin plasma with a $\sim$keV temperature. Whether the GPDE is actually diffused or arising from an unresolved populations of X-ray binary systems is an open problem. Ebisawa et al. [42], using the Chandra Advanced CCD Imaging Spectrometer Imaging array (ACIS-I), carried out a deep X-ray observation of the Galactic plane region devoid of discrete X-ray source. They concluded that because the sum of all the detected point source fluxes accounts for only $\sim$10 % of the total X-ray flux in the field of view, even hypothesizing the presence of a new population of much dimmer and numerous Galactic point sources, the total observed X-ray flux cannot be explained, hence part of the X-ray emission from the Galactic plane must truly have a diffuse origin. This diffuse component could arise not from the thermal emission of a very hot plasma but from the reprocessing by the interstellar gas of the X-ray radiation produced by luminous X-ray binary sources located in the Galaxy. Revnivtsev et al. [121] and Hong [68] however concluded that $\sim$70-80% of the GPDE flux can be explained by discrete sources such as white dwarfs and stars with active coronae, or some mixture of these sources.

### 5.2.3 Cosmic X-ray background

The cosmic X-ray background (CXB) is defined as the integrated emission of all the extra-galactic sources in the X-ray energy band (from $\sim$2 Kev up to $\sim$100 keV).
It's definition as a background comes from the very first X-ray astronomical observations [54], when an apparently diffuse background was observed together with the first extra-solar X-ray source (Sco X-1).
Today, the general consensus is that the CXB is mostly due to unresolved point sources, mostly AGN, with a small contributions from galaxy clusters and starburst galaxies [11, 20, 69, 153].

## 5.3 Data analysis

There are several approaches to extract the response to polarization from an X-ray polarimeter. The "classical" approach, as shown in Chapter 2 consists in building the modulation curve as a histogram of events based on their azimuthal distribution and fitting it with a function $N(\phi) = A + B\cos^2(\phi - \phi_0)$. A variation on this theme is based on the use of the Stokes Parameters I, Q, U, fitting the modulation curve with a function of the form $N(\phi) = I + \mu(Q\cos 2\phi + U\sin 2\phi)$. The "classical" approach to extract polarization from data by fitting the modulation curve with either a $\cos^2$ function or with the Stokes parameters has some practical shortcomings. First, it is limited by the fact that modulations and phase are treated as independent parameters, when usually they are not. Moreover, the uncertainties on the observables are not well defined when modulation is small (and unfortunately this will be the case for many scientific applications). On the other hand, the use of Stokes parameters has many practical advantages. The variables can be treated as independent and normally-distributed, at least when modulation is small and for sufficiently-large set of data. The Stokes parameters are additive, so they can be straightforwardly applied to calibrations (Rankin et al., in press) and background subtraction, with the application of an appropriate response matrix, they can be treated as fluxes. Hence, calibrations using the Stokes formalism allow the detector dependent spurious modulation subtraction that This uniforms the response of each of the three IXPE detectors, that otherwise have their own different response to polarized and unpolarized radiation because of construction differences. We are hence able to add together the results of the three detector units and improve the significance of the observations.

Finally, the Stokes formalism represent a common approach with polarimetry at other wavelengths, e.g. radio. However, the Stokes parameters do not substitute polarization degree and angle that remain relevant physical quantities in most of the cases and that can be easily derived from them. A technique based on the determination of the Stokes parameters of the single events was developed by [79].

### 5.3.1 Event-by-event Stokes approach

Thanks to the additive nature of Stokes parameters, the event-by event approach reduces to the sum of the Stokes parameters of the individual events and the subtraction of the ones characterizing the eventual background (see Section 5.2).

Last but not least, in this kind of analysis the uncertainties on the Stokes parameters are easily computed and are well behaved.

For each photon with polarization direction $\phi_k$ the individual Stokes parameters $i_k$ $q_k$ and $u_k$ are computed as:

$$i_k = 1 \quad , \tag{5.2}$$

$$q_k = \frac{2}{\mu}\cos 2\phi_k \quad , \tag{5.3}$$

$$u_k = \frac{2}{\mu}\sin 2\phi_k \quad . \tag{5.4}$$

The events can be selected in energy (and position in the case of imaging photoelectric polarimeters, see Section 3.5) so that the Stokes parameters of the $N$ selected events

is:

$$I = \sum_k i_k = N \quad , \tag{5.5}$$

$$Q = \sum_k q_i \quad , \tag{5.6}$$

$$U = \sum_k u_i \quad . \tag{5.7}$$

Introducing the normalized Stokes parameters $q$, $u$ as

$$q = \frac{1}{\mu} \frac{\sum_k q_k}{N} \quad , \tag{5.8}$$

$$u = \frac{1}{\mu} \frac{\sum_i u_k}{N} \quad , \tag{5.9}$$

the uncertainties of these quantities in given by standard deviations:

$$\sigma_q = \frac{1}{\mu} \sqrt{\frac{2 - q^2}{N - 1}} \quad , \tag{5.10}$$

$$\sigma_u = \frac{1}{\mu} \sqrt{\frac{2 - u^2}{N - 1}} \quad . \tag{5.11}$$

The polarization degree is then

$$P = \sqrt{q^2 + u^2} \quad , \tag{5.12}$$

with uncertainty

$$\sigma_P = \sqrt{\frac{2 - P^2 \mu^2}{(N - 1)\mu^2}} \quad . \tag{5.13}$$

Finally the polarization angle $\phi$ is equivalent to Eq. 2.16, with uncertainty given by:

$$\sigma_\phi = \frac{1}{\mu P \sqrt{2(N - 1)}} \quad . \tag{5.14}$$

**Background subtraction**

As previously hinted, a powerful consequence of the additivity property of the Stokes parameters is the possibility of describing not only the source events, but the background ones too.

The Stokes parameters of the latter can be simply subtracted from the ones of the source of interest in order to obtain the undiluted signal:

$$P_r = \frac{\sqrt{(Q - Q_{bkg})^2 + (U - U_{bkg})^2}}{I - I_{bkg}} \tag{5.15}$$

Where $I$, $Q$, $U$ are defined as in Eq. 5.5, 5.6, 5.7, and $I_{bkg}$, $Q_{bkg}$, $U_{bkg}$ are the background Stokes parameters. An application of this principle will be presented in Chapter 10. This works even if the backgrounds are anisotropic and thus mimic a polarization signal. The astrophysical background flux can be measured during off-source observations, looking at a dark patch of the sky close to the targeted X-ray source. The instrumental background can be instead measured during Earth occultations of the celestial sources, or while the detector is in closed position (see Chapter 6).

The uncertainties on the polarimetric observables (Eq. 5.13 and 5.14) after the subtraction of the background Stokes parameters become (see Kislat et al. [79] for the complete calculations):

$$\sigma_{P}r = \sqrt{\frac{\rho_{BS}(2 - P_r^2\mu^2)}{(\rho_{BS}(2R_B + R_S) - 2(R_B^2 + R_B R_S))T\mu^2}} \quad , \tag{5.16}$$

$$\sigma_{\phi r} = \frac{\sqrt{R_B + \frac{R_S}{2} + \rho_{BS}}}{P_r R_S \mu \sqrt{T}} \quad . \tag{5.17}$$

Where $P_r$ is the polarization degree as defined in Eq. 5.15, $R_B$ and $R_S$ are, respectively, the background and source rate, and $\rho_{BS} = \sqrt{R_B(R_B + R_S)}$.

In the next Chapter, the sources of background that can impact an X-ray polarization measurement by IXPE will be described in detail, along with the instrumental background rejection techniques that can be employed to minimize its impact.

### 5.3.2 Event weighting

Recently, attention has been dedicated to the development of "weights" that can be applied to X-ray polarimetric data to improve the sensitivity of the measurements (Di Marco et al., accepted for publication on the Astronomical Journal). The general idea is to apply to each event a weight that is representative of its quality, i.e. how well the event photoelectron track correlates to the detector response to linear polarization.

Kislat et al. [79] developed a method to weight data from instruments with non-uniform acceptance. The event-by-event Stokes parameters defined in Eq. 5.2, 5.3 5.4 are modified with the introduction of the weights as a multiplicative factor so that:

$$i_k = w_k \quad , \tag{5.18}$$

$$q_k = \frac{2}{\mu} w_k \cos 2\phi_k \quad , \tag{5.19}$$

$$u_k = \frac{2}{\mu} w_k \sin 2\phi_k \quad . \tag{5.20}$$

In the IXPE observation simulator ixpeobssim (see later Section 5.4), the weight $w_k$ is defined as the inverse of the effective area at the energy of the event $k$:

$$w_k = \frac{1}{A(E_k)} \quad . \tag{5.21}$$

The inverse of the effective area in the definition of the weights acts as an acceptance correction guaranteeing that the Stokes parameters are summed (or averaged) over the input source spectrum, as opposed to the measured count spectrum. The measured Stokes parameters over a generic subset of the events is obtained by simply summing the event-by-event quantities as in Eq. 5.5, 5.6, and 5.7. The uncertainties on the weighted normalized Stokes parameters for a large data set (i.e. N≫1) modify Eq. 5.10, 5.11 as

$$\sigma_q = \frac{1}{\mu}\sqrt{\frac{W_2}{I^2}(2-q^2)} \quad , \tag{5.22}$$

$$\sigma_u = \frac{1}{\mu}\sqrt{\frac{W_2}{I^2}(2-u^2)} \quad . \tag{5.23}$$

where here $I = \sum_k^N w_k$ and $W_2 = \sum_k^N w_k^2$.

An analysis method proposed by Muleri et al. [103] is the so-called "standard cuts". It is based on a two-step selection of the events in which 20% of them are removed. In the first step an energy cut is applied by fitting the spectrum with a Gaussian profile and removing all the events falling outside $\pm 3\sigma$ from the peak. In the second step, the remaining events are ranked in order of eccentricity of their tracks and the lower eccentricity ones are removed up to a threshold so that 20% of the initial events are removed. This method can also be considered a weighted analysis, in the sense that the "good" events are given weight $w_k = 1$ and the bad ones $w_k = 0$. However, the standard cuts were developed for the analysis with monochromatic laboratory sources, while IXPE will observe astrophysical sources with continuum spectra. Alternative approaches are currently being studied. Di Marco et al. (accepted for publication) suggests a weighted analysis based on multiple track properties: the weight is the parameter $\alpha$ defined as the difference between the longitudinal and transverse second moment of the tracks, divided by their sum, and is naturally comprised between 0 and 1. This approach assures a sensitively better MDP (equivalent to adding another telescope module to IXPE!) Others [110, e.g.], propose the application of neural networks to enhance the sensitivity of X-ray polarimeters.

### 5.3.3 Spectro-polarimetry with XSPEC

The tools developed by the IXPE community allow also to analyse X-ray polarization data through spectro-polarimetric fits. Indeed, ixpeobssim (see later Section 5.4) allows to produce binned Stokes spectra that can be used in XSPEC [6]. The Stokes I spectrum is a simple PHA (Pulse Height Amplitude) file that can be readily used for simple spectral fitting. Togheter with the binned Q and U spectra, it is possible to perform simultaneous spectro-polarimetric fitting of the data from the three IXPE DUs.

The spectrum of the I parameter is an un-weighted histogram of the event-by-event Pulse Invariant (PI) values, where the content of the bin $k$ is given by the accumulated counts:

$$I_k = \sum_{PI=k} 1 = n_k \quad \text{and} \quad \sigma_{Ik} = \sqrt{n_k} \quad . \tag{5.24}$$

On the other hand, the Q and U spectra are weighted histograms with the same binning, with weights:

$$Q_k = \sum_{PI=k} 2\cos(2\phi_k) = \sum_{PI=k} w_k^Q \quad \text{and} \quad \sigma_{Qk}^2 = \sum_{PI=k} (w_k^Q)^2 \quad , \tag{5.25}$$

$$U_k = \sum_{PI=k} 2\sin(2\phi_k) = \sum_{PI=k} w_k^U \quad \text{and} \quad \sigma_{Uk}^2 = \sum_{PI=k} (w_k^U)^2 \quad . \tag{5.26}$$

In order to obtain a fit of the polarization quantities (degree and angle) together with the spectral information, a set of phenomenological multiplicative (that is, that can be multiplied to the standard XSPEC spectral models) polarization models have been developed by the IXPE community:

- constpol: constant polarization degree and angle;

- linpol: polarization degree and angle scaling linearly with energy;

- quadpol: polarization degree and angle scaling quadratically with energy;

- powpol: polarization degree and angle with a power-law dependence on energy.

In order to obtain a correct fit of the model, if more components are present, both polarized and unpolarized, it is necessary to attach a polarimetric model to each component, eventually imposing the condition that the polarization degree of an unpolarized spectral component is zero.

In Chapter 8 I will show an application of X-ray spectropolarimetric fitting with XSPEC of simulated IXPE data.

## 5.4   ixpeobssim

"*ixpeobssim*" [111] is a Monte Carlo simulation framework specifically developed for the Imaging X-ray Polarimetry Explorer (IXPE) and based on the Python programming language. This is the software used for the simulations of the IXPE observation that will be shown throughout this thesis. The *ixpeobssim* development was led by INFN-Pisa, with contributions from all the IXPE community. It allows to produce realistic simulations of observations given as basic inputs a source model including morphological, temporal, spectral and polarimetric information, and the response functions of the detector i.e., the effective area, the energy dispersion, the point-spread function (PSF) and the modulation factor.

*ixpeobssim* produces output files that can be directly fed into the standard visualization and analysis tools used by the X-ray community, including the HEASARC XSPEC [6], DS9 [75], and ftools [17], which makes it a useful tool not only for simulating astrophysical objects, but also to develop and test end-to-end analysis pipelines.

The development of this simulator is aimed at preparing the observation plan by simulating the most promising sources and getting an estimate of what could be seen given the current best knowledge of the source models and the spacecraft performances.

In addition, *ixpeobssim* provides some basic analysis tools that allows to select and bin the simulated photon list in several flavors, e.g. performing spatial or energy cuts (with the *xpselect.py* tool), or to produce binnned polarization and significance maps (with the *xpbin.py* tool).

In Figure 5.6 the block diagram of the *ixpeobssim* data flow, from the input models and files, to the output is shown.

Here I aim to describe the inputs (Source model and instrumental response functions, Section 5.4.1) and outputs (Section 5.4.2) of the *ixpeossim* simulator.

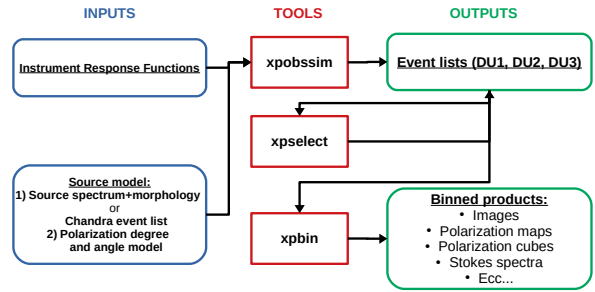

**Figure 5.6.** Block diagram of the *ixpeossim* data flow, from the source model, to the IXPE photon list simulation, to the data selection and binning.

### 5.4.1 Simulation inputs

**Source model**

Source models are specified in *ixpeossim* through Python configuration files. The input source model can be of two types. The first is Model component based, in which the user needs to provide an energy spectrum, morphology, and polarization properties of the source. The energy spectrum must be expressed in units of cm$^{-2}$ s$^{-1}$ keV$^{-1}$ and can be defined analytically (e.g. power-law, black body, etc) or can be the result of interpolations built from a series of data points or even a XSPEC spectral model. The morphology can be provided by a FITS image, of by a predefined morphology such as point-source, uniform disk, gaussian disk, etc... the polarization degree and angle are functions of energy, time, and position in the sky. The first type of source model is based on a Chandra event file accompanied by a polarization degree and angle model. The implementation of the simulation of these two types of input model will be discussed in more details in Section 5.4.2.

**The Instrument Response Functions**

The Instrument Response Functions (IRF) are the properties of an instrument that allow the mapping between the incoming photon flux and the detected events. The IXPE IRFs are stored in FITS files and are used both in the event simulation and in the analysis of the data products from the simulations, as they are fully compatible with spectral analysis tools such as XSPEC. The response functions are defined between $1-12$ keV in steps of 40 eV, and are detector-unit based so that each IXPE telescope-detector modules has its own effective area, modulation factor and PSF, informed by the most up-to date laboratory and calibration measurements. In the following, the effective area, the energy dispersion, the PSF, and the modulation factor as implemented in *ixpeossim* are briefly described.

- **Effective area.**
  The effective area (shown in Fig. 5.7 (a)) is modeled based on either analytic calculations from tabulated data or calibration data. The effective area curves of the three telescopes are fairly similar to each other, with small differences due to the MMA calibration measurement. The effect of the vignetting is also implemented in *ixpeobssim*: the vignetting, along with the relative orientation of the three IXPE detector units, defines the relative exposure across the field of view of the instrument. The vignetting function shown in Fig. 5.7 (b) comes from a preliminary study by MSFC based upon ray-tracing simulations for a perfect mirror module assembly, and is relevant for the simulation of extended sources. The effect of vignetting is relatively more important above 6 keV and for off-axis angles larger than 5'. Indeed the vignetting is dominated by the change of reflectivity with the angle: some high energy photons at energies above 6 keV at offset angles fall above the critical reflection angle and are lost. The fact that the critical angle is inversely proportional to the energy is the reason why the vignetting is larger at higher energies.

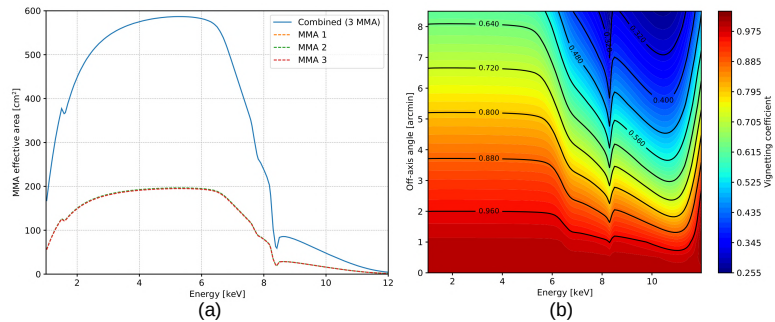

**Figure 5.7.** (a) On-axis effective area as a function of the energy for three MMA (solid blue line) and for the single modules (dashed orange, green, and yellow lines). (b) Vignetting of the optics as a function of energy and off-axis angle. Adapted from *ixpeobssim* user manual.

- **Energy dispersion.**
  The energy dispersion is described as a Gaussian profile where the energy resolution is tabulated, based on laboratory measurements and linearly interpolated in between a grid of energies. In Fig.5.8 (a) is shown the one-dimensional probability density functions at a few fixed energies, corresponding to vertical slices of Fig. 5.8 (b).

- **Point Spread Function.**
  The representation of the PSF is derived by Fabiani et al. [45], with scaling factors applied to account for the differences measured during mirror calibration. As shown in Fig. 5.9, it was found during the tests in the MSFC stray-light facility that MMA 1 has a significantly better PSF (less than 20 arcsec Half Power Diameter-HPD) than MMAs 2 and 3 (running at more than 25 arcsec HPD). Fig. 5.9 shows the profile of the PSF radius as a function of the Encircled Energy Function (EEF) at 4.51 keV.

- **Modulation Factor.**

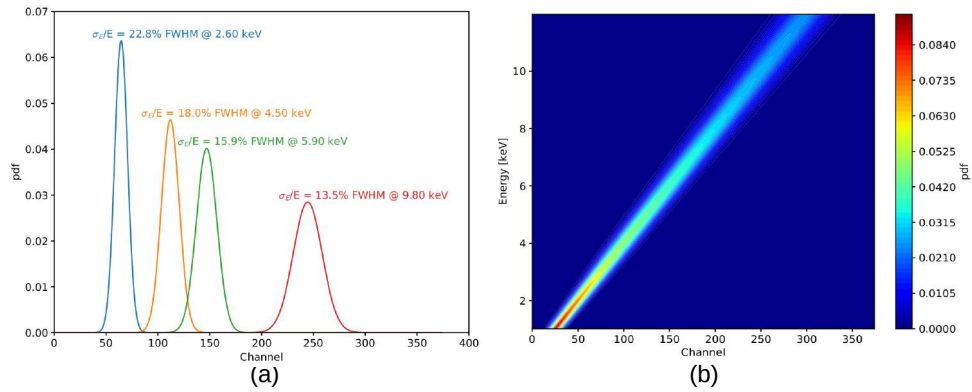

**Figure 5.8.** (a) Energy dispersion (one-dimensional probability density function) at a set of discrete energies. The FWHM (full width half maximum) energy resolution is indicated for completeness. (b) GPD Pulse invariant channel-energy response matrix. Adapted from *ixpeobssim* user manual.

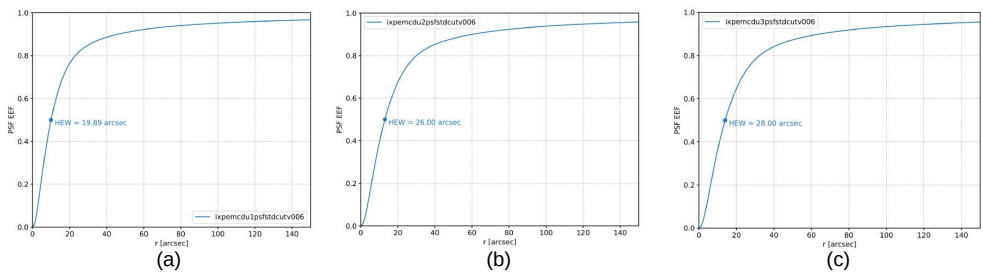

**Figure 5.9.** Encircled energy fraction (EEF) at 4.51 keV for the IXPE PSF for the MMA1 (a), MMA2 (b), and MMA3 (c). Adapted from *ixpeobssim* user manual.

In *ixpeobssim*, the modulation factor is parameterized from Monte Carlo simulations informed by the ground calibrations of the three detector units. Its value as a function of the energy for the GPD is shown in Fig. 5.10 (a). The edge around 9 keV is due to the copper K-edge, above which the extraction of photoelectrons from X-rays absorbed in the GEM becomes significantly more likely, causing an increase of effective area, accompanied by a dilution of the modulation.

In Fig. 5.10 (b), instead, is shown the modulation response function, that is the product of the effective area times the modulation factor for DU1 and MMA1.

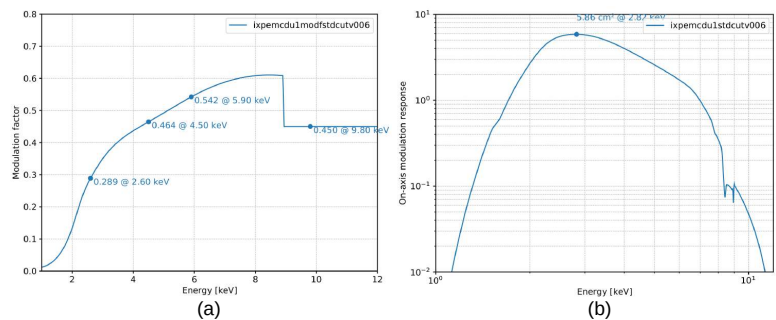

**Figure 5.10.** (a) Modulation factor as a function of the photon energy for the GPD. (b) Modulation response function (product of the effective area times the modulation factor) for a single telescope (DU1 and MMA1 in this case). Adapted from *ixpeobssim* user manual.

### 5.4.2  Simulation outputs

The Monte Carlo observation-simulation tool, *xpobssim.py*, produces a photon list in FITS format for a specific observation time, given a source model and a set of instrument response functions. The first step of the simulation work-flow is to calculate the expected number of events; this is done by convolving the input source spectrum with the instrument effective area, and then by integrating the curve in energy and in time. Using the resulting spectrum and the light curve as one-dimensional probability density functions, the event times and the true energies are randomly sampled and smeared with the energy dispersion. From the given source morphology and the instrument PSF, the simulator determines the sky-direction of the incoming photons. The final step is the generation of the angular distribution of the photo-electron emission angles based on the polarization model. If the source model is made of more components, the process is repeated for each one of them. In the special case of the simulation of the IXPE instrumental background (see Section 5.2.1), the simulation happens in detector (instead of sky) coordinates, and the resulting event list is not convolved with any of the instrument response function, as the events are internal to the detector. For source models involving more than one component, this is done for each component separately, and the different resulting photon lists are then merged and ordered in time at the end of the process.

The final outputs of the simulation are three FITS files, one for each IXPE detector unit, containing all the information for each event needed to obtain images, spectra, and polarimetric information. The simulations presented in Chapter 7, 9, and 10 are produced with this approach.

A second simulation channel allows to convert a Chandra observation event list to a corresponding IXPE observation by folding the event characteristics with the IXPE instrument response functions. The major advantage of this technique is that it allows to fully preserve the correlation between the morphology and the energy spectrum of the source. This is particularly important for the simulation of extended sources. The Chandra angular and energy resolutions are indeed much better than those of the IXPE mission (angular resolution: $\lesssim 1$ arcsec vs. $\lesssim 30''$, energy resolution: $\lesssim 5\%$ vs. $<20\%$ at 5.9 keV see Fig. 5.8 (a)), so that the Chandra measured energies and positions can be assumed as the Monte Carlo truth.

In this case, the simulation starts from a Chandra event list, which includes energies and spatial information of every event. The Chandra events are then down- or over-sampled according to preserve the ratio of the IXPE and Chandra effective areas and observation times, and then smeared with the IXPE response functions. Finally the photo-electron emission angles are generated and the output event list saved in FITS format as in the previous case. In case of definition of multiple sub-regions of the region of interest, the conversion is done separately for each of them, and the resulting photon lists are merged and ordered in time at the end of the process.

The simulations presented in Chapter 8 are produced with this approach.

# Part II

# Part 2

# Chapter 6

# On-board calibration sources

## 6.1 The Filter and Calibration Set

The Gas Pixel Detector enables the X-ray polarimetric measurements of IXPE is a novel instrument that is subject to changes over time. For instance, a secular decrease of the DME gas pressure, with a time-scale of months, was found [8]: this has the effect of provoking a slow change with time of the detector gain. Hence it is particularly important to verify the performance of the detectors and their stability during the mission lifetime. For this purpose, as anticipaded in Chapter 5, the spacecraft hosts a filter and calibration set (FCS) mounted on a Filter and Calibration Wheel (FCW). The sets are denominated as Flight Model (FM) 1, FM2, FM3, FM4 and are assigned to the three DUs installed on board of the IXPE spacecraft and to a DU that acts as a spare. The FCS includes both polarized and unpolarized calibration sources, capable of illuminating the whole detector or just a part of it, for mapping and monitoring of the GPD modulation factor (i.e. the detector response in terms of modulation to 100% polarized radiation), quantum efficiency and energy resolution at different energies. IXPE envisioned the use of a FCW since its inception for the above mentioned reason, but its presence became even more important with the discovery of instrumental effects such as gas cell pressure variation in time [8]. Another application of the on-board calibration sources is of particular interest for the extended sources part of the IXPE observing plan: the faintest ones will necessitate of long integration times, of the order of $10^6$ s, hence it will be fundamental to check during the mission lifetime the detector response. The DUs are calibrated in orbit during the X-ray sources occultations. The on-board calibration sources have also been employed during the optics-detector integration tests in the MSFC staylight facility to study the detector gain variations. In orbit calibrations enables us to check for the presence of spurious polarization and map and monitor the gain and its uniformity across the $15 \times 15$ mm$^2$ detector surface. These pieces of information will help to improve our understanding of the detector performance and asses the reliability of the scientific results.

In this Chapter, whose contents are based on the work I published in Ferrazzoli et al. [50], I present the first measurements performed on the flight models of the FCS (whose design was presented in Chapter 5) using silicon drift detectors and CCD cameras, as well as those in thermal vacuum (TV) with the flight units of the GPD. I will show that the calibration sources successfully assess and verify the functionality of the GPD and validate its scientific results in orbit; this improves our knowledge of the behavior of these detectors in X-ray polarimetry.

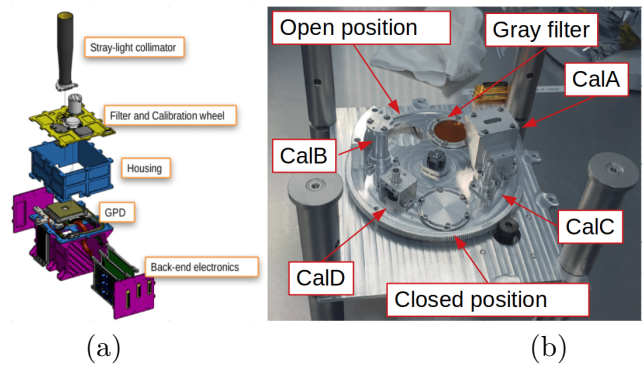

|                |                |
| :------------: | :------------: |
| (a)            | (b)            |

**Figure 6.1.** (a) Exploded view of the IXPE Detector Unit, showing the position of the Filter and Calibration wheel on top of the GPD; (b) picture of the Filter and Calibration Set installed onto the wheel in the IAPS clean room.

In Section 6.1.1), I describe the calibration sources and the filters present in the FCS. I present in Section 6.2 the experimental setups: first the setup in Clean Room conditions with a commercial Silicon Drift Detectors (SDD) and Charge-Couple Device (CCD), then in TV conditions with the Flight Models of the GPD. In Section 6.2 I describe the data analysis tools. In Section 6.3 I present the results of the tests that I discuss in Section 6.4.

## 6.1.1 The Filter and Calibration set

The FCS hosts four calibration sources: a polarized source, CalA; a collimated unpolarized source, CalB; two uncollimated, unpolarized sources, CalC and CalD. A gray filter, an open, and a closed position complete the items of the FCS. The FCS is hosted in the FCW, shown in Fig. 6.1 (b), that is placed on the top lid of the DU as shown in Fig. 6.1 (a). The FCW, by rotating on its central axis, allows us to place in front of the GPD one of the four calibration sources or a gray filter, in addition to the open and closed position, depending on the observational requirements. The FCW also hosts other elements that assure the stability and position accuracy of the calibration sources. A rotary potentiometer is used to exactly determine the wheel angular position. In addition, for redundancy, three radially placed Hall effect sensors and twelve magnets (positioned such to realize a unique binary coding for the wheel's seven positions) work as position reference points. The calibration sources can thus be positioned with an accuracy better than $\pm 500$ $\mu$m with respect to their nominal positions. The angular position of the polarized calibration source is known with an uncertainty better than 20 arcmin with respect to the DU coordinate system. The fixed parts of the FCW (e.g. the cover lid) are connected to the rotating parts (the wheel itself) by a bearing sub-assembly. Finally, a ballast mass is installed to balance the weights and the momentum of inertia on the wheel. Each calibration source inside the FCW contains a radioactive source constituted of a $^{55}$Fe nuclide that, following a K electron capture, emits X-rays at 5.9 and 6.5 keV, i.e. the Mn K$\alpha$ and Mn K$\beta$ fluorescence emission lines, respectively. The activity of $^{55}$Fe naturally decays with half life of 2.7 years, which provides a sufficient time to cover the entire operative life of IXPE. This solution removes the difficulty of having X-ray tubes on board of the spacecraft, as they would have been complex and mass- and power-demanding, especially on a moving support.

In Table 6.1 we show the rate requirements that were set internally for the laboratory measurement, and the scientific observable that can be obtained by each source. The requirements on the counting rate are set by the statistical significance required to validate the results and by the activity of the radioactive sources on board. A counting rate of at least $R > 3$ c/s assures that the counts needed to reach a Minimum Detectable Amplitude ($MDA$, see Chapter 2, i.e. $\mu \times MDP$ with $\mu$ modulation factor and $MDP$ Minimum Detectable Polarization) of at least 1% can be collected in less than a day. A number of counts for an $MDA/2$ must be actually collected in order to achieve a $1\sigma$ measurement of a certain amplitude [143]. Other important observables are the photo-electron track size and length: the track size is defined by the number of contiguous pixels in which the charge is collected for a single event, the track length is defined as the second momentum of the track. Their correct determination allows us to distinguish between noise and real events, to determine the polarization direction of the latter, and to check the pressure of the gas in the GPD.

**Table 6.1.** For each on-board calibration source, spatial and rate requirements and scientific observables are reported.

| Calibration Source | Requirements | Scientific observables |
|---|---|---|
| Cal A | Polarized beam illuminating the entire detector; rate >3 c/s at 3 keV; rate >40 c/s at 5.9 keV | Modulation factor and energy resolution at 3 and 5.9 keV, counting rate, track length and track size, gain |
| Cal B | Collimated 3 mm beam at the center of GPD; rate >30 c/s at 5.9 keV | Spurious modulation, energy resolution at 5.9 keV |
| Cal C | Uncollimated beam illuminating the whole 15×15 mm$^2$ surface of the GPD; rate >80 c/s at 5.9 keV | Gain, counting rate, energy resolution at 5.9 keV, spurious modulation |
| Cal D | Uncollimated beam illuminating the whole 15×15 mm$^2$ surface of the GPD; rate >10 c/s at 1.7 keV | Gain, counting rate, energy resolution at 1.7 keV, spurious modulation |

### Calibration source A (CalA)

CalA [101, 102], shown in Fig. 6.2, produces polarized X-ray photons with precisely-known energy and polarization state, allowing for the monitoring of the modulation factor of the instrument at two energies, 3 keV and 5.9 keV, in the IXPE energy band (2 - 8 keV). Its working principles is based on Bragg diffraction (see Section 3.2.1). Photons at 5.9 keV are emitted by the $^{55}$Fe nuclide sources and diffracted (at the second order) on a graphite flat mosaic crystal. The diffraction angle is $\sim 38°$. Photons at 3.0 keV are generated by partially absorbing the primary $^{55}$Fe emission on a silver foil, to extract its L$\alpha$1 and L$\alpha$2 fluorescence at 2.99 and L$\beta$ line at 3.15 keV. The silver foil is deposited between two polyimide foils which are 8 $\mu$m (on the side towards the $^{55}$Fe) and 2 $\mu$m (on the side of the crystal) with negligible X-ray losses. Radiation at 3.0 keV is then diffracted at nearly the same diffraction angle, i.e., $\sim 38°$, on the same graphite crystal used to diffract 5.9 keV radiation, but at the first order. As the diffraction angle is not 45°, expected polarization is lower than 100%, but the source design is robust and compact, and allows to contemporaneously generate photons at two energies. Graphite mosaic crystal has a mosaicity of 1.2 degrees. A second broad collimator is finally used to block stray-light X-rays. Since, given a point-like source, the locus of points on the plane of the crystal satisfying the Bragg condition is a circle, its projection on the detector will appear as an arc (see Fig. 6.3). We call this "Bragg arc", whose width depends on the X-ray source and crystal used. Polarization along the arc is expected to remain constant with the polarization angle tangent to the Bragg arc. In the case of the CalA, the image of the diffracted photons appears as a 4 mm wide, slightly curved strip extending across the detector (see for example Fig. 6.10). This can be exploited to study the response of the whole detector to polarized radiation by moving the wheel. We call the four sources CalA1, CalA2, CalA3 and CalA4 according to the FCS they are part of.

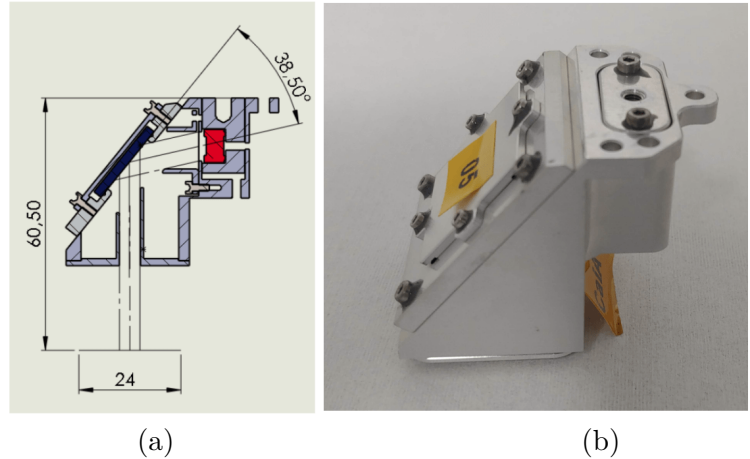

(a)                                                    (b)

**Figure 6.2.** The polarized on board calibration source Cal A allows to produce polarized X-ray photons of known energy. (a) Cut-out view of CalA: in red is the $^{55}$Fe radioactive source in front of which a thin Silver foil is placed to extract 3.0 keV fluorescence. In dark blue is the graphite mosaic crystal for Bragg diffraction. Quotes are in mm. (b) Picture of CalA.

### Calibration source B (CalB)

This source produces a collimated beam of unpolarized photons to monitor the absence of a spurious modulation. A $^{55}$Fe radioactive source is glued on a holder and screwed in a cylindrical body at the end of which, a diaphragm with an aperture of 1 mm collimates X-rays to produce a spot of about 3 mm on the GPD. Such a spot has a size which is representative of the region illuminated by the photons of a point-like source when the spacecraft pointing dithering strategy is actuated. An exploded view and a picture of the source are shown in Fig. 6.4. We call the four sources CalB1, CalB2, CalB3 and CalB4 according to the FCS they are part of.

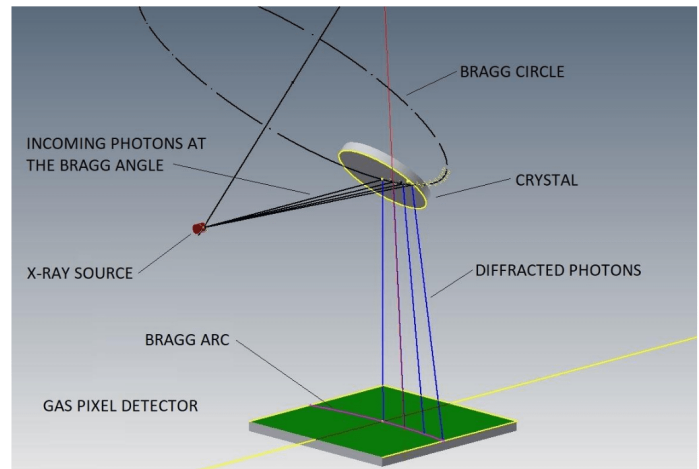

**Figure 6.3.** Geometry of Bragg diffraction for monochromatic photons.

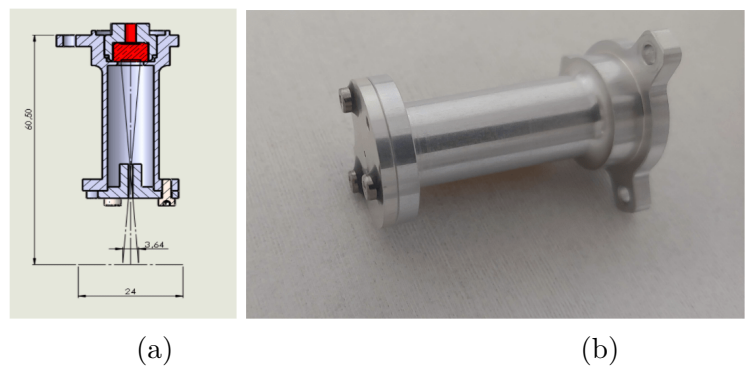

(a)                   (b)

**Figure 6.4.** CalB illuminates the GPD with monochromatic unpolarized X-ray, simulating the region illuminated by a point-like source. (a) Cut-out view of CalB: in red is the [55]Fe radioactive source. Quotes are in mm. b) Picture of CalB.

## Calibration source C (CalC)

This source illuminates all the detector sensitive area at one energy (5.9 keV) to map how the gain (i.e. the relation between the pulse height, PHA, that is proportional to the charge collected by the detector and the energy of the photons) changes as a function of the position and time. This source is composed of a $^{55}$Fe radioactive source, glued in a holder similar to CalB but with a diaphragm-less collimator that allows X-ray photons to impinge on the whole detector sensitive area. An exploded view and a picture of the source are shown in Fig. 6.5. We call the four sources CalC1, CalC2, CalC3 and CalC4 according to the FCS they are part of.

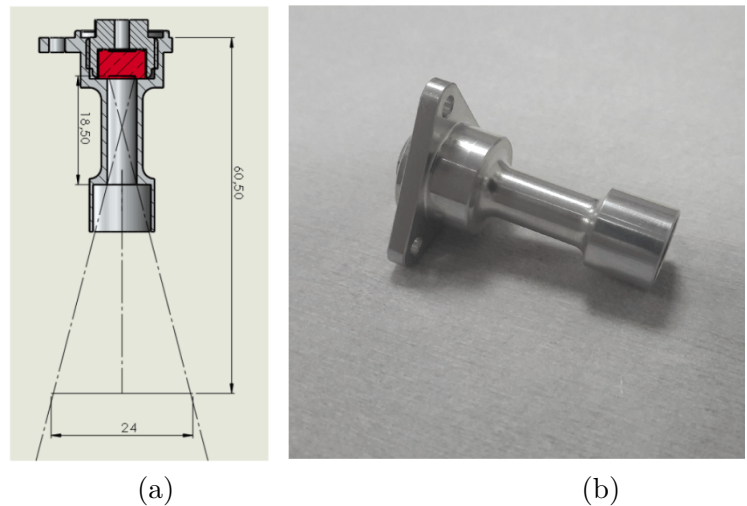

(a)                                                (b)

**Figure 6.5.** CalC illuminates the whole GPD surface with monochromatic unpolarized photons. (a) Cut-out view of CalC: in red is the $^{55}$Fe radioactive source. Quotes are in mm. (b) Picture of CalC.

## Calibration source D (CalD)

This source illuminates all the detector sensitive area as CalC, to map the gain on the whole GPD surface at a different energy. CalD is based on a $^{55}$Fe source which illuminates a silicon target to extract K fluorescence from Silicon at 1.7 keV, which impinges on the detector. Since the length of the photoelectron tracks is a function of the impinging photon energy, at 1.7 keV the CalD tracks are shorter than the 5.9 keV track from CalC. This enables us to map the gain of the GPD at a higher spatial resolution and also provide a check on the systematics at low gain. Moreover, the large energy difference between the 1.7 keV of CalD and 5.9 keV from CalC allows us to determine the calibration relation (PHA/Energy) with high accuracy. CalD is designed so that X-ray photons from $^{55}$Fe cannot directly impinge on the GPD sensitive area, avoiding detector saturation. Some of these photons are scattered by silicon and impinge on the detectors with an energy almost unchanged. An exploded view and a picture of the source are shown in Fig. 6.6. We call the four sources CalD1, CalD2, CalD3 and CalD4 according to the FCS they are part of.

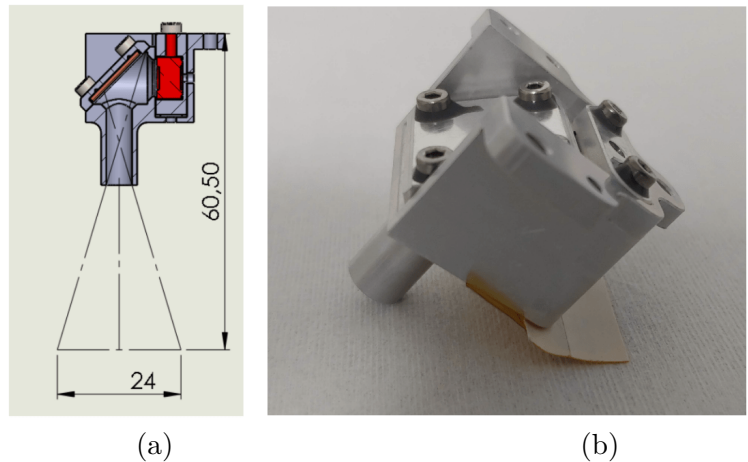

|       |       |
| :---: | :---: |
| (a)   | (b)   |

**Figure 6.6.** CalD illuminates the whole GPD surface with monochromatic unpolarized photons with lower energy than CalC. (a) exploded view of CalD: in red is the $^{55}$Fe radioactive source, in orange is the Silicon target for 1.7 keV fluorescence. (b) Picture of CalD.

**Open position, closed position, gray filter**

In the FCS, beside the four calibration sources, an Open Position, a Closed Position, and a gray Filter are also present. The Open position will be the standard science operation position. The Closed Position (i.e. a black filter) is a lid opaque to X-ray radiation (with transparency lower than $10^{-6}$ at 8 keV) and covers the detector during the internal background measurements. The gray filter will be employed when observing very bright sources (with flux higher than $\sim 4\times 10^{-8}$ erg/cm$^2$ /s, or 2 Crab in the $2-8$ keV range) that would otherwise saturate the detector because of the dead time. The gray filter is called so in analogy with the ones that in the optical band have the same function but, differently from those, the transparency of the IXPE one is strongly energy dependent. The transparency of the gray filter is of 14% at 2.6 keV such that the flux for a source with a power law with spectral index of 2 is reduced by a factor of 8 in the $1-12$ keV energy range. However, this does not affect the response to polarization: while being more opaque to softer X-ray photons, which are typically those with a lower polarization, it will have the advantage to improve the statistics at higher energies that are of major astrophysical interest.

## 6.2 Experimental setup

The experimental setup was designed in order to study the performances of the FCS prior of its installation. To this aim we used a Silicon Drift Detector (SDD) for the spectrum and a Charge Coupled Device (CCD) for images. As no commercial standard X-ray polarimeter exists, the polarimetric performances are tested with the GPD.

In Table 6.2 we report the activity and emission rate of the $^{55}$Fe radioactive sources employed during the measurements with the SDD and in the TV chamber with the DU. The sources are manufactured by Eckert & Ziegler, that provides for each nuclide a nominal activity and a measured emission rate: while the former can be affected by errors as high as 30% due to self absorption, the latter can be measured with a greater accuracy, hence representing a better estimate of the $^{55}$Fe source activity. For this reason we use the emission rate to calculate the expected flight rate.

Note that for CalC, that directly illuminates the detector, a weaker source can be employed. Soon before the integration on the spacecraft, the calibration sources where be equipped with radioactive sources of higher nominal activity: 100 mCi for CalA, 20 mCi for CalB, 0.5 mCi for CalC and 100 mCi for CalD. We refer to these as "flight radioactive sources". The emission rate of the flight radioactive sources has been measured and is also reported in Table 6.2. Because of the high radioactivity of some of these sources, we cannot directly perform tests with them in our lab. The weaker $^{55}$Fe nuclides already present in our lab are employed and the flight rate (at the beginning of the mission) is inferred by multiplying the measured rate by the ratio between the emission rate of the flight sources at the beginning of the mission operations and that of our sources at the time of the measurement. In the following, we will refer to the "expected flight rate" for the rate of the calibration sources when using the flight radioactive sources. To maximize the activity of the on-board nuclides for achieving the required counting rate, flight radioactive sources were be inserted in the instrument as late as possible in the integration flow.

**Calibration source spectra and images with SDD and CCD**

The spectrum of each source is taken with an Amptek XR-100SDD SDD (see Fig.6.7 (a)) in order to verify that X-rays are correctly emitted. The tests are performed

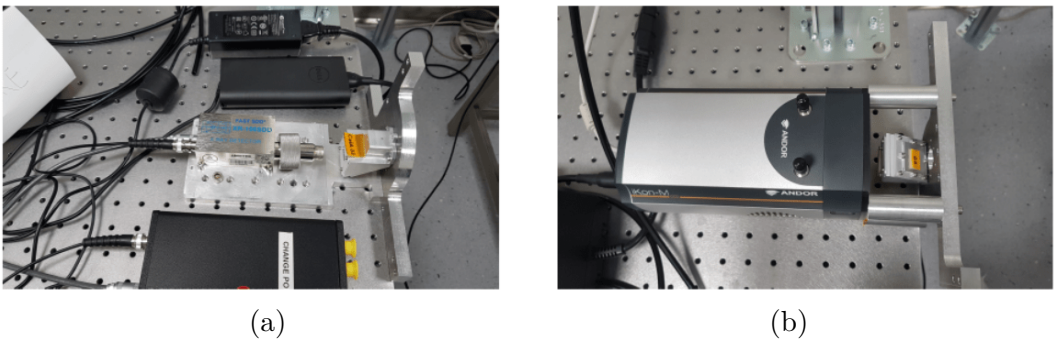

|  (a)  |  (b)  |

**Figure 6.7.** (a) CalA in front of SDD for spectrum measurement. (b) CalA in front of CCD for image acquisition.

inside an ISO 7 Clean Room with controlled temperature and humidity conditions. The measurement of spectra in air is affected by the absorption of X-rays, with transmission factor $\tau(E)$ given by the relation:

$$\tau(E) = \frac{I}{I_0} = e^{-\mu(E)\rho d} \qquad (6.1)$$

where $I/I_0$ is the fraction of photons arriving on the detector, $\mu$ is the energy-dependent total mass attenuation coefficient of air (in cm$^2$/g) obtained with the

**Table 6.2.** $^{55}$Fe radioactive nominal source activity and emission rate at the time of the tests with SDD and GPD. Within parentheses is reported the date to which the nominal activity refers to. In the last row is reported the activity of the flight radioactive sources and the emission rate of the nuclides. The emission rate of the source employed in the CalC measurements with SDD are not known.

| Measurement | CalA Activity [mCi] (date) | CalA Emission rate [s$^{-1}$strd$^{-1}$] | CalB Activity [mCi] (date) | CalB Emission rate [s$^{-1}$strd$^{-1}$] | CalC Activity [mCi] (date) | CalC Emission rate [s$^{-1}$strd$^{-1}$] | CalD Activity [mCi] | CalD Emission rate [s$^{-1}$strd$^{-1}$] |
|---|---|---|---|---|---|---|---|---|
| Nominal | 8.1 (2017/01/16) | 6.55E6 | 8.1 (2017/01/16) | 6.55E6 | 5 (2018/01/18) | - | 8.1 (2017/01/16) | 6.55E6 |
| SDD FM1 | 5.13 | 2.81E6 | 5.13 | 2.81E6 | 0.34 | - | 5.13 | 2.81E6 |
| SDD FM2,3,4 | 4.82 | 2.42E6 | 4.82 | 2.42E6 | 0.31 | - | 4.82 | 2.42E6 |
| Nominal | 8.1 (2017/01/16) | 6.55E6 | 0.50 (2019/07/01) | 2.33E5 | 0.50 (2019/07/01) | 2.18E5 | 4 (2019/07/01) | 2.01E6 |
| DU1, FCW1, TV | 4.22 | 2.11E6 | 0.49 | 2.29E5 | 0.49 | 2.15E5 | 3.95 | 2.01E6 |
| DU2, FCW2, TV | 4.09 | 2.06E6 | 0.48 | 2.22E5 | 0.48 | 2.09E5 | 3.83 | 1.93E6 |
| DU3, FCW3, TV | 4.03 | 2.02E6 | 0.47 | 2.19E5 | 0.47 | 2.06E5 | 3.78 | 1.90E6 |
| DU4, FCW4, TV | 3.96 | 1.98E6 | 0.44 | 2.02E5 | 0.44 | 2.15E5 | 3.71 | 1.86E6 |
| Flight radioactive sources | 100 (2020/04/29) | 2.81E7 (DU1) 2.82E7 (DU2) 3.16E7 (DU3) 3.06E7 (DU4) | 20 (2020/04/29) | 7.10E6 (DU1) 7.54E6 (DU2) 8.14E6 (DU3) 7.64E6 (DU4) | 0.5 (2019/12/18) | 1.53E5 (DU1) 2.07E5 (DU2) 2.06E5 (DU3) 1.70E5 (DU4) | 100 (2020/04/29) | 2.82E7 (DU1) 2.85E7 (DU2) 3.09E7 (DU3) 3.06E7 (DU4) |

National Institute of Standards and Technology online tables[132], $\rho = 1.225 \times 10^{-3}$ g/cm$^3$ is the mass density of air and $d$ is the distance between source and detector. Low energy sources like the Ag line of CalA and the Si line from CalD are the most affected by air absorption. Moreover, the Argon in the air, excited by the X-rays emitted by the $^{55}$Fe inside the sources, emits at an energy, 2.96 keV, very similar to the L line of Silver, thus contaminating the Cal A measurements. Even if the fluorescence yield of Ar is moderate (12.3% [9]) and only a limited fraction of the fluorescence photons is emitted towards the detector, this distortion must be accounted for. Knowing the distance between source and detector, fixed by the mechanical setup, and the air properties, we can derive the losses due to X-ray absorption in air. As for the Argon contamination, rather than performing complex calculations strongly dependent on geometric details, we solved this problem when we placed the DU, CalA included, in the chamber for TV tests, as described later. The counting rate measured with the SDD is then used to extrapolate the expected GPD rate through the relation

$$R_{GPD} = \sum^{E} R_{SDD}(E)\epsilon(E)\frac{1}{\tau(E)}\Big(\frac{A_{GPD}}{A_{SDD}}\Big)\Big(\frac{r_{flight}}{r_0}\Big) \qquad (6.2)$$

where $\epsilon(E)$ is the GPD efficiency at energy $E$, $\tau(E)$ is the transmission factor as defined in Eq. ( 6.1), $A_{GPD}$ is the area illuminated by the source on the GPD and $A_{SDD}$ the area illuminated on the SDD. The area ratio is needed to account for the loss of photons due to the difference in detector area, especially for CalC and CalD, as the 15×15 mm$^2$ area of the GPD is larger than the 7×7 mm$^2$ area of the SDD. For CalB the area ratio is 1, as the beam is collimated and there are no spatial losses, while for CalA we have to account for the shape of the Bragg arc that can be approximated with a 14×6 mm$^2$ rectangle on GPD, while on the SDD a 7×6 mm$^2$ region is assumed to be illuminated. The sum over energies is due to the fact that the SDD has an higher spectral resolution than the GPD, and it is able to detect both $K_\alpha$ and $K_\beta$ emission line that are not resolved by the GPD and hence considered together. Finally, $r_{flight}$ is the emission rate of the $^{55}$Fe radioactive source at the beginning of the mission as reported in Table 6.2), while $r_0$ is the emission rate at the time of the measurement. By multiplying for their ratio, the expected counting rate of each source at the beginning of the mission can be estimated. Of course due to the corrections needed and the contamination by Argon K$\alpha$ line in air, the measurements taken during TV tests provide a better estimate of the flight rate. We also recall here that we supposed what the nominal activity will be at launch, while the real activity will be determined only at a later time. Since for the source employed during the SDD tests of CalC the emission rate is unknown, we employ the nominal activity to estimate the flight rate. In Table 6.3, for each calibration source, we show the experimental parameters common to all the measurements with the SDD: the area illuminated on the SDD and the GPD, the air path, i.e. the linear distance between the source and the SDD, the energy of each line detected by the SDD, the GPD efficiency at that energy and the air mass-attenuation coefficient at that energy. The image of the X-rays coming from the sources is taken with an Andor iKon-M 934 CCD camera (see Fig. 6.7 (b)). Since the CCD surface (13×13 mm$^2$) is similar to the GPD one (15×15 mm$^2$) the image on the former is similar to the one on the latter.

**Table 6.3.** Parameters common to the SDD measurements of all the Flight Models. Energy
lines from [158], mass attenuation coefficients from [132].

| Source | Area on SSD $A_{SDD}$ [mm$^2$] | Area on GPD $A_{GPD}$ [mm$^2$] | Air path $d$ [cm] | Energy (Line) [kev] | GPD efficiency $\epsilon$ | Mass attenuation coefficient $\mu$ [cm$^2$/g] |
|---|---|---|---|---|---|---|
| CalA | 42.0 | 84.0 | 4.88 | 2.9 (Ag L$\alpha$) | 0.150 | 165.3 |
|      |      |      |      | 3.1 Ag (L$\beta$) | 0.134 | 140.4 |
|      |      |      |      | 5.9 (Mn K$\alpha$) | 0.025 | 41.74 |
|      |      |      |      | 6.5 (Mn K$\beta$) | 0.018 | 31.64 |
| CalB | 7.1 | 7.1 | 5.39 | 5.9 (Mn K$\alpha$) | 0.025 | 41.74 |
|      |     |     |      | 6.5 (Mn K$\beta$) | 0.018 | 31.64 |
| CalC | 49.0 | 225.0 | 5.39 | 5.9 (Mn K$\alpha$) | 0.025 | 41.74 |
|      |      |       |      | 6.5 (Mn K$\beta$) | 0.018 | 31.64 |
| CalD | 49.0 | 225.0 | 5.31 | 1.7 (Si K$\alpha$) | 0.217 | 777.8 |

### Thermal vacuum measurement with GPD

After the tests with SDD and CCD the calibration sources were installed in the
FCWs and integrated in the three DU Flight Models and the spare unit. We then
tested them in the INAF-IAPS TV facility in order to measure a more accurate
expected rate at launch, a detailed response of the DU to the sources installed in the
FCW and the centering of each source. During the tests we kept the temperature
fixed and the pressure inside the TV chamber at about 1E-6 mbar, much smaller
than 0.01 mbar that is the maximum pressure allowed before discharges occurs in
the DU, as derived by the Paschen law and considering a breakdown voltage of 3 kV
and a maximum distance of 2 cm to grounded elements, derived by HV board design.
We monitored the vacuum conditions at least every 8 hours. We point out that
these are the only complete set of measurements of the FCS on ground in vacuum
and hence the most representative (except for the flux). In Fig. 6.8 we show the
integrated DU ready to be tested in the TV chamber.

### Data analysis tools

We took the SDD spectra with the DPPMCA Display and Acquisition Software. We
then analyzed the spectra with custom Python scripts to fit the detected emission
lines with a gaussian+constant profile that provides the estimate for $R_{SDD}$ (see
Eq. 6.2). We acquired the CCD images with the Andor Solis software. Finally,
we acquired and analyzed the GPD measurements with the Python based ixpesw
toolkit developed by the IXPE collaboration.

## 6.3   Results

### SDD and CCD measurements results

In Figure 6.9 we show the spectra of the calibration sources of the the FCS FM2
that is representative of all the models, as the others do not present significant
differences. Since the spectrum of CalA is taken in air, the 2.98 and 3.15 keV Silver

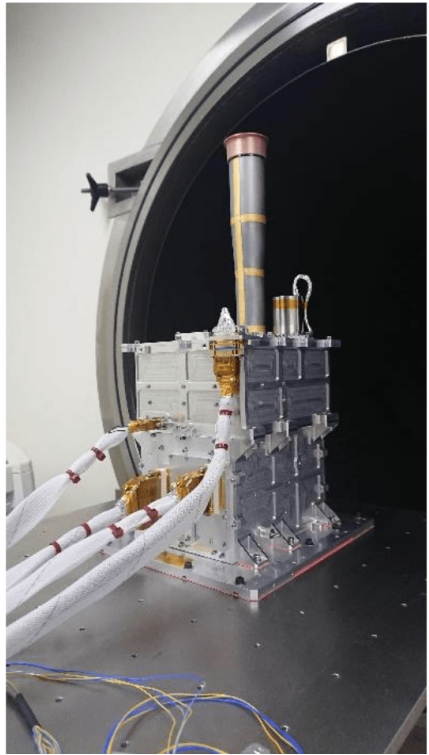

**Figure 6.8.** DU with FCS4 in front of the TV chamber before the tests.

fluorescence lines are polluted by the presence of the 2.96 and 3.19 keV lines due to the Argon in the air. At an energy of 4.2 keV, the line due to the escape of the 5.9 keV $K_\alpha$ Mn photons is well visible in the spectra of CalA, CalB and CalC . Due to the source geometry, this line is absent in the spectrum of CalD, where the Argon lines are however visible. The flux of the 1.7 keV line in the spectrum of CalD is heavily reduced by air absorption. The spectra of CalB and CalC are dominated by the 5.9 keV Mn $K_\alpha$ and 6.5 keV Mn $K_\beta$ lines. In Table 6.4 we report the expected GPD rate for the four FCSs obtained from Eq. 6.2 by considering the activity of the radioactive sources that will be employed on IXPE. In Figure 6.10 we show the images taken with the CCD of the items of FCS FM2 only since, again, this set is representative enough of the properties of the other three. As expected by the Bragg diffraction law, as seen in Fig. 6.3, the polarized diffracted photons at 3 keV and 5.9 keV appear as a curved strip across the detector surface. Another strip above the main one is interpreted as 6.5 keV photons that are diffracted at a different angle. CalB illuminates a 3 mm wide spot and CalC floods the whole detector surface. CalD, should also illuminate the whole detector area, but since the measurement is taken in air, most of the 1.7 keV photons are absorbed, and only 5.9 keV and 6.5 keV, that are diffracted at the right angle by the Silicon target, appear in the image as faint stripes.

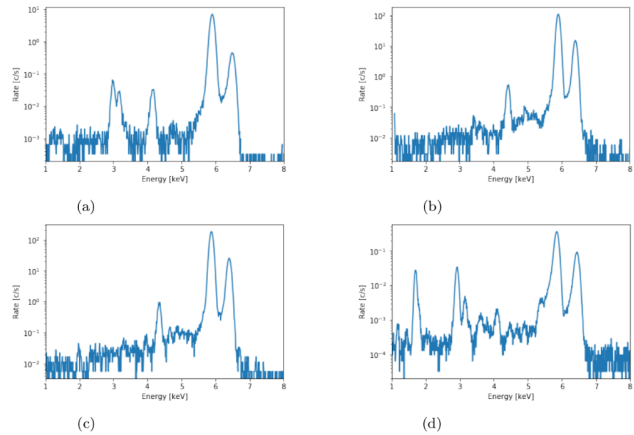

**Figure 6.9.** Spectra of the elements of the FCS FM2 taken with the Amptek XR-100SDD SDD. (a) Spectrum of CalA2: the closely packed peaks of the Ag $L_\alpha$ and $L_\beta$ lines at 2.98 keV and 3.15 keV are visible, as are the 5.9 keV Mn $K_\alpha$ line and its escape at 4.2 keV and the $K_\beta$ line at 6.5 keV. (b) Spectrum of CalB2 and (c) spectrum of CalC2: in both are visible the 5.9 keV Mn $K_\alpha$ line and its escape at 4.2 keV and the Mn $K_\beta$ line at 6.5 keV. (d) Spectrum of CalD: the lines at ∼3 keV due to Argon in the air are well visible, as is the 1.7 keV Si fluorescence.

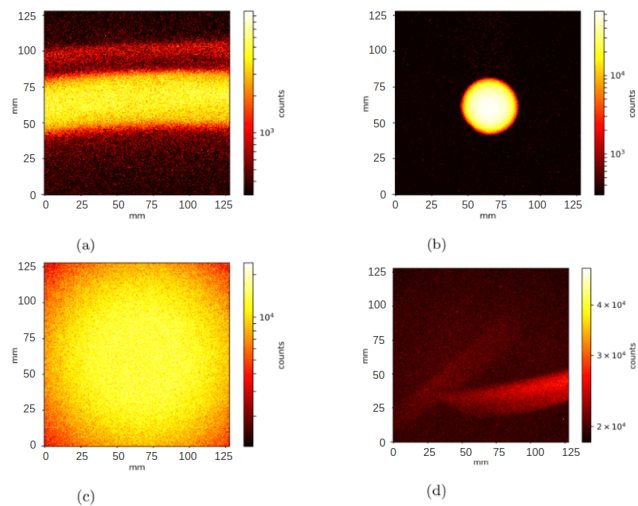

**Figure 6.10.** Images of the elements of the FCS FM2 taken with the Andor iKon-M 934 CCD camera. (a) CalA2: the central strip is due to the Bragg diffracted 3 keV and 5.9 keV polarized photons, while the upper strip is interpreted as 6.5 keV photons diffracted at a different angle. (b) CalB2: illuminates a 3 mm wide spot. (c) CalC2: illuminates the whole detector. (d) CalD2: due to heavy air absorption, the 1.7 keV fluorescence is not clear, instead two stripes are visible and interpreted as 5.9 keV and 6.4 keV photons diffracted by the Silicon target.

**Thermal vacuum results**

Since the polarized calibration source CalA is the most complex of the FCS, in Fig. 6.11 and 6.13 we show side by side the results of the four CalA in TV at 3 keV and 5.9 keV, respectively. In Figure 6.15 we show the TV results of the monocromatic sources CalB, CalC and CalD of the FCS FM2 that are representative since there are no great differences across the four sets. For all the sources we present (1) the folded modulation curve, fitted with a $\cos^2$ function, showing to which extent the source is (un)polarized; (2) the PHA spectrum; (3) the source image on the detector as a rate density map, i.e. the image of the source on the detector in units of counts per second per $mm^2$, showing the uniformity of the illumination.

Since, unlike the other sources, CalA is not a monochromatic source, its analysis is done separately for the 3 keV and 5.9 keV emission. For CalA we also show the charge density map, i.e. the plane in which the photoelectron track size (which is defined as the number of triggered pixels of the photoelectron track above a detection threshold) and PHA of each event are represented. On this plane we select a subset of events (shown as a white outline) associated to the emission line and remove the events generated close to the GPD beryllium window and to the Gas Electron Multiplier where they lose part of their energy in these passive materials. These events appear as thin strips respectively above and below the spot of the events to be selected. A $14 \times 6$ $mm^2$ region around the strip is selected to exclude from the analysis the events that occur on the edge of the detector and have incomplete photoelectron tracks. Finally, in Fig. 6.12 and 6.14 we present for the polarized calibration sources the comparative analysis of the strip produced by the diffracted photons at 3 keV and 5.9 keV, respectively: the strip is divided into ten rectangular $1.4 \times 4$ $mm^2$ regions and for each region the ADC peak, modulation and phase are plotted to check their spatial uniformity. These CalA results are corrected for the spatial differences in gain of the detector and corrected by the GPD response to unpolarized radiation, characteristics that are unique to each detector and that have been determined during the characterization campaign of the DUs.

In Table 6.4 we summarize the results of the measurements in TV of all the four FCS: in particular we report the flight rate without and with dead time correction (within parentheses), the Minimum Detectable Amplitude (MDA) reached, the measured modulation amplitude and polarization angle phase.

## 6.4 Discussion

Table 6.4 shows and compares the results of the test with SDD and with the DU in thermal vacuum. The results with the DU in TV are presented with and without (within parentheses) the dead time correction. Since the dead time of the SDD is almost negligible, the results of the extrapolation from the SDD are to be compared with the one from GPD corrected for the dead time. Dead time corrected rate will be important to establish the real activity of the radioactive sources. The correction is particularly important for CalC that is a bright source. The raw counting rates without dead time correction are used when dealing with the statistical significance of a measurement. The extrapolated GPD rate from the SDD measurements are in good agreement with the rates measured in TV with the GPD in the case of

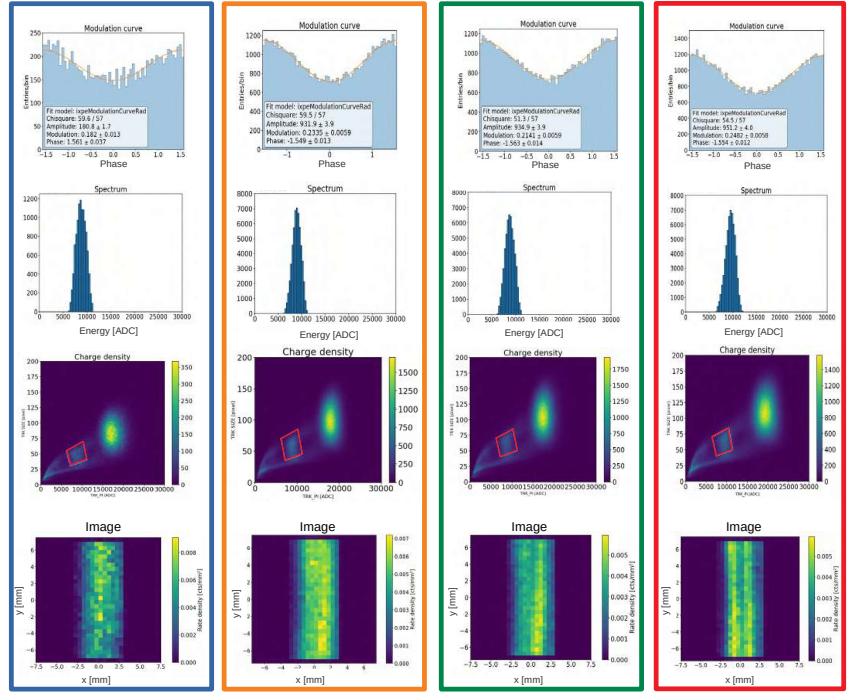

(a)          (b)          (c)          (d)

**Figure 6.11.** Comparison of the four CalA models for the Ag fluorescence emission at 3 keV in TV. From top to bottom for CalA1 (a), CalA2 (b), CalA3 (c) and CalA4 (d): modulation curve; charge density-selected source spectrum; charge density plot with the applied cut in energy and track size outlined in red; image of the source on the GPD as rate density map.

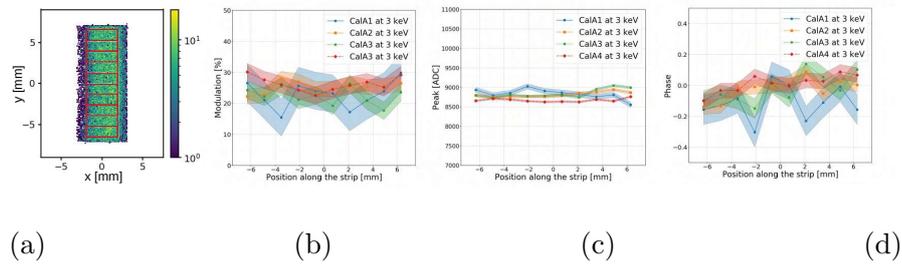

(a)          (b)          (c)          (d)

**Figure 6.12.** Comparison of the strip analysis of the four CalA models for the Ag fluorescence emission at 3 keV in TV. The shaded area represents the error on the measurement. (a) subdivisions of the strip for analysis highlighted in red; (b) modulation as a function of position along the strip; (c) energy peak in ADC as a function of position along the strip; (d) phase as a function of position along the strip.

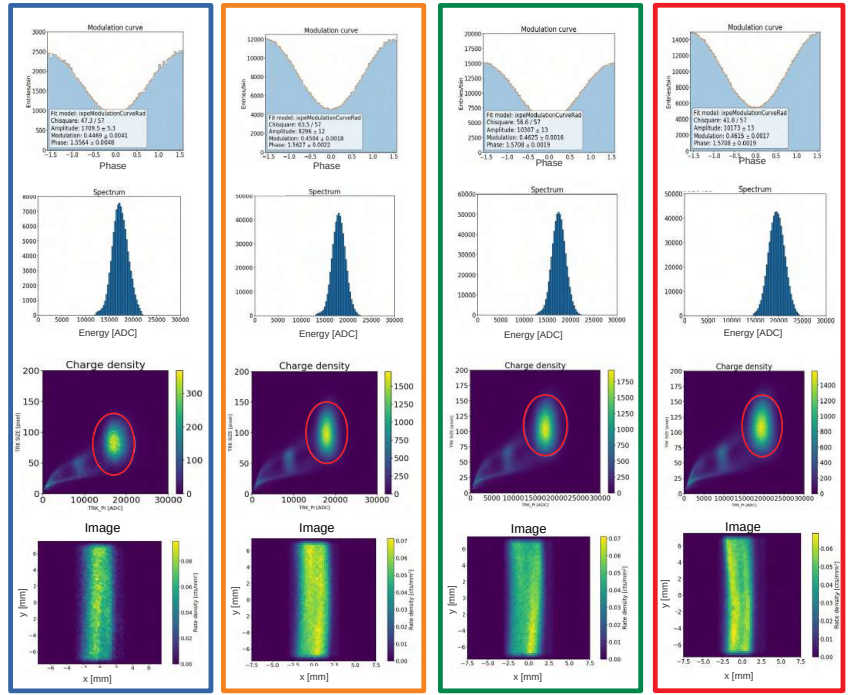

(a)            (b)            (c)            (d)

**Figure 6.13.** Same as Fig. (6.11) but for the Mn emission at 5.9 keV in TV for CalA1 (a), CalA2 (b), CalA3 (c) and CalA4 (d).

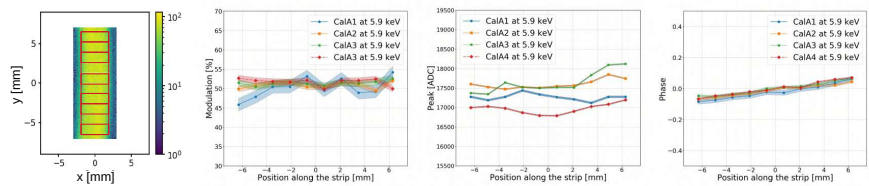

(a)            (b)            (c)            (d)

**Figure 6.14.** Same as Fig. (6.11) but for the Mn emission at 5.9 keV in TV.

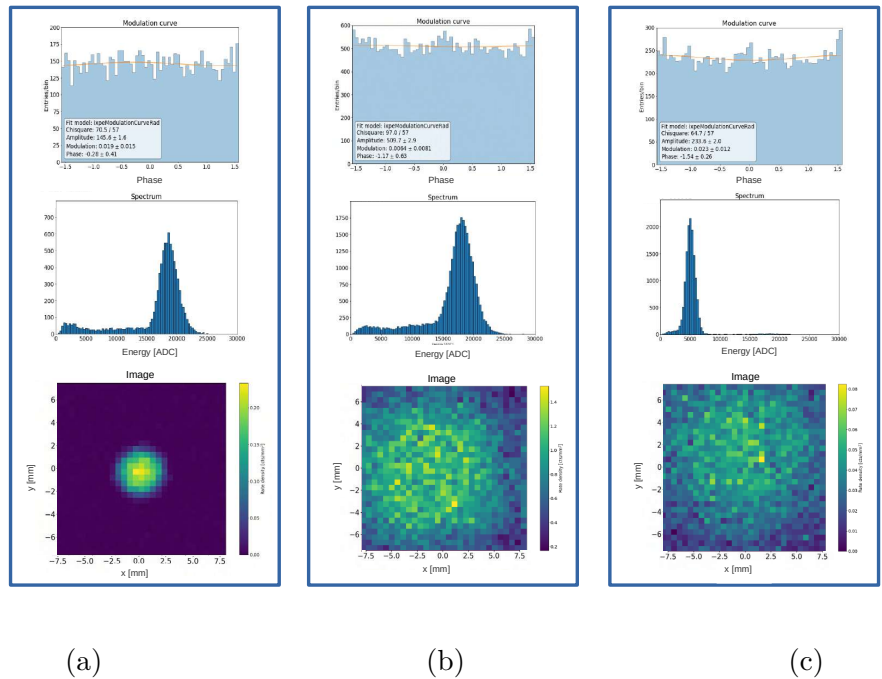

|  (a)  |  (b)  |  (c)  |

**Figure 6.15.** Results for CalB2 (a), CalC2 (b) and CalD2 (c) in TV. Top row: modulation curve; middle row: source spectrum; bottom row: image of the source on the GPD as rate density map.

CalA at 5.9 keV, CalB and CalD. The rates of CalC are overestimated because the geometric correction assumes a perfectly uniform illumination, while as it is shown from the images with CCD and GPD, the central region is more illuminated than the corners of the detector. Morover, since the real activity of the source used during the measurement is unknown, it is possible it was higher than the nominal value. The rates of CalA at 3 keV appear to be higher in the SDD spectra because of the Argon contamination. The systematic underestimation in the expected and measured rates of CalD is explained by the difficulty in estimating correctly the effect of air absorption at 1.7 keV that removes most of the photons. We were therefore confident that we would obtain the required statistical significance from the flight activities.

The flight counting rates extrapolated by the TV measurements are all above the requirements set in Table 6.1, with the differences across the sets that can be ascribed to the different energy resolution of each detector. The images of the source on the GPD (e.g. Fig. 6.11, 6.13 and 6.15 for the FCS FM2) are compatible with the one observed with the CCD (Fig.6.10). The diffraction stripes observed in CalD with the CCD can be easily removed in the analysis phase by applying a cut in energy, therefore their impact on the calibrations is null.

In Fig. 6.12 and 6.14 we also show and compare across the four FCS the analysis of the Bragg arc produced by the polarized calibration sources: the energy response of the detector, traced by the value of the peak in ADC, and the modulation amplitude are constant across the strip, with fluctuation small enough to be negligible during the in flight calibrations. Moreover the value of the modulation, the peak position in energy and the phase are in good agreement across the four sets, especially in the central region of the GPD. As expected from the law of Bragg diffraction, the polarization angle, traced by the phase, stays tangent to the Bragg arc, with the observed change in sign due to the choice of the reference frame, with zero at the center of the detector, where the polarization changes direction. The four CalA show a modulation amplitude that is consistent with radiation polarized at 67% and 69% at 3 and 5.9 keV, respectively, as expected from [67].

The unpolarized collimated sources, CalB, show modulations consistent with no polarization. The unpolarized calibration sources CalB, CalC and CalD show modulations consistent with no polarization, except for the case of CalD3, where a 6% modulation amplitude is observed with a $3\sigma$ significance level because of detector spurious modulation (see also Figure 11.1 showing the same source measured in-flight).

However, CalD will be mostly employed to monitor the gain, and only partially to check for the presence of spurious modulation.

IXPE orbits at 600 km of altitude, with an orbital period of 5.76 ks with 35% of this time affected by Earth occultations of X-ray targets depending on the source declination. This fraction of orbital period can be used for calibration purposes: during ∼2 ks long sessions, the three Detector Units will be calibrated one at time in order to guarantee that at least two DUs will remain operative and reduce the risk associated with a failure of the FCW mechanism. Moreover this strategy allow for limiting the count rate for effective downlink. Eventually, all three DUs will be exposed to all four calibration sources. These calibrations will monitor potential changes in detector gain, modulation factor, polarization response and, within the limits of statistics, spurious modulation. The results will be cross-checked with the one performed on the ground, further advancing our understanding of the technology of the Gas Pixel Detector and investigating possible secular variations of physical parameters.

In Table 6.4 we also report the number of orbits necessary to reach an MDA of 2%, corresponding to an absolute error of 1% in the determination of the modulation amplitude, since from [143] the uncertainty on the measurement of the amplitude is given by MDA/2.

This means that Calibration of the DU with CalA at 3 keV on four orbits, for a total of a three hours long calibration, would allow us to reach a level of 22 sigmas at 3 keV and 45 sigmas at 5.9 keV. Finally the measured modulation is convoluted with the GPD modulation factor to obtain the polarimetric sensitivity. In order to determine these values, the mean raw rate from TV measurements without cuts and dead time correction is used.

**Table 6.4.** Comparison of the results of the SDD tests with the ones performed with GPD in TV on the four FM of the FCS. The results of CalA at 3 keV marked with an * are affected by Argon contamination. The results within parentheses are the rates with dead time correction. The last two columns show the mean time needed to acquire enough counts, based on the mean rate from each source, to reach an absolute error of 1% on the modulation amplitude and the number of IXPE orbits needed (assuming that calibrations are performed during the 35% of the orbit during Earth occultation of celestial sources). This enables to check the polarization of the CalA sources, and to verify that the sources CalB, CalC, and CalD are unpolarized.

| Source | Energy [keV] | Flight rate from SDD [c/s] | Flight rate from DU in TV [c/s] | MDA [%] | Modulation [%] | Phase | Mean time to 1% absolute error [ks] | Orbits to to 1% absolute error |
|---|---|---|---|---|---|---|---|---|
| CalA1 | 3 | 5.5*±0.1 | 3.03±0.04 (3.05±0.04) | 2.8 | 18.2±1.3 | 1.56±0.04 | 11.488 | 4.34 |
| CalA2 | | 5.5±0.1 | 3.56±0.02 (3.58±0.02) | 2.2 | 23.4±0.6 | -1.55±0.01 | | |
| CalA3 | | 5.7±0.1 | 3.34±0.02 (3.40±0.02) | 2.2 | 21.4±0.6 | -1.56±0.01 | | |
| CalA4 | | 5.4±0.1 | 3.42±0.02 (3.47±0.02) | 2.1 | 24.8±0.6 | -1.55±0.01 | | |
| CalA1 | 5.9 | 40.2±0.2 | 40.3±0.1 (40.6±0.1) | 0.9 | 44.7±0.4 | 1.536±0.005 | 0.932 | 0.35 |
| CalA2 | | 45.6±0.4 | 41.0±0.1 (41.5±0.1) | 0.7 | 45.0±0.2 | 1.563±0.002 | | |
| CalA3 | | 48.7±0.4 | 42.0±0.1 (42.5±0.1) | 0.6 | 46.3±0.2 | 1.571±0.002 | | |
| CalA4 | | 44.7±0.4 | 41.2±0.1 (42.1±0.1) | 0.6 | 46.2±0.2 | 1.570±0.002 | | |
| CalB1 | 5.9 | 70±1 | 65±2 (71±2) | 13.4 | 7.7±4.3 | 0.5±0.3 | 0.642 | 0.24 |
| CalB2 | | 79.7±0.9 | 71.4±0.8 (76.3±0.8) | 4.7 | 1.9±1.5 | -0.3±0.4 | | |
| CalB3 | | 83±1 | 72.8±0.8 (78.5±0.9) | 5.0 | 0.6±1.6 | -1.2±1.4 | | |
| CalB4 | | 83±1 | 77.8±0.9 (83±1) | 5.0 | 0.5±1.6 | -0.7±1.5 | | |
| CalC1 | 5.9 | 256.6±0.7 | 131.7±0.6 (170.2±0.7) | 1.9 | 0.8±0.6 | 0.8±0.4 | 0.300 | 0.11 |
| CalC2 | | 279±1 | 182.8±0.5 (234.7±0.7) | 1.2 | 0.6±0.8 | -1.2±0.6 | | |
| CalC3 | | 271±1 | 176.8±0.7 (224.6±0.8) | 1.6 | 1.8±0.5 | -1.5±0.1 | | |
| CalC4 | | 276±1 | 127.9±0.4 (159.0±0.5) | 1.4 | 0.3±0.5 | 0.8±0.8 | | |
| CalD1 | 1.7 | 89.4±0.2 | 105.8±0.7 (106.8±0.7) | 2.7 | 1.1±0.9 | -0.5±0.4 | 0.408 | 0.15 |
| CalD2 | | 84.6±0.2 | 113±1 (114±1) | 3.6 | 2.3±1.2 | -1.5±0.3 | | |
| CalD3 | | 101.0±0.4 | 121±1 (122±1) | 3.6 | 6.0±2.0 | -1.3±0.1 | | |
| CalD4 | | 103.5±0.2 | 111±2 (112±2) | 6.7 | 3.2±2.3 | -1.4±0.4 | | |

# Chapter 7

# Effect of Background on IXPE observations

In imaging X-ray astronomy, the background is expected to influence the observations of very faint sources, depending on various factors, with the angular resolution being the most important. Below a certain level of source luminosity, the fluctuations on the background rate prevail on the source rate and the observation becomes background limited. The angular resolution of IXPE, in terms of half-energy width (HEW), defined as containing half of the counts from a certain direction, is < 30" (full opening angle). Projected onto the readout plane, the HEW is a round area of $\sim 0.22$ mm$^2$ with the mirror focal length of 4 m. For point-like sources the source signal to background ratio is relatively high enough, while for extended sources this ratio is usually significantly smaller and has to be carefully evaluated.

In this Chapter, I will show the effect of the sources of background described in Section 5.2 on the expected polarization degree of a selection of extended sources such as the Tycho, CasA, and SN1006 SNRs, the MSH 15-52 PWN, and finally the G0.11-0.11 molecular cloud. The dilution of the polarization degree due to each background source, and then due to their total effect, is evaluated through the complement to unity of Eq. 5.1, so that we express in the following the dilution as

$$Dilution = 1 - \frac{P_{dil}}{P_0} = 1 - \frac{1}{1 + \frac{R_B}{R_S}} \quad . \tag{7.1}$$

The lower the value of this dilution, the smaller the impact of the background on the observed signal. In Table 7.1, for each of the previously listed sources, the energy bins, the area, flux, and IXPE rate considered for the background evaluation for the rest of the Chapter are shown. For the SNR Cas A and Tycho, I consider spatially the whole source, while in energy I consider the full 2−8 keV IXPE energy band and the line-less continuum in the 4−6 keV energy range. For the SN 1006 SNR, being its spectrum mostly non-thermal, I consider the full 2−8 keV energy rim, but due to its size I select a circular region on the N-W rim. For the PWN MSH 15-52 I again select the full energy band for the same reason of SN 1006, but I consider a rectangular region inside the nebula that includes the bright lower jet. Finally, for the molecular cloud G0.11-0.11, I consider the full energy band and the 4−8 keV band, were the polarized non-thermal emission is dominant.

**Table 7.1.** Energy bins, observation duration, area of the region considered for the background evaluation, source flux, and the IXPE counting rate for each source.

| Target | Energy (keV) | Duration (s) | Area (arcmin$^2$) | Source Flux (erg/cm$^2$/s) | IXPE Rate (c/s/cm$^2$) |
|--------|--------------|--------------|-------------------|----------------------------|------------------------|
| Cas A | 2-8 | 100000 | 3.22E+1 | 1.30E-09 | 2.49E+01 |
|       | 4-6 |        |         | 4.58E-11 | 9.15E-01 |
| Tycho | 2-8 | 1000000 | 7.85E+1 | 2.00E-10 | 2.77E+00 |
|       | 4-6 |         |         | 2.04E-11 | 4.07E-02 |
| SN1006 | 2-8 | 1000000 | 1.26E+1 | 2.00E-11 | 1.79E-01 |
| MSH 15-52 | 2-8 | 1000000 | 1.13E+01 | 3.14E-11 | 2.10E+00 |
| G0.11-0.11 | 2-8 | 1000000 | 7.07E+0 | 1.17E-12 | 1.55E-02 |
|            | 4-8 |         |         | 1.10E-12 | 1.04E-02 |

## 7.1 Instrumental background effect on extended sources

In Xie et al. [178], a paper that i co-authored, we presented Monte Carlo simulations of the GPD instrumental background, developed rejection techniques to distinguish it from real celestial data, and studied its impact on real observations of point and extended X-ray sources.

**Simulation of the GPD instrumental background**

We performed the instrumental background simulations with the matter-radiation interaction Monte Carlo simulator GEANT4 [version 0.03.p011, 3]. The inputs of the simulations are the geometric model of the spacecraft, the background spectra of the orbit environment and the physics model of the interactions. Then, mimicking the data processing and analysis of real data, we applied selections in order to determine the background rates.

The geometric model is based on the latest GPD design [8] and implemented in two steps. The first step is the definition of the GPD structure.

The second step is to build on top of the first step a model of the whole satellite in order to evaluate the background in a realistic configuration. Indeed, the materials around the detector are essential for background simulation: background particles mainly come from the region outside the field of view, producing secondary particles as they cross the mass surrounding the detector. Some of them interact with the detector, leaving a stream of charges in the gas which are eventually detected by the GPD. In Fig. 7.1 the relevant DU components are shown, i.e. the GPD, the calibration wheel, the collimator, the detector shielding box, and the back-end electronics, that we reproduced for the simulation. The IXPE circular, equatorial, 600 km-altitude orbit minimizes the detector background and optimizes the observing efficiency by minimizing the passage in the South Atlantic Anomaly (SAA), achieving a minimum duty cycle of 60% from the 94.6 minute-long orbital period. In the radiation environment outside the SAA, the background sources which need to be considered are primary cosmic rays (protons, electrons, positrons and alpha particles) and secondary cosmic rays (protons, electrons, positrons). Primary cosmic rays are accelerated by celestial sources and travel through the galaxy before reaching the Earth. They are made up of an unmodulated term plus a modulated term depending on the Solar cycle, that is assumed to be at its minimum during the IXPE operations, providing a larger amount of cosmic particles arriving to the Earth magnetosphere. The dominant component ($\sim$90%) are protons. When primary particles impinge on the top of the atmosphere and interact with residual gas molecules, showers of secondary cosmic rays are produced and some of these eventually go upward and escape the atmosphere [96]. CXB is also is reflected back by the Earth's atmosphere. We also study other albedo components generated in the atmosphere, such as gamma-rays, neutrons and the extragalactic one due to scattered CXB. The latter is usually the dominant background component for large field of view (FOV) X-ray telescopes, and hence negligible for IXPE. The input spectra are originally derived from simulations for the LOFT mission proposal [26] which assumes a low Earth orbit with an altitude of 600 km and an inclination of 5°, then adapted for XIPE with a 0° inclination, and eventually used for IXPE in this work. The input spectra of these sources are detailed in Fig.7.2.

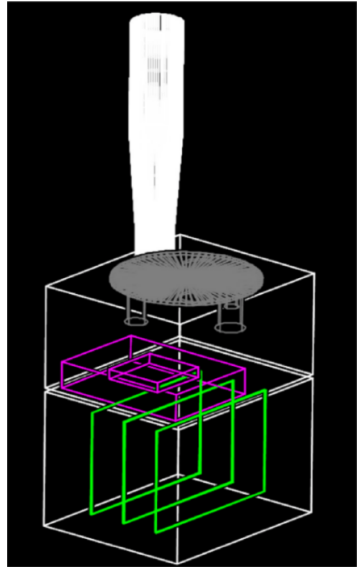

**Figure 7.1.** Detector Unit (DU) mass model employed in the GEANT4 simulations [178]. The collimator (white), calibration wheel and calibration sources (gray), GPD (magenta) and PCBs (green) are shown. Structural details are hidden for a clearer illustration.

In order to fully describe the interactions in the space environment, electromagnetic, hadronic, and decay processes are included in the simulation with a dedicated library called "G4LivermorePolarizedPhotoElectricGDModel" that properly takes into account the emission directions of the photoelectrons [37]. The estimated background rate (that from now on we define as "unrejected" instrumental background) is of $4.47\times10^{-2}$ counts s$^{-1}$ cm$^{-2}$ per DU in the 2$-$8 keV energy range. This value is about 2.9 times larger than the requirement, yet it is still negligible when observing point like sources. The PolarLight cubesat, that also hosts a GPD, found an internal background that, after accounting for the different orbit and mass distribution with respect to IXPE, is compatible with what we found [70, 183].

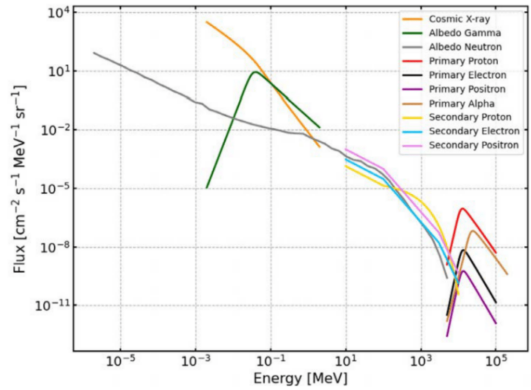

**Figure 7.2.** Sources of instrumental background implemented in the GEANT4 simulation [178].

### Instrumental background rejection methods

Background rejection methods have to be correctly defined and applied in order to remove the noise from the signal without affecting the latter. Background rejection methods are based on the fundamental differences between photoelectron and background tracks. A straightforward comparison is present in Fig. 7.3, which shows two tracks with a similar energy deposit in the detector but a different origin. The case (a) is a classical photoelectron track for a 5.9 keV X-ray photon decayed from $^{55}Fe$. From the Bethe-Bloch law, the energy loss increases as the emitted particle velocity decreases [1] [1]. Therefore, photoelectron tracks show the maximum charge density at their end, which is called the "Bragg peak". The dashed line in the figure represents the photoelectron ejection direction. The case (b) shows, instead, a track generated by an energetic proton (tens of GeV). The charge particle leaves the gas producing a long, discontinuous string of electrons by ionizing the gas medium. Background tracks do not always look like Fig. 7.3 (b), they may be also similar to photoelectron tracks, depending on the particle type, the kinetic energy, the incident direction and the interaction process. On the other hand, some photoelectron tracks also deviate from the ideal cases. The principle for background rejection is to remove background events as much as possible while keeping source photons. This is done by parametrizing the properties efficient in recognizing the background tracks. The rejection parameters are: Pulse Invariant, Track size, Skewness, Elongation, Charge density, Cluster number, and Border pixels.

The Pulse Invariant (PI) is the sum of the charge of the track, which is proportional to the deposited energy. It is conventionally expressed in ADC channels and has to be corrected for the possible non-uniformity in the detector gain. Only tracks

---

[1] As a reminder for the reader, the relativistic expression of the Bethe-Bloch law is:

$$-\left\langle \frac{dE}{dx} \right\rangle = \frac{4\pi}{m_e c^2} \cdot \frac{nq^2}{\beta^2} \cdot \left( \frac{e^2}{4\pi\epsilon_0} \right)^2 \cdot \left[ \ln\left( \frac{2m_e c^2 \beta^2}{I \cdot (1-\beta^2)} \right) - \beta^2 \right] \tag{7.2}$$

where where a particle with speed $v$, charge $q$, and energy $E$ travels a distance $x$ through a target having electron number density $n$ and mean excitation potential $I$. $c$ is the speed of light, $\epsilon_0$ the vacuum permittivity, $\beta = v/c$, and $e$ and $m_e$ are the electron charge and rest mass respectively.

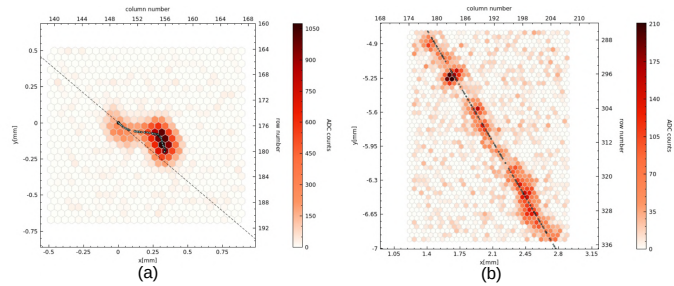

**Figure 7.3.** Imaging of tracks with a random electronic noises from the simulation [178]. The dots represent the energy deposit 2D positions inside the gas recorded by GEANT4. (a): a track from a $^{55}$Fe decay X-ray photon and dashed line is the photoelectron eject direction.(b): a track from an energetic cosmic ray proton.

with final energy in the $2-8$ keV energy range are considered in the background rate calculations.

The Track size is the number of pixels above the threshold in the main cluster, that is, in the largest group of contiguous pixels of the event. With the same energy deposit, background events usually have larger track size than real photoelectrons. Skewness, or third standardized moment, refers to the asymmetry of the energy distribution in the track along the major axis. As the mean energy loss of a charged particle varies inversely with its energy, for a photoelectron of a few keV the energy loss increases progressively towards the end point, forming a skewed track. On the contrary, for a background charged particle of the order of MeV or GeV the energy loss, and therefore the ionization density, is constant and the track has a low skewness. For example, in Fig. 7.3, photoelectron track (a) is more skewed (asymmetric) than the background track (b), allowing us to discriminate between them.

The Elongation is defined as $\sqrt{M2L/M2T}$, where $M2L$ and $M2T$ are the longitudinal and transverse second moments of the track, and their ratio refers to the eccentricity of the charge distribution.

The Charge density is defined as the energy (PI) divided by track size, which is expected to be lower for background than for photoelectrons. As an example, the relativistic background, known as the minimum-ionizing particle (MIP), has energy losses of about 2 MeV per g cm$^{-2}$ in light materials [80], while the photoelectron energy density is about 10 times larger than that of a minimum ionizing charge particle [137]. A comparison is shown in Fig. 7.4: the position of the peak from the background is lower than that of a source simulated as a power-law spectrum with a photon index of 2. Background identification benefits from this significant deviation.

The Cluster number is the number of disjoint tracks that appear after applying the clustering algorithm in the region of interest. Background particles are more likely to produce discontinuous tracks, therefore more than one cluster may be identified by the clustering algorithm. On the contrary photoelectrons are mainly grouped as single cluster.

Finally, Border pixels is the number of pixels in the track which are at the edge of the ASIC. Background particles entering in the gas from the side of the wall have a larger probability of leaving a track with pixels on the border, hence making them easily detectable. Having defined the event properties that allow to discriminate between background and real photoelectrons, we can establish a rejection method to apply to the real data in order to reduce the impact of the background.

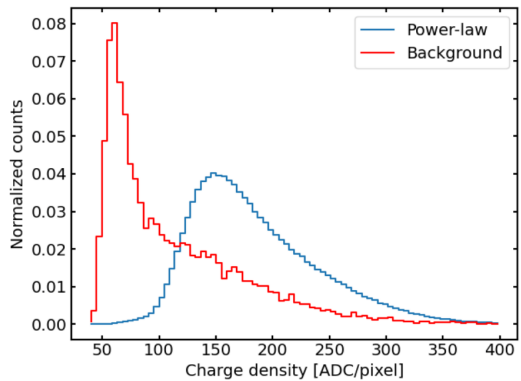

**Figure 7.4.** Comparison of the charge density distributions expected from a power-law spectrum with the photon index of 2 (in blue) and the background including all the simulated components (in red).

The first step is the selection of three energy bins in the $2-8$ keV energy range: $2-3.4$ keV, $3.4-5$ keV, and $5-8$ keV. From the instrument calibrations, the energy resolution is about 18% at 5.9 keV [8] and at the other energies resolutions are described approximately as a function that scales as $E^{-1/2}$, therefore three energy bins are a reasonable assumption.

Then, the event parameters are defined in terms of acceptance from a flat spectrum source and cuts are applied to a background simulation.

The total background reduces to $2.62 \times 10^{-2}$ counts s$^{-1}$, i.e. $1.16 \times 10^{-2}$ counts s$^{-1}$ cm$^{-2}$ in $2-8$ keV per DU, with an extra rejection efficiency of 75.0%, though it is still larger than the requirement by a factor of 2.9. The calculated instrumental background level is comparable with the one observed for the GPD on board of the PolarLight cubesat [48, 70, 183].

Applying the cuts to a simulated polarized Crab-like point source (with no background included) allows to determine the acceptance level for real events. We found that [178], for such Crab-like source in the $2-8$ keV energy band, the background rejection methods removes 14.9% of the source events.

All in all, with the background rejection methods developed in Xie et al. [178], we removed 92.6% of the simulated instrumental background events. We are left with a residual background level of $1.16 \times 10^{-2}$ counts s$^{-1}$ cm$^{-2}$ per DU in the $2-8$ keV energy range. Of course this is the result of simulations, but from late December 2021 we started collecting background data from IXPE in orbit, and we plan to apply in the near future the rejection method which we developed to confirm that the background can be rejected sensitively.

In Table 7.2 the result of the expected dilution, obtained through Eq. 7.1, for each source due to the GPD instrumental background are shown. I compare its effect in the case in which the rejection methods are applied or not. I find that for Tycho, CasA, and MSH 15-52, the difference between rejecting or not the instrumental background is generally small ($\ll 10\%$). However, for the SNR SN1006 and the MC G0.11-0.11, the rejected background in the energy ranges of interest already causes a dilution of 16% and 53-69%, respectively. This means that for these sources, instrumental background mitigation techniques will need to be applied when observing them.

**Table 7.2.** In this table are listed, for each source, the IXPE counting rate, the dilution factor from the rejected and unrejected instrumental background.

| Target | Energy (keV) | IXPE Instrumental background rate (rejected) (c/s/cm²) | IXPE Instrumental background rate (unrejected) (c/s/cm²) | Dilution (rejected) (%) | Dilution (unrejected) (%) |
|---|---|---|---|---|---|
| Cas A | 2-8 | 3.35E-02 | 1.34E-01 | 0.1 | 0.5 |
| | 4-6 | 8.42E-03 | 3.36E-02 | 0.9 | 3.5 |
| Tycho | 2-8 | 3.42E-02 | 1.36E-01 | 1.2 | 4.7 |
| | 4-6 | 8.12E-03 | 3.24E-02 | 2.0 | 7.4 |
| SN1006 | 2-8 | 3.48E-02 | 1.39E-01 | 16.3 | 43.7 |
| MSH 15-52 | 2-8 | 3.48E-02 | 1.39E-01 | 1.6 | 6.2 |
| G0.11-0.11 | 2-8 | 3.44E-02 | 1.37E-01 | 68.9 | 89.9 |
| | 4-8 | 1.17E-02 | 4.67E-02 | 53.0 | 81.8 |

## 7.2 Diffuse Galactic background effect on extended sources

In this thesis, I employ the spectral model of the GPDE presented in Ebisawa et al. [42] to estimate the IXPE counting rate due to this background component.

To obtain these estimates I extracted the blanksky-subtracted and point-source-removed spectrum from regions outside of the extended sources here considered. Where possible, I stacked the spectral data from many observations to increase the statistics. I fitted the spectrum of the background region with the model from Ebisawa et al. [42], consisting in a power-law plus gaussian emission lines relative to MgXII Ly$\alpha$, low ionized Si, SiXIII K$\alpha$, SiXIV Ly$\alpha$, SXV K$\alpha$, SXV K$\alpha$, SXVI Ly$\alpha$, Ar K$\alpha$, Ca K$\alpha$, Fe, and Ni K$\alpha$. Finally, to obtain the IXPE counting rate, I fed this spectral fit to the IXPE observation simulator ixpeobssim with a uniform disk morphology the same size as the region of interest.

In Table 7.3 the components of the spectral model from Ebisawa et al. [42] are reported. In Table 7.4 the $2-8$ keV GPDE flux and reduced $\chi^2$ fit obtained for each extended source by fitting the background spectrum with the model from Ebisawa et al. [42] presented in Table 7.3 are reported. I do not fit the MC G0.11-0.11 with this model, as the peculiar case of the MC will be discussed in grater detail in Chapter 9 and 10.

In the following sections, I describe in more detail the estimates obtained for each source.

### CasA

The spectrum from a rectangular region close to CasA, see Fig. 7.5 (a)) was extracted stacking the data from three Chandra observations (OBSID 194, 234, and 235) for a total of 6 ks. This shape is chosen to maximize the off-source area, because the Chandra observations used for the background extraction had CasA in a corner of the chip. The spectrum was fitted with XSPEC obtaining a reduced Chi2 of 0.94 (data and folded model shown in Fig. 7.6 (a)).

**Table 7.3.** Spectral components of the GPDE [adapted from 42].

| Component | Energy (keV) | Spectral Index | Notes |
|-----------|--------------|----------------|-------|
| Power Law | - | 2.5 | |
| Gaussian | 1.47 | - | MgXII, Lyα |
| Gaussian | 1.74 | - | Low ionized Si |
| Gaussian | 1.86 | - | SiXIII, Kα |
| Gaussian | 2.00 | - | SiXIV, Lyα |
| Gaussian | 2.45 | - | SXV, Kα |
| Gaussian | 2.62 | - | SXVI, Lyα |
| Gaussian | 2.9 | - | Ar, Kα |
| Gaussian | 3.12 | - | ArXVII, Kα |
| Gaussian | 3.7 | - | Ca, Kα |
| Gaussian | 6.52 | - | Fe |
| Gaussian | 7.4 | - | Ni, Kα |

**Table 7.4.** Reduced $\chi^2$ and 2-8 keV GPDE flux fitted for each extended source.

| Source | $\chi^2/dof$ | 2-8 keV background Flux ergs/cm$^2$/s |
|--------|--------------|----------------------------------------|
| CasA | 763.5/808 | 4.97E-12 |
| Tycho | 482.35/398 | 3.87E-12 |
| SN1006 | 203.47/188 | 5.67E-13 |
| MSH 15-52 | 445.96/397 | 4.47E-13 |

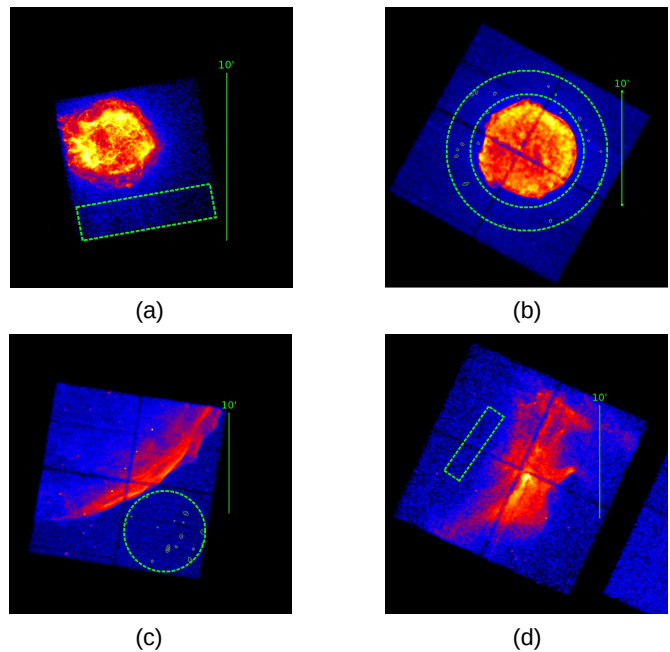

(a)          (b)

(c)          (d)

**Figure 7.5.** Diffuse Galactic plane background extraction regions (dashed lines) for (a) CasA, (b) Tycho, (c) SN 1006, and (d) MSH 15-52. 10' line shown for scale.

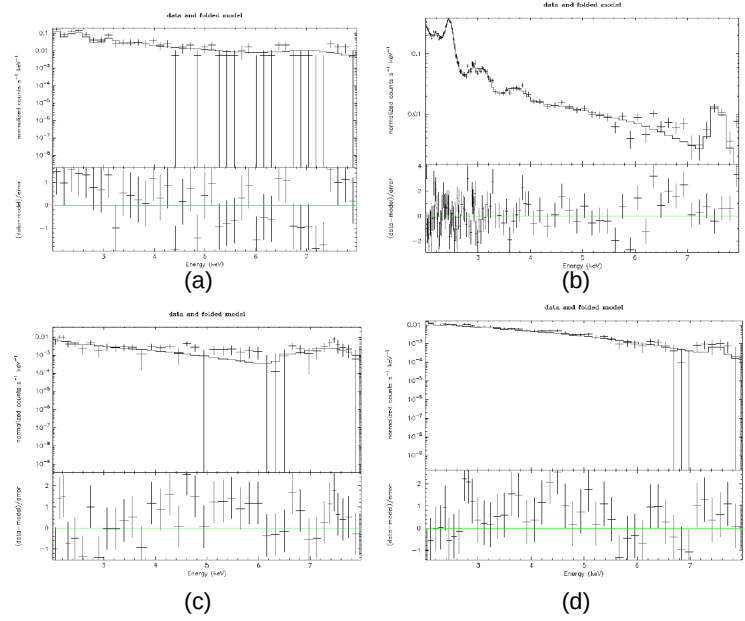

**Figure 7.6.** Diffuse Galactic plane background spectral fit and residuals for (a) CasA, (b) Tycho, (c) SN 1006, and (d) MSH 15-52.

**Table 7.5.** Flux, IXPE counting rate, and dilution factor due to the diffuse Galactic plane emission for each source.

| Target | Energy (keV) | DGPE Flux (erg/cm²/s) | IXPE DGPE Rate (c/s/cm²) | Dilution (%) |
|---|---|---|---|---|
| Cas A | 2-8 | 4.97E-12 | 2.42E-01 | 0.9 |
| | 4-6 | 8.45E-13 | 3.22E-03 | 0.4 |
| Tycho | 2-8 | 3.87E-12 | 7.70E-02 | 2.7 |
| | 4-6 | 4.26E-13 | 1.09E-03 | 2.6 |
| SN1006 | 2-8 | 5.67E-13 | 1.03E-02 | 5.5 |
| MSH 15-52 | 2-8 | 4.47E-13 | 9.03E-03 | 0.4 |
| G0.11-0.11 | 2-8 | 2.29E-12 | 1.47E-01 | 90.4 |
| | 4-8 | 1.22E-12 | 2.39E-02 | 69.7 |

The IXPE counting rate for the diffuse Galactic background in the $2-8$ keV band is 2.42E-01 c/s/cm$^2$, to be compared with the CasA counting rate of 24.9 c/s/cm$^2$. In the $4-6$ keV continuum, the IXPE counting rate for the diffuse Galactic background is 3.22E-03 c/s/cm$^2$, w.r.t. the source rate of 9.15E-1 c/s/cm$^2$. In both cases the impact of the GPDE is close to negligible, being $<1\%$ in the full energy band and $\ll 1\%$ in the continuum.

**Tycho**

The spectrum from an annular region around Tycho of area equivalent to the SNR (circle of radius 5 arcminutes, see Fig. 7.5 (b)) was extracted stacking the data from two Chandra observations (OBSID 10095 and 15998) for a total of 310 ks. The spectrum was fitted with XSPEC obtaining a reduced $\chi^2$ of 1.16 (data and folded model shown in Fig. 7.6 (b)).

The IXPE counting rate for the diffuse Galactic background in the $2-8$ keV band is 7.7E-2 c/s/cm$^2$, to be compared with the Tycho IXPE counting rate of 3 c/s/cm$^2$. In the $4-6$ keV continuum, the IXPE counting rate for the diffuse Galactic background is 1.09E-3 c/s/cm$^2$, w.r.t. the source rate of 4.07E-2 c/s/cm$^2$. In both cases dilution due to the GPDE is small, with a dilution factor of $\sim 3\%$.

**SN1006**

The spectrum from a circular region close to SN1006 of area equivalent to the region on the SNR rim (circle of radius 2 arcminutes, see Fig. 7.5 (c)) was extracted stacking the data from two Chandra observations (OBSID 13739 and 13742) for a total of 179 ks. The spectrum was fitted with XSPEC obtaining a reduced $\chi^2$ of 1.08 (data and folded model shown in Fig. 7.6 (c)).

The IXPE counting rate for the diffuse Galactic background in the $2-8$ keV band is 1.03E-02 c/s/cm$^2$, to be compared with the Tycho IXPE counting rate of 1.79E-01 c/s/cm$^2$. For this source, the dilution due to the GPDE is of $\sim 5\%$.

**MSH 15-52**

The spectrum from a rectangular region outside the MSH 15-52 plerion (see Fig. 7.5 (d)) was extracted from Chandra observation OBSID 5534 for a total of 49.5 ks. The spectrum was fitted with XSPEC obtaining a reduced $\chi^2$ of 1.12 (data and folded model shown in Fig. 7.6 (d)).

The IXPE counting rate for the DGPE in the $2-8$ keV band is 9.03E-03 c/s/cm$^2$, to be compared with the MSH 15-52 IXPE counting rate of 2.10E+00 c/s/cm$^2$. For this source, the dilution due to the GPDE is almost negligible, with a dilution factor $\ll 1\%$.

## 7.3 Cosmic X-ray background effect on extended sources

As CXB model for my simulations, I employ the estimate of Moretti et al. [98], that determined the flux and spectrum of the CXB using Swift-XRT archival data. They model the CXB as a power-law of index $\sim 1.5$ with normalization $3.7 \times 10^{-3}$ keV$^{-1}$

cm$^{-2}$ deg$^{-2}$ in the 1.5−7 keV range. I rescaled this model for the size of the IXPE FOV, and simulated it in ixpeobssim as a uniform source as large as the FOV.

In Table 7.6 are reported the flux, IXPE counting rate, and dilution expected from the CXB for the extended sources here considered. The dilution is negligible in all cases except for the MC G0.11-0.11 when the whole 2−8 keV band is considered. In this case the source polarization degree is diluted by the CXB by ∼9% in the 2−8 keV band and ∼3% in the 4−8 keV band.

**Table 7.6.** In this table are listed, for each source, the flux, IXPE counting rate, and dilution factor from the CXB.

| Target | Energy (keV) | CXB Flux (erg/cm$^2$/s) | IXPE CXB Rate (c/s/cm$^2$) | Dilution (%) |
|---|---|---|---|---|
| Cas A | 2-8 | 7.80E-13 | 1.34E-01 | 0.01 |
|  | 4-6 | 1.89E-13 | 7.80E-1 | 0.03 |
| Tycho | 2-8 | 7.80E-13 | 1.36E-01 | 0.05 |
|  | 4-6 | 1.89E-13 | 3.24E-02 | 0.06 |
| SN1006 | 2-8 | 7.80E-13 | 1.39E-01 | 0.08 |
| MSH 15-52 | 2-8 | 7.80E-13 | 1.39E-01 | 0.07 |
| G0.11-0.11 | 2-8 | 7.80E-13 | 1.37E-01 | 9.4 |
|  | 4-8 | 3.50E-13 | 4.67E-02 | 2.3 |

## 7.4 Discussion

In Table 7.7 are reported for each source the total dilution of the polarization degree due to the main sources of background: the instrumental background, the GPDE, and the CXB. For the brightest sources here considered, the CasA and Tycho SNR, and the MSH 15-52 PWN, the combined effect of these three sources of background is such that the dilution is ≪10%, so that the diluted polarization is till >90% of its intrinsic, undiluted value.

However, for fainter sources such as the SN 1006 SNR and the Molecular clouds in the Galactic center, the total dilution ranges to ∼20% for the former, to ∼80-90% for the latter.

In these cases, background subtraction techniques will be fundamental to correctly evaluate the polarimetric information. Rejection algorithms, dark, and Earth occultations in particular will be fundamental to identify and remove the GPD instrumental background.

In Chapter 10 I will show data analysis techniques that could be applied to the case of the Molecular clouds in the Galactic center to recover the intrinsic X-ray polarization degree, undiluted by the unpolarized backgrounds.

**Table 7.7.** In this table are listed, for each source, the total background counting rate (with and without instrumental background rejection), and dilution factor (with and without instrumental background rejection).

| Target | Energy (keV) | Total IXPE background rate (c/s/cm$^2$) | Total IXPE background rate (unrejected) (c/s/cm$^2$) | Dilution (%) | Dilution (unrejected) (%) |
|--------|--------|--------|--------|--------|--------|
| Cas A | 2-8 | 2.77E-01 | 3.77E-01 | 1.1 | 1.5 |
|       | 4-6 | 1.42E-02 | 3.94E-02 | 1.5 | 4.1 |
| Tycho | 2-8 | 1.13E-01 | 2.15E-01 | 3.9 | 7.2 |
|       | 4-6 | 9.45E-03 | 3.37E-02 | 2.3 | 7.6 |
| SN1006 | 2-8 | 4.66E-02 | 1.51E-01 | 20.7 | 45.7 |
| MSH 15-52 | 2-8 | 4.53E-02 | 1.49E-01 | 2.1 | 6.6 |
| G0.11-0.11 | 2-8 | 1.83E-01 | 2.86E-01 | 92.2 | 94.9 |
|            | 4-8 | 3.59E-02 | 7.08E-02 | 77.6 | 87.2 |

# Chapter 8

# Simulation of an IXPE observation of the Tycho SNR

## 8.1   Introduction

Supernova Remnants (SNRs) are the result of the death of a star: the stellar matter ejected from the supernova explosion shocks the swept-up interstellar matter producing a diffuse emission observable from the Radio to the X and $\gamma$ rays. SNRs are strong particle accelerators and are though to be the source of the Galactic Cosmic Rays. Historically, the measurement of X-ray polarization from the Crab PWN confirmed the hypothesis that its X-ray emission was due to synchrotron radiation. Synchrotron emission from accelerated electrons is present also in shell-structure SNR that do not host a pulsar injecting energy with its wind, such as Tycho, as indicated by their power law spectra and TeV emission. Yet, questions remains on the mechanism through which the magnetic fields are amplified and whether the fields in the shock are turbulent. As described in Chapter 4, X-ray polarimetry will be an important diagnostic tool to answer these questions, and IXPE will allow us to assess the polarization properties of SNRs as a function of their magnetic field strength and geometry.

The Tycho SNR, together with CasA and SN1066 will be part of the IXPE first year observation plan (see Table 5.2). The three remnants will be observed for 1 Ms each and all of them entirely (or in part in the case of SN 1006) fit IXPE's field of view. In this Chapter, I will focus on the Tycho SNR, whose IXPE observation I will lead. Tycho is a young SNR that was observed in modern times by its name-giver in 1572. Its X-ray detection arrived in 1984 with the UHURU satellite [177], the spectrum of the remnant suggesting that its origin was a type Ia explosion, as confirmed in 2008 by Krause et al. [85]. It is known to be polarized in the radio band, with a linear polarization degree ranging from 0 at the center, to 8% at the outer rim [86] with orientation suggesting a large scale radial magnetic field structure.

Fig.8.1 shows a false color X-ray image of Tycho's SNR constructed from a merge of Chandra observations: in red is the Fe L emission ($0.8-0.95$ keV), in green the Si XIII ($1.75-1.95$ keV) and blue is the $4-6$ keV band in which the emission lines are absent. The first two energy bands trace the thermal emission of the out-of-equilibrium plasma in the remnant, while most of the emission in the latter band is synchrotron-dominated. In particular, Tycho exhibits peculiar structures in the 4−6 keV band, such as the so called "stripes" in the western rim (see Figure 4.1) that were first identified by Eriksen et al. [44]. These stripes, along with the entire SNR rim, are thought to be cosmic ray acceleration sites. As testified by the rich theoretical literature [22, 23, 24, 25] these regions are expected to be highly polarized in the X-rays. Hence, Tycho is one of the most interesting targets that IXPE can study through spatially resolved X-ray polarimetry. In this Chapter, I will present a realistic simulation of a 1 Ms IXPE observation of the Tycho SNR including all the sources of background described in Section 5.2, I will focus on the possibility of detecting polarization in the aforementioned stripes region, distinguishing between different polarization models and geometries, and I will reconstruct the intrinsic polarization degree of the synchrotron emission, undiluted by the thermal emission. Last but not least, I will evaluate the MDP that can be achieved in a realistic exposure time.

In Section 8.2 I will describe the source model employed in the simulations, in Section 8.3 I will describe the simulations and the data analysis, and finally in Section 8.4 I will show and discuss the results.

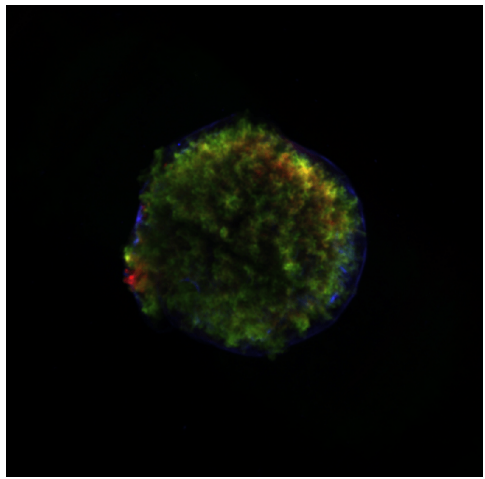

**Figure 8.1.** RGB image of the Tycho SNR: in red is the Fe L emission (0.8−0.95 keV), in green the Si XIII (1.75−1.95 keV) and blue is the 4−6 keV band.

## 8.2   Model

The source model for the simulation of the Tycho SNR comes from a 33.3 ks long Chandra observation (OBSID 8551) that was reprocessed with the CIAO tool chandra_repro. As described in Section 5.4, the use of a Chandra event-list as a basis for an IXPE observation simulation allows to preserve the spectral and morphological distribution of the source, that is then smeared and down-sampled to the IXPE response. The Chandra observation already includes the GPDE and the CXB, so in the source model I add the instrumental background component as described by Xie et al. [178] after the application of the rejection methods.

This way, the model includes all the spectral and morphological components IXPE

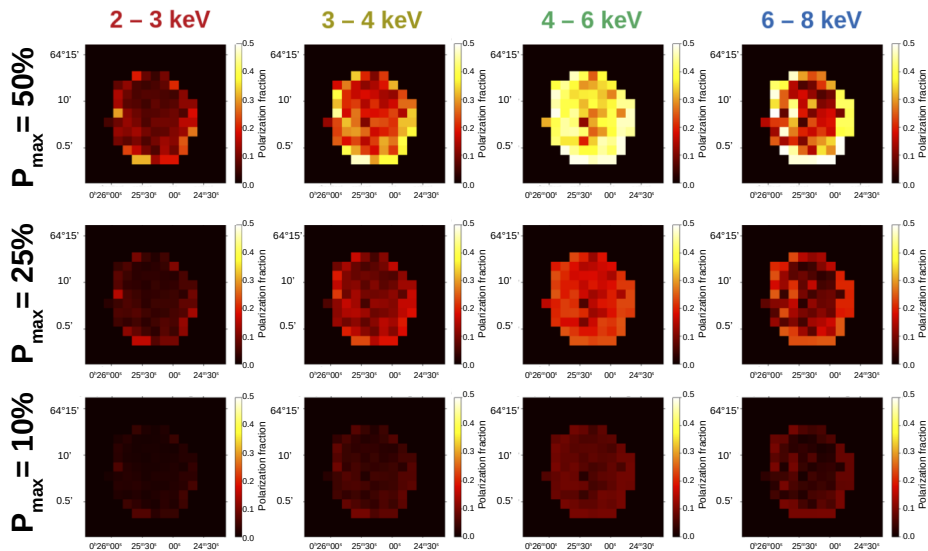

**Figure 8.2.** Polarization degree models for the Tycho SNR: from top row to the bottom row, polarization models for four energy bins corresponding to a maximum synchrotron polarization of 50%, 25%, and 10%, respectively.

is expected to observe in a real observation.

In Fig. 8.2 are shown the input polarization degree maps for the Tycho SNR in the case of a maximum synchrotron polarization of 50%, 25%, and 10%. The 50% polarization map was first produced by the SNR TWG Chair Pat Slane (private comunication) starting from a template of a radial polarization map normalized to a polarization degree of 1.0. This map can then be multiplied by a factor (e.g. 0.50 to consider the case of 50% maximum polarization), that we call $P_S$ as it corresponds only to the synchrotron radiation, that is, the quantity we are actually trying to derive from IXPE observations. From merged Chandra data to get a deep observation, spectra are extracted from a 15×15 grid of 40″ boxes. The spectra are fitted in each region independently, using a template spectral model consisting of a power law and two vpshock components for the SNR ejecta. The power law describes the synchrotron emission. The constant temperature, plane-parallel shock vpshock models, instead, describe the thermal out-of-equilibrium plasma of the SNR [19]. One vpshock model is set to be Fe rich and high temperature ($kT \sim 8$ keV), the other is set as Si rich and low temperature ($kT \sim 2.4$ keV) From these spectral fits, the synchrotron fraction of the flux for each region in five energy bands (2−3 keV, 3−4 keV, 4−6 keV, and 6−8 keV) is derived as

$$F(E_i) = \frac{F_S(E_i)}{F_{tot}(E_i)} \quad , \tag{8.1}$$

where $F_S(E_i)$ is the synchrotron flux and $F_{tot}$ is the total flux. The synchrotron fraction map for these energy bands and for the full 2−8 keV band is shown in

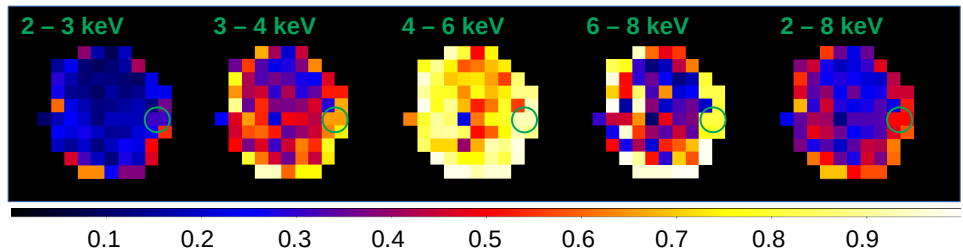

**Figure 8.3.** Synchrotron fraction maps for the Tycho SNR for the for four different energy bands and for the whole 2−8 keV band. The stripes region is highlighted with a green circle. The color bar indicated the synchrotron fraction

Fig. 8.3. Thus, $F(E_i)$ is a 15×15 map of the contribution of the synchrotron flux relative to the total flux in energy band $E_i$. Finally, in each energy band the effective polarization map is

$$P_{eff}(E_i) = P_S \times F(E_i) \qquad (8.2)$$

This is what IXPE will observe: i.e. the synchrotron polarization diluted by the unpolarized thermal emission. The polarization angle map is rotated to match a tangential polarization field, that from Chapter 4 is the most probable large scale polarization morphology expected from a young SNR such as Tycho.

I also consider an alternative polarization angle model in which the polarization direction is perpendicular to the highly aligned non-thermal structures observed in the so-called stripes region. In this latter case, the polarization angle is assumed constant $\phi = 55°$. In Fig. 8.4 are shown in the (a) panel the purely tangential polarization field model, corresponding to an underlying radial magnetic field, and in panel (b) the latter model with stripes region modified to have a polarization angle perpendicular to the stripes. I will test the possibility of distinguishing between these two models with an IXPE observation.

## 8.3  Methods

For each polarization model shown in Fig. 8.2 I ran a 1 Ms long simulation with ixpeobssim (version 16.10), that according to the IXPE observing plan (see Table 5.2) will be the integration time dedicated to Tycho.

I binned the full simulated IXPE event list in order to obtain the IXPE count map and polarization map, the latter applying a spatial binning of 30" (∼1 PSF) and 70" (∼3 PSF). Then I selected from the simulated IXPE event list the stripes region as a ∼1.4' diameter circular region centered on the fk5 coordinates (RA, DEC: 6.190667812,64.12856983), and an annular region ∼1.4' thick containing the rim of the SNR. The event selection was done with the ixpeobssim xpselect tool. I binned the selected events with the ixpeobssim xpbin tool using the PCUBE and PHA1, PHA1Q, PHA1U algorithms. The former binning algorithm produces data products that contains the polarization information. The latter three algorithms produce the Stokes spectra that can be fed to XPSEC to perform a spectro-polarimetric fitting of the data. In order to evaluate the background, the same selection and binning procedure is applied to a region external to the SNR of the same size as the stripes one. From the counting rate from the background region, I obtain the correct MDP (see Eq. 2.26), and I evaluate the dilution factor with Eq. 7.1 allowing me to correct the observed polarization degree. To reconstruct the intrinsic synchrotron polarization, undiluted by the unpolarized thermal plasma component of the SNR, I used Eq. 8.1 and 8.2, therefore dividing the background-corrected observed polarization fraction by the synchrotron fraction in the relevant energy band estimated from the synchrotron fraction maps, as shown in Fig. 8.3.

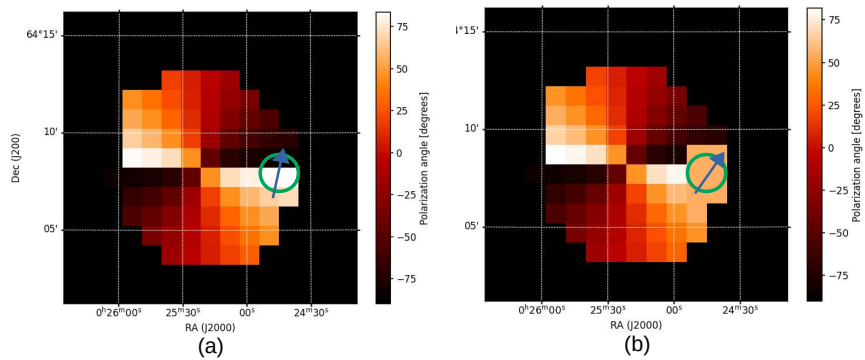

**Figure 8.4.** Polarization angle models for the Tycho SNR: (a) purely tangential polarization field, corresponding to a radial magnetic field topology; (b) polarization angle perpendicular to non-thermal structures in the "stripes" region (highlighted in both figures by the green circle). The blue arrows in the stripes regions is illustrative of the average polarization direction in the two cases.

### 8.3.1 Spectro-polarimetric model and fitting

The I, Q. and U spectra are binned with the HEASOFT tool *ftgrouppha* in order to apply the $\chi^2$ statistics. For the I spectrum I require a minimum of 50 counts in each spectral channel. For the Q and U Stokes spectra I apply a constant grouping of 10 channels. I fitted the spectra of Stokes parameters of the three DUs simultaneously (for a total of nine spectra: three for each Stokes parameter) with XSPEC [6]. I used the model *TBabs(constpol\*vpshock + constpol\*vpshock + constpol\*powerlaw)*, where the spectral parameters were fixed for the best fit of the spectrum from the input Chandra observation of the stripes region. The *vpshock* models are constant-temperature plane-parallel shock plasma model that describe the unpolarized thermal emission from the SNR. The first one has a temperature fixed at 7.97 keV and the Fe abundance is set to $Z_{Fe} = 580.6$. The other elemental abundances are fixed at $Z = 1$. The second one has a temperature fixed at 2.42451 keV and $Z_{Si} = 134.699$, $Z_S = 82.5564$, $Z_{Ar} = 51.6188$, $Z_{Ca} = 61.9644$, and $Z_{Fe} = 0$. The other abundances are left at the Solar one. The *powerlaw* component describes the non-thermal, polarized, synchrotron emission. It is convolved with the multiplicative spectro-polarimetric model *constpol* that describes a polarization constant in energy, as the polarization of synchrotron emission should be. The polarization degree and angle are left as free parameters to be fitted. The *vpshock* models are also convolved with *constpol*, but in this case the polarization fraction parameter is set to zero, as the thermal plasma is assumed to be unpolarized.

## 8.4 Results

### 8.4.1 Stripes result

The polarization and significance maps for these two spatial binnings and for the full $2-8$ keV band are shown in Fig. 8.5 and 8.6, respectively. The maps are relative to the model with 50% polarization of the synchrotron emission, and a $3\sigma$ threshold significance is applied. Instead, in Fig 8.7 the same maps are shown for three energy

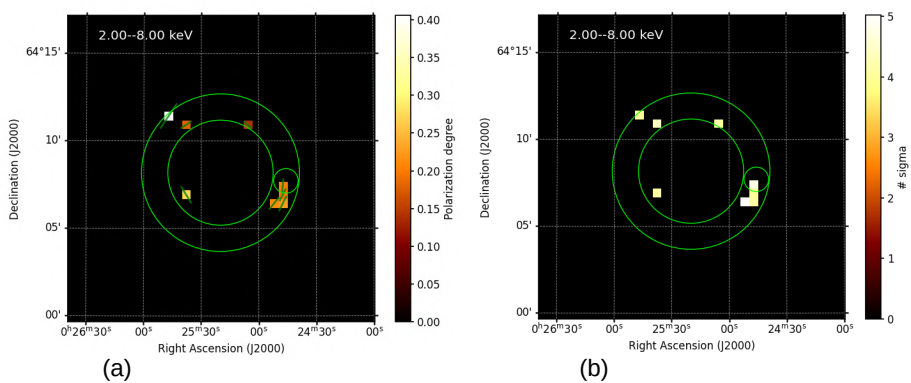

**Figure 8.5.** Polarization (a) and significance (b) map of the Tycho SNR in the 2-8 keV energy band for a 1 Ms-long IXPE observation with the tangential polarization model and 50% synchrotron polarization. Highlighted are the rim region and the stripes region. The maps are binned on a spatial scale of 30".

binnings: $2-3$ keV, $3-4$ keV, and $4-6$ keV.

In Table 8.1 are shown the results of the simulation of an IXPE 1 Ms long observation of the stripes region of Tycho for three synchrotron polarization models (50%, 25%, and 25%) and for two magnetic field topologies: radial (i.e. a purely tangential polarization field) and aligned to the stripes (i.e. polarization field perpendicular to the stripes).

The first result that can be noted is that the MDP for a 1 Ms IXPE observation of the Tycho stripes region, once the background contribution is taken into account, is 4.8%. This allows for a marginal $3\sigma$ detection of polarization in the whole $2-8$ keV energy band in the case in which the synchrotron component has a maximum polarization of 10%. In the most optimistic case of a maximum synchrotron polarization degree of 50%, significant detection of polarization is also possible for multiple energy bands, except for the 6−8 keV band where not only the SNR is fainter, but the emission is dominated by the unpolarized Fe K$\alpha$ line and the hard tail of the instrumental background.

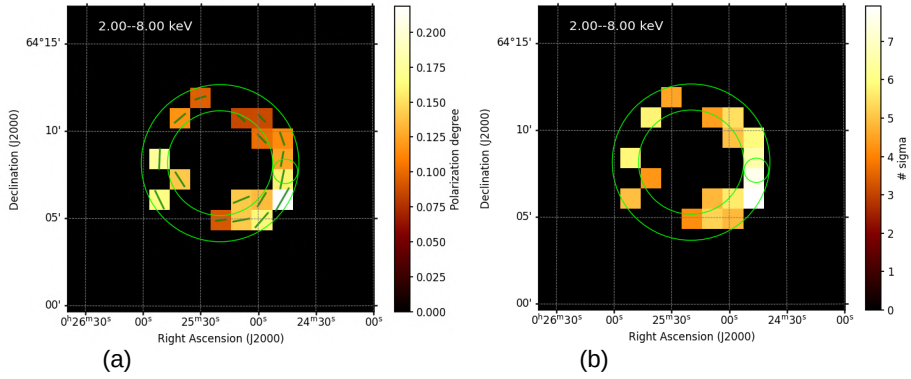

**Figure 8.6.** Polarization (a) and significance (b) map of the Tycho SNR in the 2−8 keV energy band for a 1 Ms-long IXPE observation with the tangential polarization model and 50% synchrotron polarization. Highlighted are the rim region and the stripes region. The maps are binned on a spatial scale of 70″.

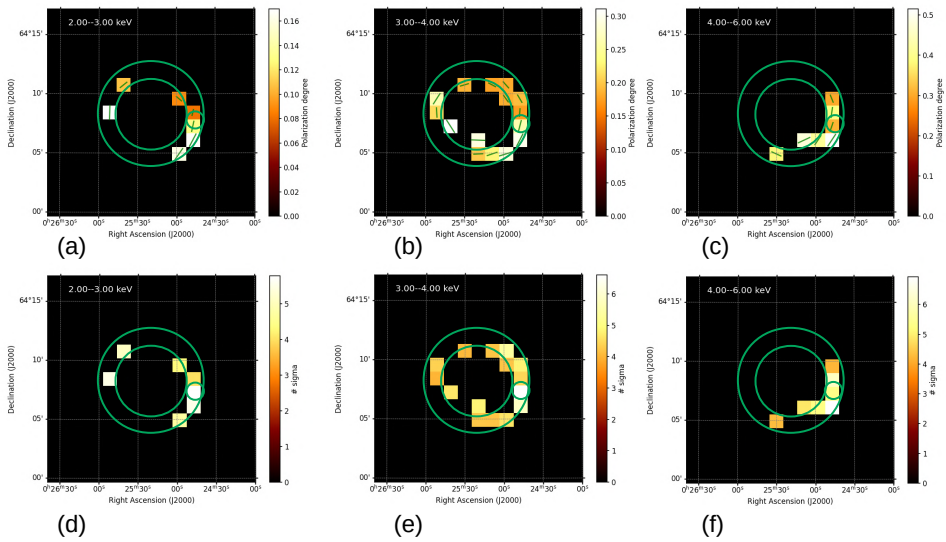

**Figure 8.7.** Polarization (top row) and significance (bottom row) map of the Tycho SNR in the 2−3 (first column), 3−4 (second column), and 4−6 (third column) keV energy bands for a 1 Ms-long IXPE observation with the tangential polarization model and 50% synchrotron polarization. Highlighted are the rim region and the stripes region. The maps are binned on a spatial scale of 70″.

**Table 8.1.** Results of the simulation of an IXPE observation of the Tycho SNR stripes region. For different polarization models and each energy band are reported the source and background rate, the background dilution factor, the synchrotron fraction, the MDP, the observed polarization $P_{PCUBE}$, the reconstructed intrinsic polarization $P_{syn}$, and the measured polarization angle for the tangential ($\phi_{tan}$) and perpendicular ($\phi_{\perp}$) model.

| $P_{model}$ | Energy band (keV) | Source rate (c/s/cm²) | Bkg rate (c/s/cm²) | Dilution (%) | Sync fraction (%) | MDP (%) | $P_{PCUBE}$ (%) | $P_{syn}$ (%) | $\phi_{tan}$ (deg) | $\phi_{\perp}$ (deg) |
|---|---|---|---|---|---|---|---|---|---|---|
| | 2-3 | 3.70E+00 | 5.47E-02 | 1.5 | 31.7 | 6.7 | 15.3±2.2 | 49±15 | 72±4 | 54±5 |
| | 3-4 | 4.40E-01 | 1.09E-02 | 2.4 | 60.5 | 11.5 | 25.2±3.8 | 42±10 | 76±4 | 63±4 |
| 50% | 4-6 | 1.41E-01 | 8.92E-03 | 6.6 | 86.1 | 16.9 | 37.7±5.7 | 44±9 | 76±4 | 53±4 |
| | 6-8 | 2.02E-02 | 4.25E-03 | 17.6 | 67.7 | 39.0 | 25.4±13.9 | 38±23 | 85±17 | 60±17 |
| | 2-8 | 4.309E+00 | 7.88E-02 | 1.8 | 40.6 | 4.8 | 18.6±2.1 | 46±11 | 75±3 | 56±3 |
| | 2-3 | 3.77E+00 | 5.43E-02 | 1.4 | 31.7 | 6.7 | 7.3±2.2 | 24±11 | 72±9 | 65±7 |
| | 3-4 | 4.54E-01 | 1.08E-02 | 2.6 | 60.5 | 11.5 | 12.5±3.7 | 21±8 | 70±9 | 53±7 |
| 25% | 4-6 | 1.45E-01 | 8.73E-03 | 6.0 | 86.1 | 17.2 | 19.5±5.6 | 24±8 | 70±9 | 59±10 |
| | 6-8 | 2.56E-02 | 3.92E-03 | 15.3 | 67.7 | 37.5 | 12.0±11.9 | 21±24 | 19±34 | 19±35 |
| | 2-8 | 4.30E+00 | 7.77E-02 | 1.8 | 40.6 | 4.8 | 9.8±2.0 | 25±8 | 70±6 | 60±6 |
| | 2-3 | 3.75E+00 | 5.38E-02 | 1.4 | 31.7 | 6.7 | 4.1±2.2 | 13±9 | 79±16 | 42±27 |
| | 3-4 | 4.52E-01 | 1.09E-02 | 2.4 | 60.5 | 11.5 | 10.3±3.8 | 17±8 | 67±11 | 82±28 |
| 10% | 4-6 | 1.48E-01 | 8.40E-03 | 5.7 | 86.1 | 16.6 | 8.3±5.4 | 11±8 | 62±18 | 44±29 |
| | 6-8 | 2.43E-02 | 5.14E-03 | 21.2 | 67.7 | 41.3 | 21.5±12.2 | 37±27 | 95±21 | 62±41 |
| | 2-8 | 4.237E+00 | 7.99E-02 | 98.2 | 40.6 | 4.8 | 5.7±2.0 | 14±7 | 76±10 | 51±22 |

In both the case of 50% and 25% polarization (albeit in the latter with lower significance) it is possible to distinguish between different polarization angles, and hence magnetic field topologies. In the 50% case, the simulated observation returns in the 2−8 keV band a polarization angle of 75±3° for the tangential model (true value: 75°), and 56±3° for the perpendicular model (true value: 55°). In the 25% case, the simulated observation returns in the 2−8 keV band a polarization angle of 70±6° for the tangential model, and 60±6° for the perpendicular model. In the case of 10% polarization, the significance is not enough to distinguish between the two models. In all polarization models, the correction for the synchrotron fraction allows to recover the intrinsic polarization degree from the PCUBEs: in the 2−8 keV energy band, 47±5% (model 50%), 25±5% (model 25%), and 14±5% (model 10%). Also spectro-polarimetric fitting returns polarization degrees and angles consistent with the models. In Fig. 8.8 are shown the data and folded model of the simultaneous spectro polarimetric fitting of the Stokes parameters of the Tycho stripes region in the case of tangential polarization and synchrotron polarization degree of 50%, 25%, and 10%. In Table 8.2 are reported the results of the fit with XSPEC for the three polarization degree models and two polarization angle models for the stripes region. In Tables 8.3 and 8.4 are reported the spectral fit parameters for the 50% polarization case of the stripes for the tangential and perpendicular field model, respectively. With the exception of the most extreme case of 10% polarization, it is possible to distinguish between a purely radial and perpendicular polarization field. In all cases, the fit allows to reconstruct the intrinsic polarization degree in the 2−8 keV energy band: $55^{+11}_{-10}$% and $54^{+11}_{-10}$% (model 50%, tangential and perpendicular, respectively), $21^{+10}_{-7}$% and 35±11% (model 25%, tangential and perpendicular, respectively), and 13±4% and 19±14% (model 10%, tangential and perpendicular, respectively). The polarization degree value models selected for the simulations correspond to particular values of the field variance $\sigma^2$, and hence the magnetic field turbulence level, using Eq. 4.3 (i.e., assuming that nothing else reduces the polarization degree due to the synchrotron emission below its theoretical maximum of 75%). Hence the assumed polarization values of 50%, 25%, and 10% correspond to the variance in Table 8.5. In all three cases the variance, and hence turbulence, of the magnetic field on the scale of the stripes region of Tycho can be estimated within the uncertainty as consistent with the value coming from the input model. Because we can reconstruct the model value of the polarization degree, a 1 Ms IXPE observation is sensitive to the associated turbulence levels on the scale of the stripes structures region. In the event of no detection, the MDP in the stripes region would correspond to a constrain at 99% confidence level on the magnetic field variance of $\sigma^2 \geq 46.8\%$. A non detection could be indicative of a change of morphology of the magnetic field in the region, from radial to tangential, or to a non uniformity due to high level of turbulence.

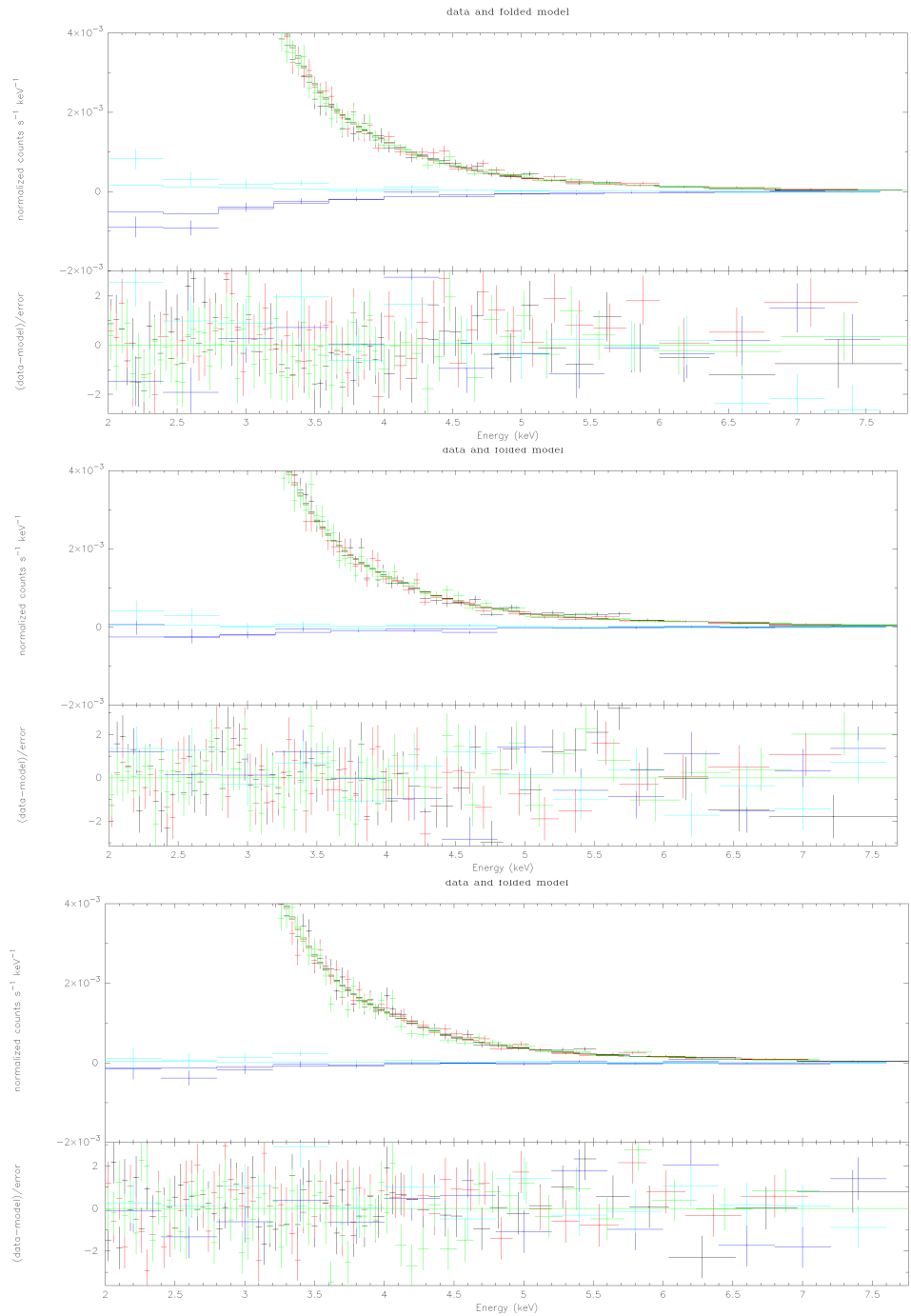

**Figure 8.8.** Data and folded model of the spectro-polarimetric fit of the stripes region of Tycho assuming a tangential polarization model with maximum polarization of 50% (top panel), 25% (middle panel), and 10% (bottom panel).

**Table 8.2.** Results of the spectro-polarimetric fitting of the Tycho stripes region.

| Model | | $P_{XSPEC}$ | $\phi$ | $\chi/d.o.f.$ |
|---|---|---|---|---|
| $P_{model}$ | $\phi$ | (%) | (°) | |
| 50% | tangential | $55^{+11}_{-10}$ | $75^{+6}_{-4}$ | 1.17 |
| | perpendicular | $54^{+11}_{-10}$ | $55\pm3$ | 1.40 |
| 25% | tangential | $21^{+10}_{-7}$ | $70^{+19}_{-10}$ | 1.10 |
| | perpendicular | $35\pm11$ | $50^{+9}_{-5}$ | 1.10 |
| 10% | tangential | $13\pm4$ | $65^{+20}_{-15}$ | 1.20 |
| | perpendicular | $19\pm14$ | $46^{+9}_{-7}$ | 1.10 |

**Table 8.3.** Best fit parameters for the spectro-polarimetric fit of the Stokes spectra of the Stripes region in the case of tangential polarization field and 50% synchrotron polarization.

| Component | Parameter | Unit | Value |
|---|---|---|---|
| TBabs | nH | $10^{22}$ | 1.77741 |
| constpol | poldeg | | 0.0 frozen |
| | polang | deg | 45.0000 |
| | kT | keV | 8.0 frozen |
| | H, He, C, N, O, Ne, Mg, Si, S, Ar, Ca, Ni | | 1.00000 frozen |
| | Fe | | 580.6 frozen |
| vpshock | $Tau_l$ | s/cm$^3$ | 0.0 frozen |
| | $Tau_u$ | s/cm$^3$ | 1.3E+10 frozen |
| | Redshift | | 0.0 frozen |
| | norm | | 7.0E-06 |
| constpol | poldeg | | 0.0 frozen |
| | polang | deg | 45.0 |
| | kT | keV | 2.4 frozen |
| | H, He, C, N, O, Ne, Mg, Ni | | 1.00000 frozen |
| | Si | | 134.7 frozen |
| | S | | 82.6 frozen |
| | Ar | | 51.6 frozen |
| | Ca | | 62.0 frozen |
| vpshock | Fe | | 3.1e-14 frozen |
| | $Tau_l$ | s/cm$^3$ | 0.0 frozen |
| | $Tau_u$ | s/cm$^3$ | 7.4E+10 frozen |
| | Redshift | | 0.0 frozen |
| | norm | | 4.0E-04 |
| constpol | poldeg | | 0.55 |
| | polang | deg | 75 |
| powerlaw | PhoIndex | | 2.95 frozen |
| | norm | | 7.6E-03 |

**Table 8.4.** Best fit parameters for the spectro-polarimetric fit of the Stokes spectra of the Stripes region in the case of polarization field perpendicular to the stripes and 50% synchrotron polarization.

| Component | Parameter | Unit | Value |
|---|---|---|---|
| TBabs | nH | $10^{22}$ | 1.50144 |
| constpol | poldeg | | 0.0 frozen |
| | polang | deg | 45.0000 |
| | kT | keV | 8 frozen |
| | H, He, C, N, O, Ne, Mg, Si, S, Ar, Ca, Ni | | 1.00000 frozen |
| | Fe | | 580.600 frozen |
| vpshock | $Tau_l$ | s/cm$^3$ | 0.0 frozen |
| | $Tau_u$ | s/cm$^3$ | 1.28051E+10 frozen |
| | Redshift | | 0.0 frozen |
| | norm | | 8.2043e-06 |
| constpol | poldeg | | 0.0 frozen |
| | polang | deg | 45.0000 |
| | kT | keV | 2.42451 frozen |
| | H, He, C, N, O, Ne, Mg, Ni | | 1.00000 frozen |
| | Si | | 134.7 frozen |
| | S | | 82.6 frozen |
| vpshock | Ar | | 51.6 frozen |
| | Ca | | 62.0 frozen |
| | Fe | | 3.1e-14 frozen |
| | $Tau_l$ | s/cm$^3$ | 0.0 frozen |
| | $Tau_u$ | s/cm$^3$ | 7.4E+10 frozen |
| | Redshift | | 0.0 frozen |
| | norm | | 3.6-04 |
| constpol | poldeg | | 0.55 |
| | polang | deg | 54.8 |
| powerlaw | PhoIndex | | 2.95 frozen |
| | norm | | 7.40-03 |

**Table 8.5.** Magnetic field variance estimated from the simulations. $\sigma^2_{synch}$ is the variance estimated from the intrinsic polarization obtained from the synchtrotron fraction, $P_{synch}$. $\sigma^2_{XSPEC}$ is the variance estimated from the intrinsic polarization obtained from the spectro-polarimetric fit, $P_{XSPEC}$. Shown are also the input polarization model and the expected true value of the variance $\sigma^2_{model}$.

| $P_{model}$ (%) | $P_{synch}$ (%) | $P_{XSPEC}$ (%) | $\sigma^2_{synch}$ | $\sigma^2_{XSPEC}$ | $\sigma^2_{model}$ |
|---|---|---|---|---|---|
| 50 | 46±11 | $55^{+11}_{-10}$ | 0.19±0.07 | 0.13±0.07 | 0.16 |
| 25 | 25±8 | $21^{+10}_{-7}$ | 0.34±0.05 | 0.36±0.06 | 0.33 |
| 10 | 14±7 | 13±4 | 0.45±0.04 | 0.41±0.12 | 0.44 |

### 8.4.2   Rim result

If the SNR magnetic field is radial, the only way to get a detectable signal from the shell is to use small regions, as the polarization degree from larger areas would be washed out by the different polarization angles as a function of azimuth.

On the other hand, if the source isn't sufficiently bright in the selected region, we won't have enough signal to obtain a statistically significant polarization measurement. A possible solution is to extract data from larger arcs along the rim, but this will dilute the signal because the angle changes from one end of the arc to the other. Ideally, to test for a radial field one would have to correct the angle for each event so that it is measured relative to the radial direction at that point. A tool called xpphialign, provided with ixpeobssim, allows to process an IXPE event list and rotate the polarization angle values so that, on an event-by-event basis, the zero for the measurement of the photoelectron direction is aligned to a given input model at the position of the event. Hence, I can select the events in the rim region of Tycho and test if they are compatible with a, e.g. tangential polarization model.

I also test the mixed model in which only in the stripes region the field is perpendicular to their direction, while in the rest of the remnant has a tangential morphology as in Fig. 8.4 (b). The results of the application of the xphialign tool to the rim region are shown in Table 8.6. As an exercise, I also align the stripes region, testing it against the tangential model for the two previously considered polarization field morphology. For the tangential model, the reconstructed, aligned polarization angle is consistent with zero regardless of the synchrotron polarization level. However, when testing the mixed case against the tangential model, we found that the reconstructed, aligned polarization angle is no longer consistent with zero. This anomaly in the aligned polarization angle allows us to claim the presence of patches with radial magnetic fields in the SNR. The possibility of distinguishing within the uncertainty the purely radial magnetic field from an anomalous one depends on the significance of the measurements, and hence on the strength of the underlying polarization.

**Table 8.6.** Polarization field morphology in the rim tested with the xpphialigned tool for a tangential or perpendicular model and for maximum synchrotron polarization degree of 50%, 25%, and 10%, and for the stripes fo 50% polarization.

| Region | $P_{model}$ (%) | $\phi_{model}$ | $\phi_{aligned}$ (degree) |
|--------|-----------------|----------------|---------------------------|
| Rim | 50 | tangential | 0±1 |
| | | perpendicular | -4.5±1 |
| | 25 | tangential | 0±3 |
| | | perpendicular | -3.4±2.5 |
| | 10 | tangential | -1.6±5 |
| | | perpendicular | -7±6 |
| Stripes | 50 | tangential | -3±3 |
| | | perpendicular | -21±4 |

For the stripes region only, the anomaly is even more clear, with the reconstructed aligned angle, -21±4° giving indication of the difference from the radial magnetic field case. I remind the reader that the expected polarization angle in the case of a radial field is ∼75°, while for the constant model is ∼55°. Hence, the anomaly in compatible with the difference between the two models.

## 8.5 Discussion

From these simulations, we can conclude that a 1 Ms observation of Tycho is sufficient to detect in the stripes region polarization as low as 5% with at least $3\sigma$ significance. This allows to distinguish between purely tangential, or radial, polarization fields, and more complex geometries as well, such as a polarization field perpendicular to the stripes. We should be able to establish that the magnetic field is primarily radial with confidence as long as the synchrotron polarization degree is >10%. However, if the field is not relatively radial, that is if it becomes tangential just before the shock, for example, or if it is not uniform over regions of the order of ~30″, it would become challenging to detect anything unless the polarization degree is considerably higher. The polarization maps shown in this chapter underscore the possibility of performing spatially resolved X-ray polarimetry across the rim region of the Tycho SNR. If the synchrotron polarization fraction is high enough, it is also possible to have significant spatially resolved results for multiple energy bins. A 1 Ms IXPE observation of the stripes region of the Tycho SNR allows to estimate the value of the magnetic field variance up to values of $\sigma^2 = 46.8\%$. On the particle diffusion length-scales sampled by the X-rays (of the order of the shell thickness), CR-driven magnetic field amplification models predicts high turbulence, with magnetic field disorder levels of $\delta B/B \sim 1$ Baring [10], for which the synchrotron X-ray polarization is expected to be $< 10\%$. If IXPE measures X-ray polarization levels higher than ~20%, it would open interesting scenarios, with the need to revise the theories on how turbulence is generated in SNR shocks. Hence, X-ray polarimetry of SNR will advance our understanding of MHD turbulence in their shocks, and the Tycho SNR observation with IXPE will undoubtedly yield many information and surprises. The analysis techniques presented in this Chapter will form the basis for the actual analysis procedure to be performed on the IXPE observations of the Tycho SNR.

# Chapter 9

# IXPE and eXTP polarimetric archaeology of the reflection nebulae in the Galactic Center

## 9.1 Introduction

The X-ray polarization properties of the reflection nebulae in the Galactic Center inform us about the direction of the illuminating source (through the polarization angle) and the cloud position along the line of sight (through the polarization degree) as seen in Section 4.2. However, the detected polarization degree is expected to be lowered because of the mixing of the polarized emission of the clouds with the unpolarized diffuse emission that permeates the Galactic Center region (see Chapter 7). In addition, in a real observation, also the morphological smearing of the source due to the point spread function and the unpolarized instrumental background contribute in diluting the polarization degree. So far, these effects have never been included in the estimation of the dilution. In this Chapter, based on a paper by Di Gesu, L. et al. [38], that I coauthored, we evaluate the detectability of the X-ray polarization predicted by Marin et al. [90] for the MC2, Bridge-B2, G0.11-0.11, Sgr B2, Sgr C1, Sgr C2 and Sgr C3 molecular clouds with IXPE, and with the Enhanced X-ray Timing and Polarimetry mission (eXTP) which is planned for launch in 2028. The scientific rationale for observing this region was illustrated in 4.2. We simulate IXPE observations of individual candidate targets. With the aid of Chandra maps and spectra, we consider (when possible) the polarization, spectral, and spatial properties of all the emission components (i.e., the cloud, soft plasma, and hard plasma) in each target field. Chandra images are most suitable for this work because the spatial resolution of Chandra is infinitely good from the IXPE point of view. In addition, we include a realistic model for the instrumental background and for the CXB. In this way, we are able to quantify how much the polarization degree of the molecular clouds is diluted in the unpolarized environmental radiation. In the ideal case of a detector with an infinite spatial resolution and zero background, the dilution factor is just the ratio between the reflection flux and the total flux [as in, e.g., 90]. In a real observation, both the morphological smearing due to a finite PSF and the unpolarized background contribute in increasing the dilution.

Our simulation strategy allows to quantify also this extra dilution. The Chapter is organized as follows: in section 9.2 we describe the selection and the preparation of the Chandra data, while in section 9.3 we present our simulation procedure; finally, in section 9.4 we discuss the results.

## 9.2 Chandra data preparation

### 9.2.1 Chandra data selection

We consider as candidate targets for a X-ray polarimetry observation the molecular clouds for which Marin et al. [90] computed the polarization properties expected in a theoretical scenario where the source of illumination is a past unpolarized outburst of Sgr A*. The molecular clouds MC1, MC2, Bridge D, Bridge E, Bridge B2 and G0.11-0.11 belong to the Sgr A complex. The Sgr B complex comprises two substructures named Sgr B1 and Sgr B2. Conversely, the clouds Sgr C1, Sgr C2 and Sgr C3 are substructures of the Sgr C complex. The morphology of the molecular clouds is known from extensive Chandra and XMM-Newton observational campaigns that were carried out over the last 20 years. The extension of the clouds [see e.g., 150] is typically larger than the nominal PSF of IXPE. Furthermore, the diffuse plasma in which the clouds are embedded has a non-homogenous morphology. Thus, in some of the simulations that follow, we use Chandra maps to define the extended spatial morphology of the cloud, the soft-plasma and the hard-plasma component. Moreover, Chandra spectra are used to input the spectral shape of each emission component. As a first step in the preparation of the IXPE simulations, we retrieved from the public archive the Chandra observations of the Sgr A, Sgr B and Sgr C complexes. We selected in the archive all the Chandra ACIS-I observations that were taken since 1999 without any gratings in place. For the Sgr A field, the total Chandra exposure time is ∼ 2.4 Ms. Thanks to the long exposure time, we were able to compute Chandra images of the cloud, the soft-plasma and the hard-plasma in this region (Sect. 9.2.2). The Chandra field of Sgr B comprises only Sgr B2, while there are no Chandra observations including Sgr B1, which is thus excluded from the present analysis. The Chandra field of Sgr C comprises Sgr C1 and Sgr C2, while Sgr C3 is included in Chandra Obs-ID 7040. For the Sgr B and Sgr C region the total exposure time of the data available in the Chandra archive is insufficient to produce sensible maps of the emission components separately. Thus, for these clouds we use the most recent Chandra observation available for the spectral analysis (Sect. 9.2.3). We list in the Appendix of Di Gesu, L. et al. [38], all the Chandra observations that we used.

### 9.2.2 Chandra maps of the Sgr A field

We processed the Chandra data using the CIAO software [53], version 4.11, in combination with the version 4.8.2 of the Chandra calibration database (CALDB). For each observation, we ran the chandra_repro routine to create the clean level 2 event file. Hence, for the Sgr A region, we created background and continuum-subtracted counts maps of the soft-plasma, the hard-plasma and the clouds. For all the images, we kept the native ACIS pixel size (i.e. ∼0.5 arcsec). We proceeded as follows. For each observation, we created the background event-file using the blank-sky event files that are provided in the Chandra CALDB. For this, we used the blanksky CIAO routine, that customizes a blanksky background file for the input event file, finding the instrument-specific background files in the CALDB and combining and reprojecting them to match the input coordinates. Hence, for each observation we ran the blanksky-image script to create background-subtracted Chandra count maps of each emission component. For the soft-plasma, we created a map in the $2.35-3.22$ keV energy band, which comprises the $S_{XV}$ and $Ar_{xvii}$ emission lines. For the hard-plasma, we created a map centered on the $Fe_{xxv}$-He$\alpha$ line ($6.62-6.78$ keV). The morphology of the molecular gas is given by a Chandra map centered on the Fe-K$\alpha$ line ($6.32-6.48$ keV). Finally, for the continuum we use the $4.0-6.32$ keV band [e.g. 34], that is line-free. As a final step, for each emission component, we merged all the images using the CIAO script reproject_image_grid routine, which reprojects all the input images to a common coordinates grid. Using the spectra of the four targets that we select for our simulations we found that a model including an absorbed power-law and a 1 keV brehmstrahlung component interpolates adequately the continuum spectral shape underlying the $S_{xv}$, $Ar_{xvii}$, $Fe_{xxv}$-He$\alpha$ and Fe-K$\alpha$ line. By averaging the results of this continuum model for the four targets of interest, we derive the scaling factors (0.38, 0.10, 0.09 for the soft-plasma, hard-plasma, and clouds, respectively) that we use to rescale the continuum images in the band of each emission component. Thus, these scaling factors are optimized for the regions used in the following simulations. The final images of each emission component are obtained by subtracting the rescaled continuum count-maps from the signal count-maps. We normalized all the maps dividing for the maximum value. We display the final background and continuum-subtracted maps of the three emission components in Fig 9.1. We searched for the targets analyzed in Marin et al. [90] in the background and continuum subtracted Fe K$\alpha$ map of the Sgr A field (Fig. 9.1, first panel). We excluded from our search and thus from the IXPE simulations the MC1 and Bridge D cloud, because they are predicted to be basically unpolarized. We identified MC 2, Bridge B2, Bridge E and G0.11-0.11, that are displayed as circular regions in Fig. 9.1. In Table 9.1 we list the central coordinates, the radius, and the projected distance from Sgr A* of each cloud. The cloud size are the same of Marin et al. [90]. As a final step in the preparation of the maps for the simulations of the MC2, Bridge B2, Bridge E and G0.11-0.11 clouds, we created, for each emission component, smaller Chandra maps cut in the region of interest (i.e. the region listed in Table 9.1). This is because, in this case, we simulate IXPE observations of each target individually and on axis. We note, however, that the IXPE field of view (FOV) is ∼6.4 arcmin in radius and thus a single IXPE pointing of the Sgr A field will catch more than one target. A simulation mapping the entire IXPE FOV will be presented in Chapter 10. Here, we simulate each cloud individually, with the aim of collecting useful information in order to decide which is the best target to point at. We centered each map on the brightest Fe Kα patch. Since the morphology of the clouds varies with time, these coordinates are shifted with respect to those used in Marin et al. [90]. This does not affect the expected polarization degree, because it depends mainly on the galactic depth (Eq. 4.6). The expected polarization angle may be affected, but changes are expected to be less than one degree (F. Marin, private communication). In the case of Sgr B1, Sgr C1, Sgr C2, and Sgr C3, we could not create the Fe-Kα map to search for position of the clouds. Thus, for these clouds we use the same regions of Marin et al. [90] to extract the spectra from the most recent Chandra observations. The regions used for Sgr B2, Sgr C1, Sgr C2, and Sgr C3 are also listed in Table 9.1. Finally, we list in Table 9.1 all the other cloud data that we input in the IXPE simulations i.e. the polarization degrees and angle resulting from the model of Marin et al. [90] that were computed assuming a position $d_{los}$ along the line of sight of the clouds. The assumed distance is the key parameter determining the polarization degree and hence the IXPE detectability. We will explore the impact of the assumed distances for our simulations in Sect. 9.4.

**Table 9.1.** Input data for IXPE simulations of molecular clouds. (a) Data of the regions for the spectral analysis and IXPE simulations. Positive and negative projected distances mean east and west of the GC. (b) Cross identification with the target names used in Marin et al. [90]. (c) Distance along the line of sight assumed in Marin et al. [90]. See references therein. Positive and negative means behind and in front of the Galactic plane. (d) Polarization properties from the model of Marin et al. [90].

| Region (a) | | | Identification (b) | $d_{los}$ (c) | Polarization properties (d) | |
|---|---|---|---|---|---|---|
| Center, | radius, | $d_{proj}$ | | | P | $\phi$ |
| (hh:mm:ss.s, dd:mm:ss.s) | (arcsec) | (pc) | | (pc) | (%) | (°) |
| 17:46:00.6, -28:56:49.2 | 49 | -14 | MC2 | -17 | 25.8% | 73.8° |
| 17:46:05.5, -28:55:40.8 | 44 | -18 | Bridge B2 | -60 | 15.8% | 77.8° |
| 17:46:12.1, -28:53:20.3 | 49 | -25 | Bridge E | -60 | 12.7% | 67.9° |
| 17:46:21.6, -28:54:52.1 | 90 | -27 | G0.11-0.11 | -17 | 55.8% | 61.6° |
| 17:47:30.6, -28:26:36.6 | 121 | -100 | Sgr B2 | -17 | 65.0% | 88.3° |
| 17:44:30.6, -29:27:22.6 | 100 | 71 | Sgr C1 | -74 | 31.1% | 94.6° |
| 17:44:54.9, -29:28:30.4 | 115 | 66 | Sgr C2 | 58 | 34.9% | 99.1° |
| 17:45:12.2, -29:22:22.0 | 146 | 50 | Sgr C3 | -53 | 32.9% | 106.4° |

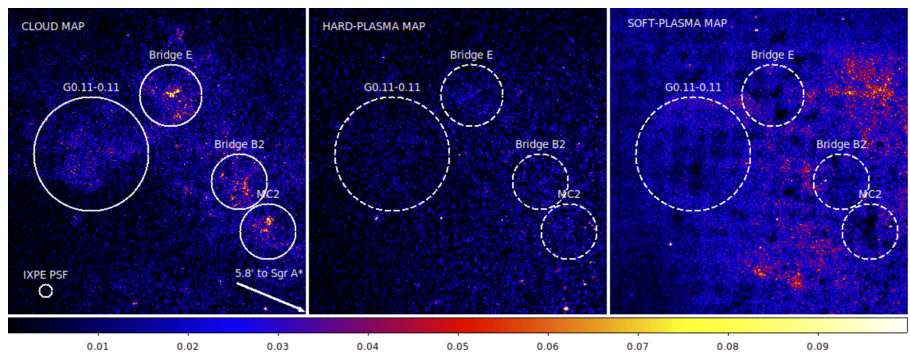

**Figure 9.1.** From the left to the right: background and continuum subtracted Chandra maps of the cloud, the hard-plasma, and the soft-plasma component in the Sgr A region. Images are smoothed using a 3 pixel Gaussian kernel. The color bar displayed on the bottom has adimensional units because the images are normalized to the maximum value. The regions comprising the targets selected for IXPE simulations (i.e. MC2, Bridge-B2, Bridge-E, and G0.11-0.11) are shown. In the first panel, a circle having the size of the IXPE PSF is shown for comparison. The direction of Sgr A* is indicated with an arrow.

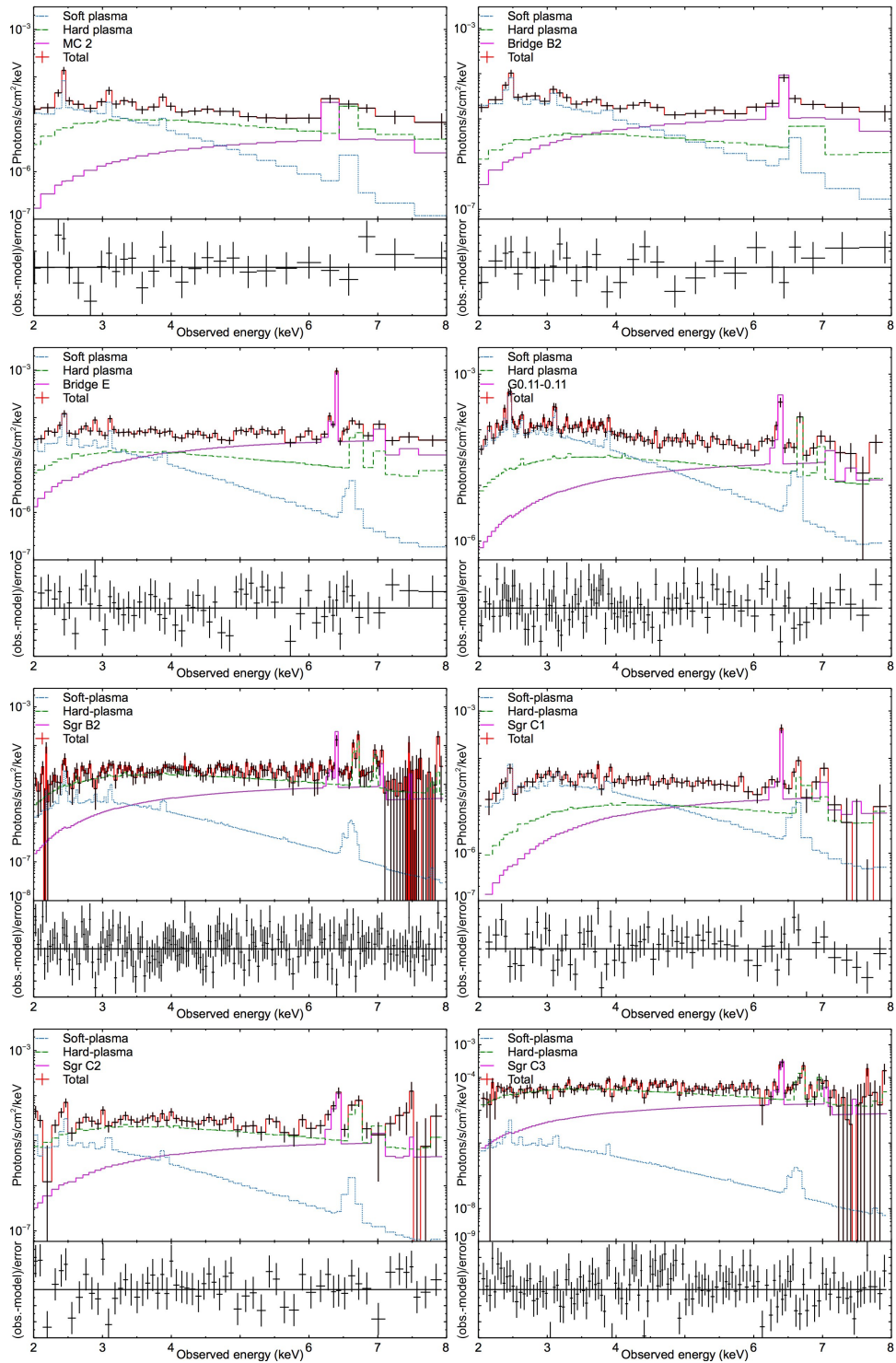

**Figure 9.2.** From the top to the bottom: unfolded spectra and the residuals to the best-fit model for MC2, Bridge B2, Bridge E, G0.11-0.11, Sgr B2, Sgr C1, Sgr C2, Sgr C3. The total best-fit model and the reflection component are displayed as a solid line. The spectrum of the hard-plasma is displayed as a dashed-line. The spectrum of the soft-plasma is displayed as a dotted line.

### 9.2.3   Chandra spectral analysis

The last necessary ingredient for simulating IXPE observations of the selected targets is the spectral shape of each emission component. For all the regions listed in Table 9.1, we extracted the spectrum from the most recent Chandra observation available. We checked that in the extraction regions there is no contamination of known bright X-ray sources (listed in e.g. Terrier et al. 150). To extract the spectra, we used the CIAO script specxtract, which creates the source and background spectra and the necessary weighted response matrices. We used the customized blank-sky event file to extract the background spectrum in the same region. We binned the spectra requiring that a minimum of 30 counts is reached in each spectral bin. We fitted all the spectra in the 2.0-8.0 keV band with Xspec version 12.10.1. We used a model including the Galactic absorption, the soft-plasma, hard-plasma and the cloud emission. For the Galactic absorption we used the phabs model, with the hydrogen column density $N_H$ as a free parameter. For the plasma components, we used a collisionally-ionized plasma model [APEC, 135] with a temperature set to 1.0 and 6.5 keV for the soft-plasma and the hard-plasma, respectively. We consider solar abundances and we set the redshift to zero. For the molecular clouds, we used the neutral reflection PEXMON model [107], where we set (as in e.g. Ponti et al. 113) the photon index $\Gamma$ to 2, the Sgr A* disk-inclination to 60°, the cutoff energy to 150 keV. Hence, the free parameters of our fits are the Galactic $N_H$ and the normalization of each emission component. We show the spectra of all the clouds in Fig. 9.2. We list the parameters and errors resulting from our spectral analysis in Table 9.2. All the spectral fits are statistically acceptable ($\frac{\chi^2}{\text{d.o.f}} \leq 1.3$).

## 9.3   Simulation of IXPE observations

We simulate IXPE observations of the targets listed in Table 9.1 with *ixpeobssim* (see Section 5.4. For each target, we perform the simulation in the region listed in Table 9.1 and we centered the FOV on the coordinates of the target. Within the regions of interest, we simulate all the components that contribute to the diffuse X-ray emission. Thus, in addition to the polarized emission of the molecular clouds, we include in our simulations the soft-plasma, the hard-plasma, the cosmic X-ray background and the instrumental background. For each emission component, we input in the simulation the spectrum, the polarization degree, the polarization position angle and, when possible, the spatial morphology. We took the polarization degree and polarization angle of each molecular cloud from the model of Marin et al. [90], as listed in Table 9.1. We consider a polarization degree that is constant with energy, but null at the energy of the fluorescence Fe K$\alpha$ line ($6.32 - 6.48$ keV). Indeed, the fluorescent lines from spherically symmetrical orbitals are unpolarized. Conversely, for the plasma components we consider a null polarization. In the case of MC2, Bridge B2, Bridge E, and G0.11-0.11 we could input in the simulator the real morphology of the hard-plasma, the soft-plasma and the clouds using the Chandra maps described in Sect. 9.2.2. In the case of the clouds in the Sgr B and Sgr C region, the Chandra data quality does not allow us to compute separated maps of each emission component. Thus, for these clouds we assume a uniform morphology of all the components over the region of interest. For both the instrumental and the sky background we simulate a null polarization. Indeed, the internal polarization of the detector is below 1%, and thus, negligible. For the instrumental background, we took the spectrum from the measurement of the non X-ray background of the Neon filled detector that flew on board of OSO-8 [21]. The gas mixture and absorption coefficient of the OSO-8 detector were similar to the one of the IXPE GPD.

**Table 9.2.** Results of the spectral analysis of the molecular clouds described in Sect. 9.2.3. (a) Galactic hydrogen column density. (b) Fluxes of each model component in the quoted bands.

| Target | $N_H$ (a) | Model component fluxes (b) | |
|---|---|---|---|
| | | Soft plasma: 2.0-4.0 keV | 4.0-8.0 keV |
| | | Hard plasma: 2.0-4.0 keV | 4.0-8.0 keV |
| | | Cloud: 2.0-4.0 keV | 4.0-8.0 keV |
| | $(10^{22}$ cm$^{-2})$ | $(10^{-13}$ erg s$^{-1}$ cm$^{-2})$ | |
| MC2 | $6 \pm 4$ | $2 \pm 1$ | $5 \pm 4$ |
| | | $1.3 \pm 0.2$ | $3.9 \pm 0.8$ |
| | | $0.13 \pm 0.06$ | $1.9 \pm 0.6$ |
| Bridge B2 | $\leq 8$ | $2.2 \pm 1.4$ | $0.3 \pm 0.2$ |
| | | $1.7 \pm 0.5$ | $1.8 \pm 0.8$ |
| | | $0.5 \pm 0 - 9$ | $4.7 \pm 0.8$ |
| Bridge E | $4 \pm 1$ | $2.0 \pm 0.9$ | $0.5 \pm 0.2$ |
| | | $1.9 \pm 0.3$ | $4.7 \pm 0.8$ |
| | | $1.19 \pm 0.09$ | $14 \pm 1$ |
| G0.11-0.11 | $7.0 \pm 0.3$ | $8 \pm 1$ | $3.2 \pm 0.4$ |
| | | $2.7 \pm 0.4$ | $9 \pm 1$ |
| | | $0.71 \pm 0.08$ | $11 \pm 1$ |
| Sgr B2 | $8 \pm 2$ | $0.4 \pm 0.3$ | $0.2 \pm 0.1$ |
| | | $1.4 \pm 0.1$ | $5.4 \pm 0.5$ |
| | | $0.21 \pm 0.03$ | $3.7 \pm 0.4$ |
| Sgr C1 | $12 \pm 1$ | $2.5 \pm 0.7$ | $1.8 \pm 0.5$ |
| | | $0.6 \pm 0.2$ | $4 \pm 1$ |
| | | $0.25 \pm 0.03$t | $6.1 \pm 0.8$ |
| Sgr C2 | $7 \pm 2$ | $0.7 \leq 0.4$ | $0.3 \pm 0.2$ |
| | | $1.7 \pm 0.2$ | $5.6 \pm 0.7$ |
| | | $0.26 \pm 0.05$ | $3.9 \pm 0.7$ |
| Sgr C3 | $7 \pm 2$ | $0.09 \pm 0.05$ | $0.03 \pm 0.02$ |
| | | $3.7 \pm 0.4$ | $12 \pm 1$ |
| | | $0.46 \pm 0.07$ | $8 \pm 1$ |

For the instrumental background, we simulate a uniform morphology on the detector. In the simulation, the instrumental background is internal to the detector, thus it is not convolved with the instrumental response functions.

Finally, for the sky background, we use the parameters of the CXB spectrum of Moretti et al. [98], and we renormalize it to match the simulated sky area. We simulate it as a sky source with a uniform morphology.

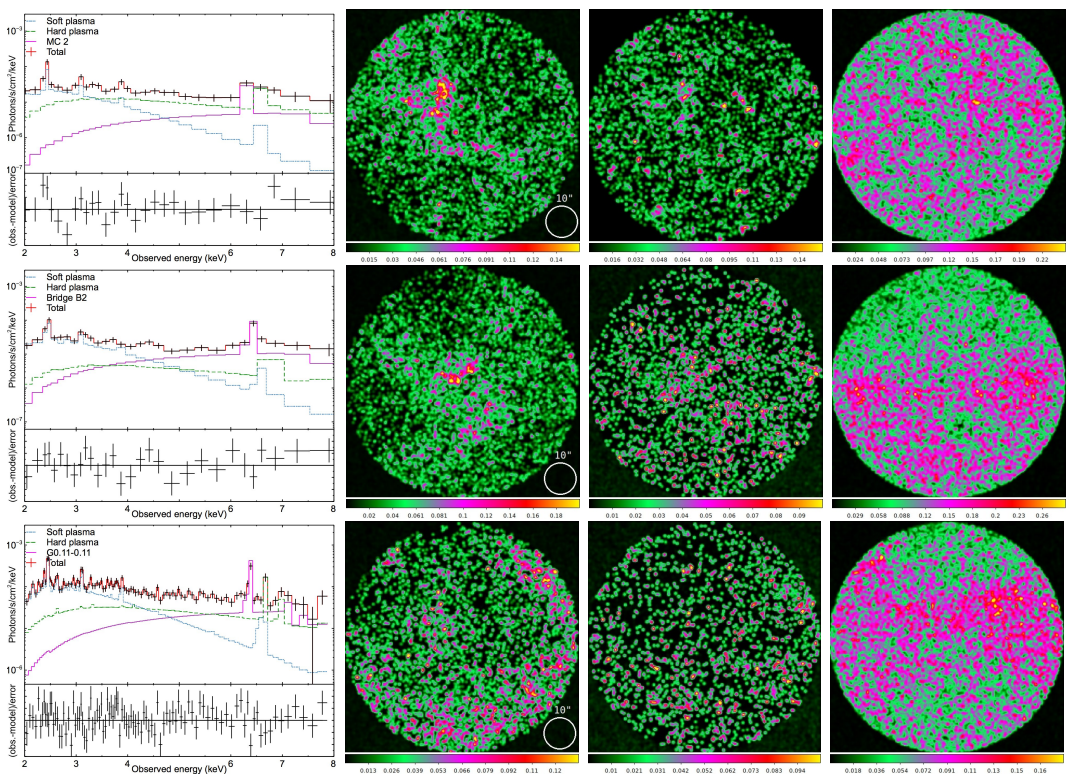

**Figure 9.3.** Chandra spectrum and normalized count maps of the molecular clouds MC2 (top panel), Bridge B2 (middle panel) and G0.11-0.11 (bottom panel). In each panel, from the left to the right, we show the unfolded spectrum and the residuals to the best-fit model, the "cloud" map, the "hard-plasma" map and the "soft-plasma" map. The total best-fit model and the reflection component are displayed as a solid line. The spectrum of the hard-plasma is displayed as a dashed-line. The spectrum of the soft-plasma is displayed as a dotted line. The images have been Gaussian-smoothed using a 2 pixels kernel. For comparison, we show a circle with the radius of the IXPE PSF.

**Table 9.3.** Simulations results for the reflection nebulae considered in this chapter. (a) Obtained from the fluxes and errors listed in Table 9.1. (b) Obtained from "mock" simulations reaching a MDP of 1%. By design, the absolute error on the diluted polarization degree is of 1% or lower. (c) Obtained for 2 Ms exposure time. (d) Minimum flux detectable by IXPE in 2 Ms with a signal to noise of at least 3. * Simulation performed using Chandra maps to define the morphology of all the components. ** Simulation performed assuming a uniform morphology for all the components.

| Target | Scaled P (a) | | Diluted P (b) | | MDP (2 Ms) (c) | | $F_{\min}$ (d) |
|---|---|---|---|---|---|---|---|
| | 2.0-4.0 keV | 4.0-8.0 keV | 2.0-4.0 keV | 4.0-8.0 keV | 2.0-4.0 keV | 4.0-8.0 keV | 4.0-8.0 keV |
| | (%) | | (%) | | (%) | | ($10^{-13}$erg s$^{-1}$ cm$^{-2}$) |
| MC2 * | 0.8%−1.6% | 5%−10% | ≤ 1% | 5% | 15% | 19% | 0.2 |
| Bridge B2 * | 1.9%−2.7% | 9%−12% | 3% | 8% | 14% | 20% | 0.1 |
| Bridge E * | 2.6%−3.1% | 8.5%−9.9% | 3% | 7% | 11% | 12% | 0.3 |
| G0.11-0.11 * | 3.1%−3.9% | 23%−29% | 3% | 16% | 7% | 9% | 0.5 |
| Sgr B2** | 6%−8% | 23%−29% | 13% | 26% | 26% | 21% | 3.5 |
| Sgr C1** | 3.5%−4.6% | 18%−23% | 1% | 10% | 13% | 14% | 0.7 |
| Sgr C2** | 4%−6% | 12%−27% | 4% | 10% | 15% | 15% | 1.1 |
| Sgr C3** | 3%−4% | 10%−14% | 3% | 8% | 12% | 11% | 2.3 |

## 9.4 Results and discussion

Using the input ingredients described in Sect. 9.2 and the procedure described in Sect. 9.3 we simulate IXPE observations of all the targets. We extract from the simulations two main quantities: how much the polarization degree is diluted by the ambient and background radiation and which MDP can be reached in a realistic exposure time. These pieces of information serve to evaluate the detectability of the considered targets in a X-ray polarimetric study of the GC.

In order to obtain a measurement of the "diluted" polarization degree, we proceeded as follows. For all the targets, we ran "mock" simulations of observations reaching a MDP of at least 1%. Thus, the mock exposure time (i.e. 100 Ms) was chosen to obtain that the absolute error on the polarization degree is of 1% or lower. This mimics an ideal case where the statistical uncertainty of the determined polarization degree is negligible. Thus, in these simulations, any observed difference between the determined polarization degree and the theoretical one must be caused by the mixing between polarized and unpolarized components. We note indeed that in simulations including no unpolarized sources in the FOV, the theoretical polarization degree is always recovered within a 3% (absolute value) or less, when the MDP of the simulation is at least 1%. In Table 9.3, we list the "diluted" polarization degrees resulting from the simulations and we compare them with the "scaled" polarization degrees that result from a simple rescaling using the ratio between the reflection flux and the total flux (e.g. Marin et al. 90). We consider that the scaled polarization degrees are affected by the uncertainty of the spectral decomposition. The ranges given in Table 9.3 are obtained as $P \times (F_{\mathrm{cloud}} \pm eF_{\mathrm{cloud}})/F_{\mathrm{tot}}$ where $F_{\mathrm{cloud}}$ and $eF_{\mathrm{cloud}}$ are the flux and error, respectively, for the cloud component, $F_{\mathrm{tot}}$ is total flux and $P$ is the theoretical polarization degree. We observe that the diluted polarization degrees are, in some cases, lower than the scaled polarization degrees. This extra dilution must be induced by the morphological smearing of the source due to the finitePSF. We illustrate this point in Fig. 9.4. We ran 100 simulations of G0.11-0.11 for an ideal case of an instrument with infinite spatial resolution and zero background and 100 "normal" simulations, where the convolution with the instrumental PSF is considered. In this exercise, we consider a "mock" exposure time of 100 Ms, so that the statistical fluctuations of the simulated polarization degree are within a 1%. In Fig. 9.4 we compare the distribution of the polarization degree obtained in the two cases. We found that an instrument with infinite spatial resolution would observe a polarization degree of ∼21%, consistent with what is predicted by a simple rescaling of the flux. An instrument having the spatial resolution of IXPE would observe an extra systematic dilution of ∼ 5%. This difference is not explained by the statistical fluctuations of the result of the simulation because that is, by design, less of 1% in our simulations. We found that the finite spatial resolution of the polarimeter can add an extra dilution depending on the extension and on the morphological details of the source. The quality of the imaging output plays a significant role for an X-ray polarimetric study of the GC region, where the polarized regions have to be resolved out of the surrounding unpolarized emission.

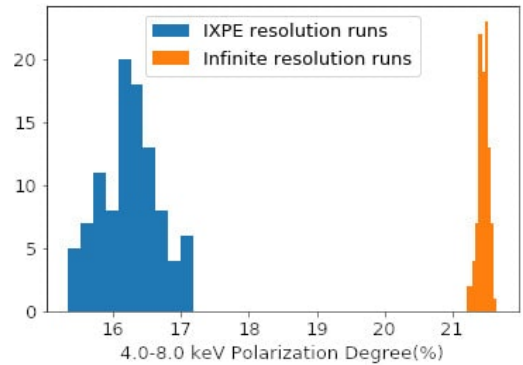

**Figure 9.4.** Histograms showing the distribution of the polarization degree in the 4.0-8.0 keV band obtained simulating the cloud G0.11-0.11 for different instrumental resolution. Orange histogram: infinite spatial resolution case. Blue histogram: IXPE resolution case.

The diluted polarization degrees have to be compared with the MDP attainable in a realistic exposure time. From our IXPE simulations, we compute the MDP in the 2.0−4.0 keV and 4.0−8.0 keV band by running "realistic" simulations with an exposure time of 2 Ms. We note that polarimetry is a photon-starving science and ∼Ms long exposure time may be required for faint or lowly-polarized sources (e.g. for extragalactic sources like AGN). Even for bright Galactic sources or extragalactic blazars exposure times of the order of hundred ks are typically required. From the MDP listed in Table 9.3 a first indication of the preferable targets for IXPE can be derived. We found that the most suitable energy band for searching for polarization signatures is the 4.0−8.0 keV band, where the emission of the molecular clouds dominates the flux output. This exercise indicates that the most promising targets for IXPE observation are G0.11-0.11 and Sgr B2. For these two targets, we found that the diluted polarization degree in the 4.0−8.0 keV band, is larger than the MDP attainable in a 2 Ms long IXPE observation. Thus, our simulations confirm the preferable targets that were already individuated in Marin et al. [90].

However, there are some caveats that must be considered in the planning of a X-ray polarimetric study of the GC. The first issue that we investigate concerns the flux variability on ∼years timescale of the molecular clouds. The flux levels considered in our simulations are those of 2017 for MC2, Bridge B2, Bridge E and G0.11-0.11, of 2010 for Sgr B2, of 2014 for Sgr C1 and Sgr C2 and of 2007 for Sgr C3. Our simulations indicate that at these flux levels, a IXPE observation of any of these targets will always be source dominated. For instance, we find that for the faintest target of the pools considered here (i.e. Sgr C3), the instrumental background accounts for the 2% of the total counts, while the CXB accounts for 3% of the total counts. Nonetheless, by the time of the IXPE observation, the flux of the flux of the molecular clouds may be higher or lower than those considered here. In a recent study of the long-term flux variability of the molecular clouds Terrier et al. [150] found that MC2, G0.11-0.11, and Sgr B are fading while the Bridge is brightening up. The trend for Sgr C is more stable, although within a larger uncertainty. It is therefore useful to compute, for each target, the minimum flux that would be detectable by IXPE in 2 Ms with a signal-to-noise of at least three. Exploiting our estimations of the background contribution we determine these flux thresholds and we list them in Table 9.3. We found that the targets in the Sgr A field remain detectable unless the total flux lowers by one (e.g., for MC 2 and Bridge B2) or even two orders of magnitude (e.g., for Bridge E and G0.11-0.11) with respect to the level considered in our simulations. In the case of Sgr B2, the total flux should lower by a factor 3, with respect to the level observed in 2010 (i.e. $1.1 \times 10^{-12}$ erg s$^{-1}$ cm$^{-2}$), to fall below the detection threshold. In addition to the variability in flux, the molecular clouds in the GC also exhibit variability in morphology. For instance, in Sgr C2, the brightest centroid underwent a displacement of 1.6 arcmin in 12 years [150]. We investigate the impact of the morphology for the result of our simulations. At first, we assess the effect of well positioning the simulated IXPE pointing onto the brightest Fe K$\alpha$ patch. We test this issue using the 2 Ms long simulation of the Bridge-B2 cloud, that displays a well defined bright knot. We find that shifting the IXPE pointing just $\sim$20 arcsec away from the brightest patch causes a loss of $\sim$300 counts and a worsening of the MDP of 1%. This suggests that it is convenient to center the IXPE pointing on a bright knot in order to maximize the counts collected and thus the chance of detecting a significant polarization. Hence, we evaluate the effect of the morphology in determining the diluted polarization over the region of interest. In figure 9.5 we show, as an example, the simulated IXPE polarization maps of the two best targets. These are produced from the "mock" simulations. In these maps, the colored arrows indicate the direction of the polarization angle. In the case of a reflection nebula, this is normal to the projected direction of the illuminating source. In the simulated map of Sgr B2 the nebula is uniform in color/polarization degree, because it was simulated assuming a uniform morphology for all the components. In the simulated map of G0.11-0.11, that was obtained starting from the Chandra maps of the different components, the irregular distribution of polarization fraction/color within the nebula reflects the different level of mixing between polarized and unpolarized emission. Nonetheless, the dilution of the polarization degree averaged over the regions of interest depends mildly on the internal morphology, likely because the substructures are on scale smaller than the IXPE PSF. We checked this point by running simulations of G0.11-0.11 field assuming a uniform morphology for all the components and "mock" exposure time of 100 Ms. The results for the diluted polarization degree are the same, within the uncertainty, as in the run using the Chandra maps. Thus, a posteriori, we are confident that our estimations of the polarization dilution in Sgr B2, Sgr C1, Sgr C2, and Sgr C3 are reasonable.

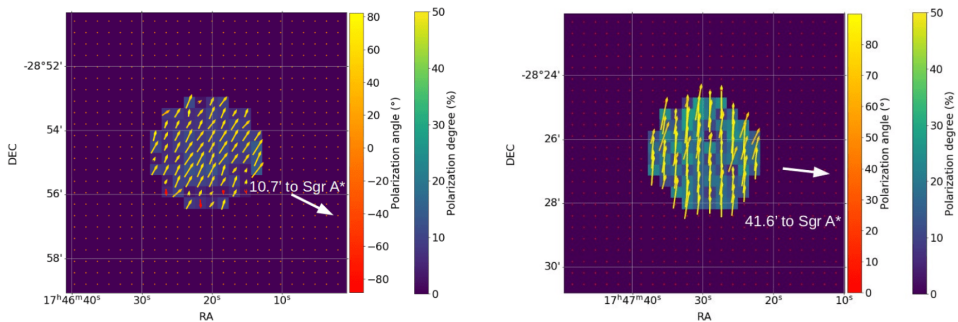

**Figure 9.5.** Simulated IXPE polarization maps of G0.11-0.11 (left panel) and Sgr B2 (right panel). The background is color-scaled according to the polarization degree. The colored arrows represent the direction of the polarization angle and are colour-scaled accordingly. The color scales for the polarization degree and angle are shown on the right of each figure. The direction of Sgr A* is also indicated as a comparison.

All in all, we remark that it would be useful that a X-ray observation of the GC is performed prior the IXPE pointing. With e.g. the Spectrum Röentgen Gamma (SRG) on board of eROSITA) it is possible to check the flux level of the candidate targets. With Chandra or XMM-Newton it is possible to check which patches are currently illuminated which would help in deciding the best pointing.

Finally, in Table 9.4, we investigate the most critical uncertainty that affects the evaluation of the detectability of the polarization of the molecular cloud. The theoretical polarization degree, relies on the poorly constrained line-of-sight distance of the cloud and shall be corrected in case a more robust determination of $d_{los}$ is found. We search in the literature for determinations of the line-of-sight distance of the clouds different from those assumed in Marin et al. [90] (listed as $d_{los}^{other}$ in Table 9.4). These are obtained in works where the scattering angle is derived from a modeling of the reflection spectrum [27, 30, 163] and are often loosely constrained. Starting from the range of $d_{los}^{other}$, we use equations 4.5 and 4.6 to compute the correspondent range of polarization degree ($P^{other}$) and we use the dilution factors in the 4.0-8.0 keV band that can be inferred from Table 9.3 to determine the correspondent range in diluted polarization degree ($P_{dil}^{other}$). Thus, we are able to check whether, for a different assumption on $d_{los}$, the diluted polarization degree of our targets rises above/drops below the MDP obtainable by IXPE in the 4.0-8.0 keV band in 2 Ms. The values listed in Table 9.4 confirm the detectability of G011-0.11 and Sgr B2 also for other possible values of distance reported in the literature. The molecular clouds Bridge B2, Bridge E and Sgr C1 could be detectable if their real distance along the line of sight lies in the upper bound of the range determined by Capelli et al. [27] and Chuard et al. [30]. We also investigate how the enhanced sensitivity of eXTP allows to enlarge the pool of suitable targets. The effective area of eXTP will be larger factor ∼4 which implies (using equation 2.26) that the MDP for the case of eXTP are lower than those of IXPE of a factor 0.51. Thus, applying this factor to the values of MDP listed in Table 9.3 implies that G0.11-0.11, Sgr B2, Sgr C1, Sgr C2, and Sgr C3 are potential targets for eXTP in the 4.0−8.0 keV band.

**Table 9.4.** Polarization obtained for alternative values of $d_{los}$ reported in literature. (a) Range of $d_{los}$ from the quoted references. (b) Range of polarization degree range correspondent to $d_{los}$, obtained from Eq. 4.5 and 4.6. (c) Range of diluted polarization degree obtained from the values of Table 9.3. (d) A: Capelli et al. [27], B: Walls et al. [163] C: Chuard et al. [30].

| Target | $d_{los}^{other}$ (a) (pc) | $P^{other}$ (b) (%) | $P_{dil}^{other}$ (c) (%) | Ref. |
|---|---|---|---|---|
| MC2 | -29.7−7.3 | 50−53 | 9−10 | A |
| Bridge B2 | -6.9−6.9 | $\geq 84$ | $\geq 42$ | A |
| Bridge E | -13.7−13.7 | $\geq 83$ | $\geq 45$ | A |
| G0.11-0.11 | -3.1−3.1 | $\geq 93$ | $\geq 26$ | A |
| Sgr B2 | -50−47 | 61−83 | 24−33 | B |
| Sgr C1 | -0.61−47 | 50−99.9 | 16−32 | C |
| Sgr C2 | -38−−25 | 50−54 | 14−16 | C |

The ranges of diluted polarization degrees obtained in Table 9.4 by relaxing the constraints on $d_{los}$ offer a window of eXTP detectability virtually for all the targets. More sensitive telescopes like for instance the X-ray Polarimetry Probe [XPP, 73] or the New Generation X-ray Polarimeter [NGXP, 141] mission concept would allow detection of the X-ray polarization of the molecular clouds with shorter exposure time. Moreover, the extended energy band of NGXP towards higher energies would permit to make negligible the contribution of the hot plasma, while detecting the reflection component.

In conclusion, an X-ray polarimetric study of the CMZ is a challenging experiment because of the dynamic behavior of the reflection emission and because of the complex gaseous environment in which the nebulae are embedded. We set up a simulation method that allows to realistically assess how some critical factors (i.e., the variability in flux and morphology of the clouds, the dilution of the polarization degree in the unpolarized ambient and background radiation) affect the detectability of a reflection nebula observed on axis. Since the time required to make a significant measurement of the reflection nebulae in the GC is of the order of ∼Ms, the impact on the planning of IXPE observations is significant. Therefore, our realistic predictions are important to inform the decision of including these observations in the planning.

# Chapter 10

# Prospects for a polarimetric mapping of the Sgr A molecular cloud complex with IXPE

## 10.1 Introduction

The X-ray polarization degree of the molecular clouds that surround Sgr A*, as we saw in Capter 9, is expected to be greatly lowered because the polarized reflection emission is mixed with the unpolarized thermal emission that pervades the Galactic center region. For this reason, this observation is a challenging experiment for IXPE. In this Chapter, based on the work I published in Ferrazzoli, R. et al. [51], I aim at expand upon the previous work by simulating with ixpeobssim a realistic, long-lasting IXPE observation of the entire Sgr A field of view (FOV), rather than individual clouds on axis. Indeed, a single IXPE pointing of the Sgr A complex will capture more than one cloud at different off-axis positions, allowing to determine the detectability of four molecular clouds of the Sgr A complex: MC2, Bridge B2, Bridge E, and G0.11-0.11. It is therefore relevant to address the issue of how the detectability changes when a cloud is not at center of the FOV. We do this by assessing the Minimum Detectable Polarization increase when a molecular cloud is off axis. In addition, our simulation method has the advantage of treating all the components that contribute to GC emission separately, each one with its own spectral and morphological property, that are well known thanks to the legacy of Chandra. Indeed, besides the clouds, there are two thermal components that contribute to the $2-8$ keV emission in the GC: a $\sim$1 keV "soft" plasma, and a thermal component that is often modeled as a 6.5 keV "hard" plasma [e.g., 129]. These thermal components permeate the GC region and are unpolarized. Thus, the detected polarization degree of the MC is lower than the intrinsic value by a factor that depends mainly on the amount of plasma contamination in the surrounding environment. For instance, in Di Gesu, L. et al. [38] we found a diluting effect of the plasma as high as 90% in the $2-4$ keV band, and 60% in the $4-8$ keV band for the MC2 cloud.

It is reasonable to assume that the diffuse plasma in the GC does not change in spectrum and morphology over time. This implies that it is possible to exploit our simulations to create simulated products of the diluting components with the aim of combining them with real data to recover the undiluted polarization degree of the MC. Here, we test two methods to achieve this goal. The Chapter is organized as follows: in Sect. 10.2 we describe the setup of our simulations, in Sect. 10.3 we present a simulated MDP map to identify the detectable targets, and we discuss how the detectability changes with the position of the targets in the FOV. In Sect. 10.4 we present two methods to recover the intrinsic polarization degree of the MC using simulated products of the diluting components. Finally, we discuss our findings in Sect. 10.5.

## 10.2 Method

### 10.2.1 Source model

In this work, we simulate IXPE observations of the Sgr A molecular complex and we investigate the detectability of the MC MC2, Bridge B2, G0.11-0.11, and Bridge E. As in Di Gesu, L. et al. [38], we do not consider the MC Bridge D and MC1 because they are expected to be basically unpolarized according to the model from Marin et al. [90].

Throughout the chapter, we use as a main pointing the 12.8'×12.8' IXPE FOV centered on coordinates fk5 17:46:02.4020, -28:53:23.981, shown in Fig. 10.1.

This position is centered on the X-ray reflection feature known as "the bridge" that is considered one of the most promising targets for X-ray polarimetric observations because its average emission has been persistently bright in the last ten years [32, 150]. This is the region of interest of our baseline simulation. Hence, in Sect. 10.3.2, we test other possible pointings centered on each MC to see how the detectability changes along with the location of the target in the FOV.

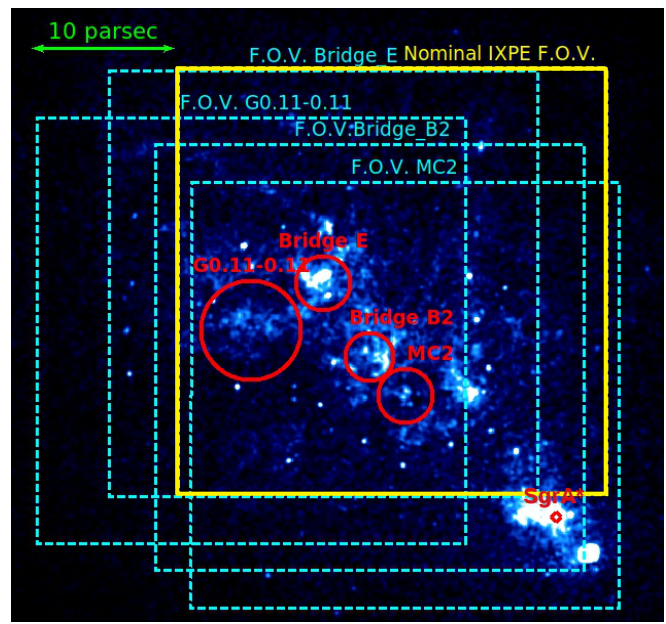

**Figure 10.1.** Chandra Fe K$\alpha$ map of the Sgr A complex. The dashed boxes are the FOVs considered in the analysis of Sect. 10.3.2, while the solid box represent the nominal IXPE FOV of our baseline simulation. The circles display the clouds considered in this work, while the diamond marks the position of Sgr A*. The double-headed arrow represents a distance of 10 pc.

Following Di Gesu, L. et al. [38] (see also Chapter 9), in the region(s) of interest we simulate all the diffuse components that contribute to the emission in the GC region. The soft and hard plasma components are simulated over the entire FOV. In order to account for the morphology of the plasma, we created background and continuum-subtracted Chandra maps. For the soft plasma, we used the $1.7-3.3$ keV energy band that comprises the $S_X V$ and $Ar_{XVIII}$ emission lines, while for the hard plasma we created a map centered on the energy of the $Fe_{XXV}$-He$\alpha$ emission line $(6.62 - 6.78$ keV). We created the maps using the procedure outlined in Di Gesu, L. et al. [38] to combine 2.4 Ms of archival Chandra-ACIS data. We show the soft and hard plasma maps in the first and second panel of Fig. 10.2. We extracted the spectrum of the plasma components for all the IXPE FOV from the latest available Chandra observation that contains the IXPE nominal pointing (i.e., Chandra OBS ID 20808 from 8 October 2017). After subtracting the blanksky and removing the point sources, that we identified through the CIAO tool wavdetect with 2 and 4 pixel scales and $10^{-6}$ signal threshold, we extracted the spectra over the whole FOV centered at the nominal pointing. We note that, in the regions that we used for the MC, there are no point sources [38]. Thus, there is no need to remove the points sources from the Chandra maps because they have no impact for our regions of interest. A transient appearing by chance in our region of interest during the IXPE observation should have a flux above $4 \times 10^{-13}$ erg/s/cm$^2$ (i.e the uncertainty on the total flux of G0.11.011 see Table 10.1) to cause a sensible contamination. In a real observation, the transient can be removed either by cutting a PSF-large region from the maps or by removing the contaminated time intervals from the event files.

We fitted the spectrum with XSPEC [6, version 12.10.1] in the $2.0-8.0$ keV band obtaining a reduced $\frac{\chi^2}{\text{d.o.f}} \sim 1.5$. We model the Galactic absorption with the phabs model. We fitted the plasma components with a collisionally ionized plasma model [APEC, 135] with a temperature set to 1.0 keV for the soft plasma and 6.5 keV for the hard plasma, and solar abundances. For the reflection component, we used the neutral reflection model PEXMON [107], which consistently models both the continuum and the Fe K$\alpha$ emission. These spectral models are commonly used for fitting the GC diffuse X-ray emission [see e.g., 99, 113, 128, 129]. The model spectra of the plasma derived from this fit (first and second panel of Fig. 10.3) serve as input in our simulations. When simulating other pointings, we extract the spectrum again to match the new coordinates. We note that the flux of the plasma does not change significantly from a pointing to another. We consider all the plasma components as unpolarized.

The reflection component of the MC is simulated over circular regions as listed in Table 10.1. For their morphology and spectral properties, we use the same Chandra maps and spectra of Di Gesu, L. et al. [38]. These are continuum- and background-subtracted Chandra maps centered on the Fe-K$\alpha$ line $(6.32-6.48$ keV) and cut over circular regions having the radius of the cloud listed in Table 10.1. In the third panel of Fig. 10.2 the cloud regions are shown superimposed to the Fe K$\alpha$ map of the whole FOV. The model spectra of each MC is shown in Fig. 10.3. We take the polarization properties from the modeling of Marin et al. [90]. We consider the polarization degree of the reflection component as constant with energy, but null at the energy of the fluorescence Fe-K$\alpha$ line, because the fluorescent lines from spherically symmetrical orbitals are unpolarized. In Table 10.1 we list, for each cloud region and for the entire FOV, the polarization properties of all the spectral components and the flux contributions in each region. The polarization degree values that we assume in our simulations were derived in Marin et al. [90] assuming the distance along the line of sight $d_{los}$ that are listed in Table 10.1. In addition, we list in Table 10.1 other possible values of $\vec{d}_{los}$ [27] and the correspondent polarization degree resulting from Eq.4.5 and 4.6. The ranges of distances calculated by Capelli et al. [27] include $d_{los} = 0$. Thus, they represent an upper limit for the absolute value of the distance along the line-of-sight, from which Eq. 4.5 and 4.6 returns the maximum theoretical polarization degree of the clouds. Because the value of the theoretical polarization degree depends strongly on the assumption of $\vec{d}_{los}$, we consider also these alternative polarization degree values in the discussion of the cloud detectability in Sect. 10.3. Finally, we include in our model the Cosmic X-ray background (CXB) and the IXPE instrumental background. The CXB is simulated as a uniform source over the entire FOV with the spectrum of Moretti et al. [98]. In Di Gesu, L. et al. [38] the instrumental background was based on the one measured for the Neon filled detector on board of the OSO-8 experiment [21]. We now employ a realistic instrumental background spectrum that is based on the estimates of Xie et al. [178]. They found for the IXPE detector a background level of $1.16 \times 10^{-2}$ counts s$^{-1}$ cm$^2$ in $2-8$ keV. The effect and removal of the instrumental background was discussed in Chapter 7 and Section 5.3.

### 10.2.2   Simulation outputs

We simulate IXPE observations of the Sgr A MC complex with *ixpeobssim*. In the following analysis, we make use of MDP map cubes and polarization map cubes that can be produced trough *ixpeobssim*. The MDP map cubes are data structures that contain the information needed for the calculation of the MDP (i.e., mean energy, counts, effective modulation factor) binned in sky coordinates for each energy bin considered. We employ them to produce the MDP maps (see Sect. 10.3.1). Conversely, the polarization map cubes hold polarization information binned in sky coordinates and contain image extension for the Stokes parameters $I$, $U$, $Q$, and for the polarization degree and angle. We use the Stokes parameters map contained in the polarization map cubes for the retrieval of the polarization degree of the MC (see Sect. 10.4.1 and 10.4.2).

To mimic a real IXPE observation (Sect. 10.3.1, 10.3.2, 10.4.1, 10.4.2) we ran simulations with exposure time of 2 Ms (that hereafter we label as "realistic" simulations), which is a realistic estimation of the time that IXPE will dedicate to the observation of the GC during the first two years of operations. In order to create synthetic polarization products accounting for the effects of the unpolarized components (Sect. 10.4.1, 10.4.2), we ran simulations of 200 Ms (that hereafter we label as "ideal" simulations) that reaches a MDP of less than 1% over the entire

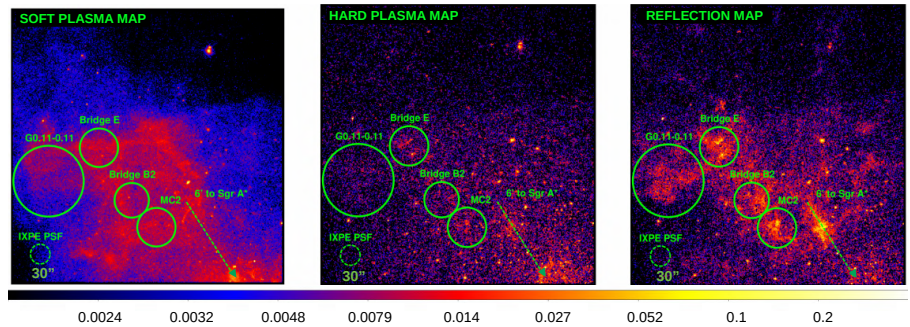

**Figure 10.2.** Background and continuum-subtracted merged Chandra maps of the soft plasma, the hard plasma, and the reflection components in the Sgr A MC complex region centered on the nominal IXPE pointing (shown from left to right). The images are smoothed using a 3 pixel Gaussian kernel. The color bar displayed on the bottom has adimensional units because the images are normalized to the maximum value. The regions shown in the solid circles are the MC considered for IXPE simulations (i.e., MC2, Bridge B2, Bridge E, and G0.11-0.11). A dashed circle having the size of the IXPE PSF is shown for comparison. The direction to Sgr A* is indicated with a dashed arrow.

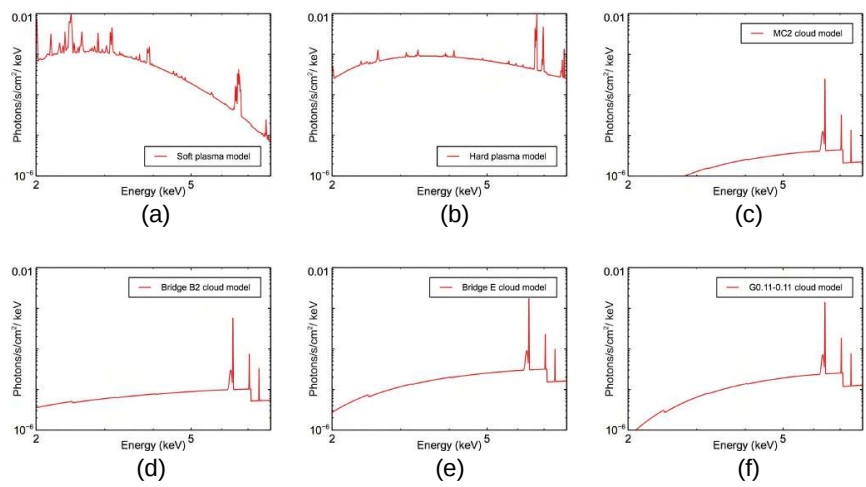

**Figure 10.3.** Spectral models for the emission components of our simulations: the soft (a) and the hard (b) plasma in the nominal FOV, and the reflection in the MC2 (c), Bridge B2 (d), Bridge E (e), and G0.11-0.11 (f) region. The models were obtained from the Chandra spectral analysis performed in the present work and in Di Gesu, L. et al. [38] .

FOV. This long exposure time serves to minimize the statistical uncertainty (error on $P \ll 1\%$) of the result of the simulation. With this simulation setup, we convert the model of the unpolarized components into IXPE data products without adding uncertainty. Thus, the synthetic maps are affected only by the uncertainty that derives from the spectral fit of the Chandra data on which the input model is based (Sect. 10.2.1).

## 10.3 Target detectability

### 10.3.1 MDP map

We created the MDP map for a 2 Ms-long IXPE observation centered on the nominal pointing. The MDP map allows us to identify the regions for which the MDP reaches the lowest value. In Fig. 10.4, we show the MDP map in the $4-8$ keV energy band, where the polarized reflection component outshines the plasma emission. The maps are obtained from the MDP map cubes described in Sect. 10.2.2. We bin the map with a sky pixel size of $\sim$1.5'. This corresponds to three times the IXPE PSF and to the typical size of the MC (Table 10.1). Minima of MDP are observed in the region of the MC G0.11-0.11 and Bridge E. The MDP values relative to each MC region in the $4-8$ keV energy range are listed in Table 10.2. The MDP map confirms what found in Di Gesu, L. et al. [38]: the cloud G0.11-0.11 has the lowest MDP, followed by Bridge E, Bridge B2, and MC2. We note that in Di Gesu, L. et al. [38] the MDP found for G0.11-0.11 in the $4-8$ keV energy range in 2 Ms was 9%, while the current value is equal to 12.5%. The likeliest reason for this difference is twofold. The first reason is that, as explained in Sect. 10.2.1 the assumed instrumental background is higher than the model considered in Di Gesu, L. et al. [38]. This results in an increase of the MDP according to Eq. 2.26. The second reason, as we discuss here below, is that in our simulations the clouds are not placed in the center of the FOV.

### 10.3.2 Off-axis detectability

We study how the MDP of the MC changes as a function of the off-axis distance. For this, we ran simulations putting each time the clouds MC2, Bridge B2, G0.11-0.11, Bridge E, and the nominal IXPE pointing at the center of the FOV. In Fig. 10.1, we show the regions covered by each pointing. For each MC, we measure the MDP that can be achieved in a 2 Ms-long observation in each pointing configuration. For this exercise, the values assumed for the polarization properties are irrelevant, as we are only interested in how the MDP changes with the distance from the center of the FOV. In Fig. 10.5 we show for each MC the MDP as a function of the distance from the center of the FOV in the $4-8$ keV energy band. We observe that the MDP of each cloud when observed at the nominal pointing increases by a factor of $\sim$1% for MC2, $\sim$2% for Bridge B2, $\sim$15% for G0.11-0.11, and $\sim$6% for Bridge E, with respect to the case of an on axis observation. The cause of the differences in MDP across the FOV is mainly the vignetting. The vignetting defines the relative exposure across the FOV and causes a drop of the effective area especially above 6 keV in energy and at 5 arcmins, in distance from the center of the FOV, resulting in a loss of counts for a target off-axis. We find that the effect of vignetting is more significant in the case of G0.11-0.11 and Bridge E, as they display a steeper increase of the MDP as a function of the off-axis distance. This is likely to be due to the fact that they exhibit a harder spectrum with respect to MC2 and Bridge B2 (see Fig. 10.3) and are generally farther from the center of the FOV when the other clouds are pointed. To assess the detectability of the clouds, the MDP has to be compared with the expected polarization degree diluted by unpolarized ambient radiation. As a visual comparison, in Fig. 10.5 we show also horizontal lines corresponding to the theoretical polarization degree of Marin et al. [90] and the range achievable assuming other line-of-sight distances. These values are also listed in Table 10.2, together with the $4.0-8.0$ keV MDP corresponding to the case of the nominal pointing. They differ from the values reported in Di Gesu, L. et al. [38], mainly because of the different background used in the present work and, at a second order, because of the updated instrumental response functions in the *ixpeobssim* simulator. We find that, when assuming a 2 Ms-long IXPE observation and the Marin et al. [90] model, only the cloud G0.11-0.11 is detectable in the $4-8$ keV energy band, even when observed off-axis. If we assume the alternative distances of Table 10.1, significant detection of polarization from the MC2 cloud remains unlikely regardless of its position in the FOV. On the other hand, polarization degree detection at confidence level of 99% is possible for the clouds Bridge B2 and Bridge E, as their expected diluted values of polarization degree is larger than the MDP in 2 Ms. We note that in the case of a non-detection of a cloud, the nominal MDP prescribes an upper limit to the cloud distance along the line-of-sight (Eq. 4.5 and 4.6). We list these $\tilde{d}_{los}^{MDP}$ in Table 10.2. These will be valuable constraints to mitigate our uncertainty in the knowledge of the 3D position of the MC in the GC region.

**Table 10.1.** Input data for IXPE simulations of the nominal FOV. (a) Region name: the MC are cross identified with the targets listed in Marin et al. [90]. (b) Geometrical dimensions of the region over which each spectral components is simulated, as described in Sect. 10.2. (c) Projected distance from Sgr A*. Negative values: MC East of the GC. (d) Line-of-sight distance $\bar{d}_{los}$ from Marin et al. [90]. Negative values: MC in front of the Galactic plane. (e) Range of other line-of-sight distances $d_{los}$ from Capelli et al. [27]. Because these ranges of distance include $d_{los} = 0$, they prescribe upper limits for the maximum theoretical polarization degree of the clouds. (f) Polarization degree $P$ from Marin et al. [90]. (g) Range of polarization degree correspondent to $d_{los}^{other}$, obtained from Eq. 4.5 and 4.6. (h) Flux contribution of each spectral component in the given region. Fluxes are taken from the spectral fits of Di Gesu, L. et al. [38] and Sect. 10.2.

| Region (a) | Region (b) center (hh:mm:ss.s, dd:mm:ss.s) | size | $\bar{d}_{proj}$(c) (pc) | $\bar{d}_{los}$(d) (pc) | $d_{los}^{other}$(e) (pc) | $P_{model}$(f) (%) | $P_{other}$(g) (%) | Model component fluxes(h) Soft plasma: 4.0-8.0 keV / Hard plasma: 4.0-8.0 keV / Reflection continuum: 4.0-8.0 keV / Fe Kα: 4.0-8.0 keV ($10^{-13}$ ergs s$^{-1}$ cm$^{-2}$) |
|---|---|---|---|---|---|---|---|---|
| MC2 | circle: (17:46:00.6, -28:56:49.2) | (radius) 0.82 | -14 | -17 | -29.7–7.3 | 25.8 | ≥10 | 5 ± 4 / 3.9 ± 0.8 / 0.2 ± 0.1 / 1.7 ± 0.7 |
| Bridge B2 | circle: (17:46:05.5, -28:55:40.8) | (radius) 0.73 | -18 | -60 | -6.9–6.9 | 15.8 | ≥77.3 | 0.3 ± 0.2 / 1.9 ± 0.6 / 0.4 ± 0.1 / 4.3 ± 0.7 |
| G0.11-0.11 | circle: (17:46:21.6, -28:53:52.1) | (radius) 1.5 | -27 | -17 | -3.1–3.1 | 55.8 | ≥62.5 | 3.2 ± 0.4 / 9 ± 1 / 1.0 ± 0.1 / 10.0 ± 0.9 |
| Bridge E | circle: (17:46:12.1, -28:53:20.3) | (radius) 0.82 | -25 | -60 | -13.7–13.7 | 12.7 | ≥97.4 | 0.5 ± 0.2 / 4.7 ± 0.8 / 1.3 ± 0.1 / 12.7 ± 0.9 |
| Field of view | box: (17:46:02.4, -28:53:23.08) | (side) 12.8 | - | - | - | - | - | 18.9 ± 1.4 / 71.3 ± 1.4 / - / - |

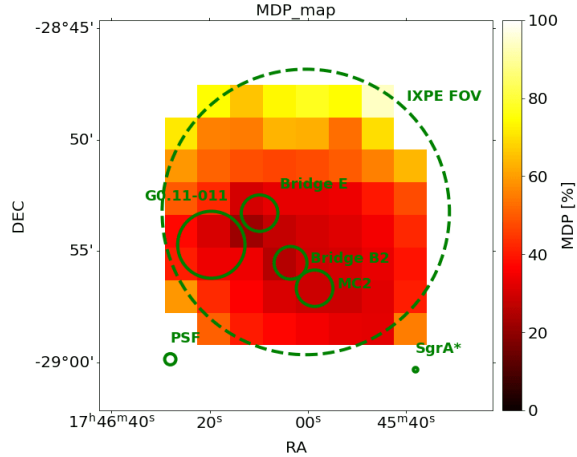

**Figure 10.4.** MDP map with 1.5', spatial binning and 2 Ms exposure in the $4-8$ keV energy band. The dashed circle represents the IXPE FOV. The solid circles are the MC regions considered for IXPE simulations (i.e., MC2, Bridge-B2, Bridge-E, and G0.11-0.11). Smaller circles indicate the size of the IXPE PSF and the position of Sgr A*.

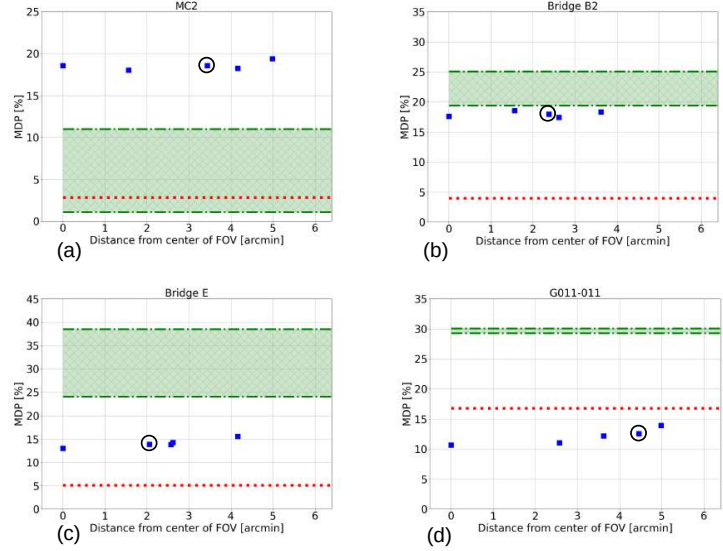

**Figure 10.5.** MDP in the $4.0-8.0$ keV band (square data points) for a 2 Ms-long observation as a function of the distance from the center of the FOV for the clouds MC2 (a), Bridge B2 (b), Bridge E (c), and G0.11-0.11 (d). The circled data points refer to the MC distance from the center of the FOV in the nominal IXPE pointing. The dashed line is the polarization degree from the model of Marin et al. [90] diluted by the unpolarized emission in the $4-8$ keV energy range. The shaded region within the dash-dotted lines covers the polarization degree range predicted for other line-of-sight distances for each MC reported in Table 10.1, diluted by the unpolarized emission in the $4-8$ keV energy range.

**Table 10.2.** (a) Minimum Detectable Polarization, expected diluted polarization, and $|d_{los}|$ relative to the MDP for the MC2, Bridge B2, G0.11-0.11, and Bridge E clouds in the $4-8$ keV band. Includes effect of instrumental background. (b) Expected polarization degree from model of Marin et al. [90] after environmental dilution. (c) Expected polarization degrees for other line-of-sight distances of the MC [27] after dilution. (d) Absolute value of the line-of-sight distance that corresponds to the MDP.

* For the MC2 cloud the MDP achievable in 2 Ms is not low enough to exclude the case $\vec{d}_{los}=0$ [27] because it corresponds to $P_{other,diluted}=11\%$, see also Fig. 10.5. Hence, for this target, the lower limit given by $\vec{d}_{los}^{MDP}$ in the case of a non detection cannot be taken at face value.

| Cloud | Energy (keV) | $MDP$(a) (%) | $P_{model,diluted}$(b) (%) | $P_{other,diluted}$(c) (%) | $|\vec{d}_{los}|^{MDP}$(d) (pc) |
|---|---|---|---|---|---|
| MC2 | $4-8$ | 17.9 | 3 | $1-11$ | $\geq 21$ * |
| Bridge B2 | $4-8$ | 17.9 | 4 | $19-25$ | $\geq 27$ |
| G0.11-0.11 | $4-8$ | 12.5 | 17 | $29-30$ | $\geq 50$ |
| Bridge E | $4-8$ | 13.5 | 5 | $24-39$ | $\geq 45$ |

## 10.4   Reconstruction of the intrinsic cloud polarization

The dilution of the polarization degree due to environmental and instrumental effects hampers the detectability of the clouds and, hence the possibility to derive their line-of-sight distance from the polarimetric data.

We tested two possible methods to create polarization products of the diluting components. We then combined them to a test data set simulated for a realistic observing time of 2 Ms (Fig. 10.6 (a)) to test whether it is possible to reconstruct the intrinsic polarization degree of the cloud and, in turn, to derive the distance of the cloud along the line-of-sight. This test data set will be replaced by real IXPE data once they will be available.

Both these methods are applicable only to the case of diluting components that are unpolarized. The case of dilution from polarized components is beyond the scope of this paper. As a visual comparison, we computed a map of $P_{reflection}$ (10.6 (b)), which is the polarization degree map that IXPE would observe with no unpolarized sources in the field of view. This was created by running a realistic simulation including only the polarized reflection continuum component.

### 10.4.1   Dilution map method

The first technique that we propose for reconstructing the intrinsic polarization degree of the MC consists of creating a map of the dilution factor over the entire FOV, simply referred to as the dilution map. Hence, the undiluted polarization degree, $P_{dmapcorr}(x, y)$, map is obtained by dividing, pixel by pixel, the observed polarization degree map $P_{obs}(x, y)$ by the dilution map $D(x, y)$:

$$P_{dmapcorr}(x, y) = \frac{P_{obs}(x, y)}{D(x, y)} \quad . \tag{10.1}$$

To create the dilution map, we proceeded as follows. We set up an ideal simulation as explained in Sect. 10.2 and we assign a polarization degree of 100% to all the molecular clouds. We produce a map of the Stokes parameters in the $4.0-8.0$ keV band range. We bin the Stokes maps so that each pixel has the size of the IXPE PSF ($\sim 30$"). We produce a polarization map in which the polarization degree is calculated in each spatial bin from the Stokes parameters from Eq. 2.15. In this way, the resulting, simulated polarization map is de facto a map of the dilution factor due to the unpolarized components. The dilution map is shown in Fig. 10.6 (c).

Hence, in order to test whether this technique is effective in recovering the intrinsic polarization degree of the clouds, we created the undiluted polarization map by using the formula reported in Eq. 10.1. This is shown in Fig. 10.6 (d). The polarization properties of individual MC are defined as the average of the values of the pixels inside the MC regions weighted by their intensity:

$$P_{dmapcorr}^{cloud} = \frac{\sum^{cloud} P_{dmapcorr}(x, y) I(x, y)}{\sum^{cloud} I(x, y)} \quad . \tag{10.2}$$

These are listed in Table 10.3 together with the values of $D$ and $P_{dmapcorr}$ for each cloud with their uncertainties. The uncertainty on the value of $D$ is obtained from the uncertainties of the spectral fit of the Chandra data of each cloud in the following way:

$$\frac{\sigma_D}{D} = \frac{\sigma_{Fcrefl/Ftot}}{F_{crefl}/F_{tot}} \tag{10.3}$$

where $F_{crefl}$ is the flux of the polarized reflection continuum, $F_{tot}$ is total flux, and $\sigma_{Fcrefl/Ftot}$ is the uncertainty on the ratio of the polarized and total fluxes ($F_{crefl}/F_{tot}$). The uncertainty on $P_{obs}$ is obtained using equation 5.16 that includes the effect of the uncertainty in the knowledge of the subtracted background. Thus, the uncertainty on $P_{dmapcorr}$ is obtained propagating the errors of $D$ and $P_{obs}$.

The results of the dilution technique are summarized in Table 10.3, where we report for each cloud the reconstructed intrinsic polarization degree and $\vec{d}_{los}$, the latter calculated from Eq. 4.5 and 4.6. The only target for which we are able to constrain the corrected polarization degree is G0.11-0.11, with a polarization degree of 49% ± 20% in the 4−8 keV energy band, which has to be compared to a 55.8% polarization degree model.

This is expected because in our simulations, G0.11-0.11 is the only MC for which the observed (diluted) polarization degree is larger than the MDP in the 4.0−8.0 keV band (see Table 10.2) and hence detectable in the first place at a confidence level of 99% in a 2 Ms-long observation. For the clouds MC2, Bridge B2, and Bridge E we can set upper limits to their polarization degree, and hence distances (see Table 6.4). The line-of-sight distance is found from Eq. 4.5 and Eq. 4.6. For G0.11-0.11, we obtain a distance of ±19±8 pc pc in the 4−8 keV energy band consistent with the -17 pc of the model. We can then break the geometric degeneracy by studying the shape at low energies of the continuum reflection [31].

### 10.4.2 Subtraction method

As an alternative, we tested a second technique which exploits the additivity of the Stokes parameters. For an unpolarized component such as the diffuse thermal emission, the $Q$ and $U$ Stokes parameters are zero. The only relevant contribution to the dilution of the polarization degree is given by the unpolarized Intensity $I_{unpol}$. By subtracting the contribution of the unpolarized emission from the observed Stokes Intensity map, what remains is the Stokes parameters of the polarized component only, from which the polarization degree can be computed as in Eq. 2.15.

We create an intensity $I_{unpol}(x, y)$ map of the unpolarized components, that are the soft and hard plasma, and the Fe K$\alpha$ line. For this, we ran an ideal simulation including the aforementioned components only. From the simulated polarization map cube, we extracted the Stokes parameter maps and we rescale them by a realistic exposure time of 2 Ms (Fig. 10.6(d)). Then, to mimic a real IXPE observation and create the maps of $I_{obs}(x, y)$, $Q_{obs}(x, y)$, and $U_{obs}(x, y)$, we ran a 2 Ms-long simulation including all the components, and with the polarization degree of the clouds set to the values of Table 10.1. The map of the intrinsic polarization degree of the MC $P_{subcorr}(x, y)$ can be obtained using Eq. 2.15:

$$P_{subcorr}(x, y) = \frac{\sqrt{Q_{obs}^2(x, y) + U_{obs}^2(x, y)}}{I_{obs}(x, y) - I_{unpol}(x, y)} \quad , \tag{10.4}$$

Here $Q_{obs}$ and $U_{obs}$ are relative to the clouds as they are the only polarized components present. We obtain the final $P_{subcorr}(x, y)$ map (Fig. 10.6(f)), by replacing pixel-by-pixel, in the observed polarization map cubes, $I_{obs}(x, y)$ with $I_{obs}(x, y) - I_{unpol}(x, y)$. The final reconstructed value of the polarization degree is the average weighted by the intensity over each cloud region (Table 10.3).

We estimated the error for $P_{subcorr}$ by propagating the error for $I_{obs}$, $Q_{obs}$, $U_{obs}$, and $I_{unpol}$. We note that in our cases the Stokes parameters can be treated as independent variables because it is generally true that $P\mu < 0.3$ [79]. The uncertainty on all the observed Stokes parameters are an output of the realistic simulation and include the uncertainty in the knowledge of the subtracted background. The uncertainty on $I_{unpol}$ derives from the the uncertainties of the fits of the Chandra data in the following way:

$$\frac{\sigma_{Iunpol}}{I_{unpol}} = \frac{\sigma_{(Fsoft\,plasma + Fhard\,plasma + FK\alpha)}}{F_{soft\,plasma} + F_{hard\,plasma} + F_{K\alpha}} \quad , \qquad (10.5)$$

where $F_{soft\,plasma}$, $F_{hard\,plasma}$, and $F_{K\alpha}$ are the fluxes of the soft plasma, the hard plasma, and the Fe K$\alpha$ line, respectively. We note that it is critical to determine correctly the Fe K$\alpha$ contribution because, as shown in Table 10.1, its flux is always one order of magnitude larger than the continuum.

We checked that in 2 Ms the contribution to $Q_{obs}$ and $U_{obs}$ of the random fluctuation of the unpolarized component is smaller than the uncertainty on $Q_{obs}$ and $U_{obs}$.

The reconstructed polarization degree map obtained with this method is shown in Fig. 10.6 (f) while the polarization degree averaged in each cloud region are listed in Table 10.3.

We find that this procedure returns a constrained value for G0.11-0.11 in the $4-8$ keV energy band, with a reconstructed polarization degree of $53\% \pm 13$ %. We calculated the $\vec{d}_{los}$ resulting from the reconstructed polarization with this method, and we list their values in Table 10.3. For G0.11-0.11 we obtain a distance of $\pm 18 \pm 4$ pc in the $4-8$ keV energy band, consistent with the -17 pc assumed in the model. Again, the geometrical degeneracy can be removed thanks to the shape of the reflection continuum [31].

**Table 10.3.** Results of the reconstruction of the polarization degree of the MC of the SgrA complex in the $4-8$ keV energy band with the dilution and subtraction methods. (a) Observed polarization degree with a 2 Ms-long observation. The uncertainties on the observed polarization degree include the effect of the subtraction of the instrumental background. (b) Dilution factor obtained from the dilution map described in Sect. 10.4.1. (c) Polarization degree recovered through the dilution method. (d) Polarization degree recovered through the subtraction method. (e) Absolute value of the line-of-sight distance derived from the polarization degree corrected with the dilution method. (f) Absolute value of the line-of-sight distance derived from the polarization degree corrected with the subtraction method.

| Cloud | Energy (keV) | $P_{obs}$(a) (%) | $D$(b) (%) | $P_{dmapcorr}$(c) (%) | $P_{subcorr}$(d) (%) | $|\vec{d}_{los}|^{dmapcorr}$(e) (pc) | $|\vec{d}_{los}|^{subcorr}$(f) (pc) |
|---|---|---|---|---|---|---|---|
| MC2 | $4-8$ | $\leq 17.9$ | $11\pm 9$ | $\leq 53$ | $\leq 30$ | $\geq 9$ | $\geq 15$ |
| Bridge B2 | $4-8$ | $\leq 17.9$ | $25\pm 6$ | $\leq 45$ | $\leq 34$ | $\geq 13$ | $\geq 18$ |
| G0.11-0.11 | $4-8$ | $15\pm 6$ | $30\pm 3$ | $49\pm 20$ | $53\pm 13$ | $19\pm 7$ | $18\pm 4$ |
| Bridge E | $4-8$ | $\leq 13.5$ | $39\pm 4$ | $\leq 31$ | $\leq 20$ | $\geq 26$ | $\geq 35$ |

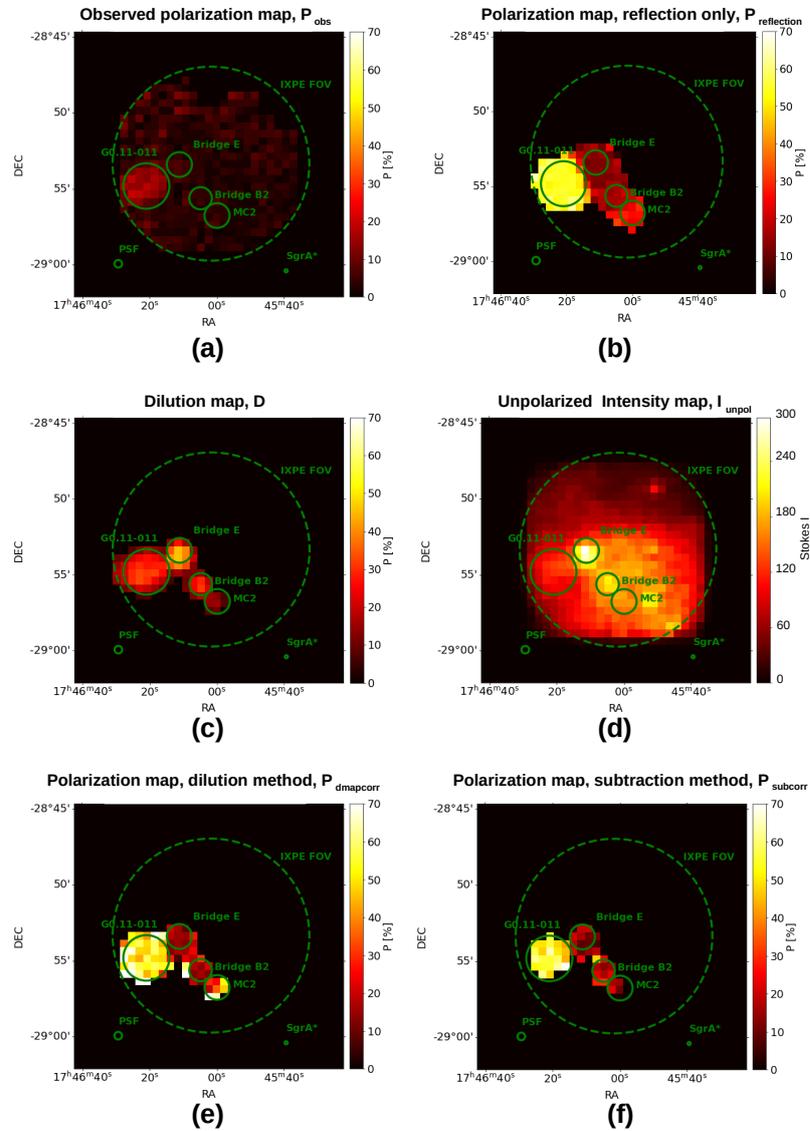

**Figure 10.6.** (a) Observed polarization map. For clarity, we show the results for an average of 100 simulations. (b) Polarization map when only the MC reflection component is considered. (c) Dilution map. (d) Unpolarized intensity Stokes parameter map. (e) Reconstructed polarization map from dilution map method. (f) Reconstructed polarization map from subtraction method. The IXPE FOV and the cloud regions are displayed as in Fig. 10.4. For the maps (a), (b), (c), (e), and (f) the color-bar displays the polarization degree, for map (d) it displays the Stokes intensity parameter. The maps have a 30 " spatial binning and are obtained in the 4−8 keV energy band.

## 10.5   Results and discussion

We estimated the MDP reached in a 2 Ms observation with IXPE of the MC of the Sgr A complex in the 4−8 keV energy band. Our estimations considered two additional factors that we did not take into account in our previous work [38]. The first is the effect of the vignetting of the telescope optics, that causes a loss of counts, and, hence, an increase of the MDP. The second is the updated background model, as derived by Xie et al. [178], for the IXPE detectors that is larger by a factor of three than the one based on Bunner [21]. The scientific case considered here in one of the few for which the instrumental background is a potential confounding factor because of the faintness of the MC.

We found that the MDP of the MC obtained in the case of the nominal IXPE pointing of the region, because of vignetting, increases by a factor in the range of 1−15% with respect to the case in which each of them is observed on-axis. Vignetting effects will be negligible for most sources that IXPE is meant to observe, because the telescope pointing will be dithered around the center of the FOV. The observation of the MC in the GC is one of the few cases in which vignetting will have an observable effect.

Assuming the model of Marin et al. [90], in a 2 Ms-long IXPE observation, G0.11-0.11 is the only MC detectable. However, the prediction of the polarization degree depends strongly on the assumed line-of-sight distance. When changing the assumption on $\vec{d}_{los}$ (Table 10.1), for MC2, Bridge B2, and Bridge E we find a higher polarization degree. In this alternative scenario, the latter two clouds are detectable in a 2 Ms-long observation for all the pointings considered in this work. MC2 is undetectable in a 2 Ms-long observation for any polarization model. This is because the cloud is the faintest among the ones we considered in this work, and it is affected by the worst environmental dilution, ∼ 90% (see Table 10.1).

With the assumed distances (and, hence, the polarization degree, see Eq. 4.5, 4.6) of the clouds, only G0.11-0.11 appears to be a candidate for a statistically significant measurement of the X-ray polarization in the Sgr A complex. The possibility of recovering the polarization degree depends on a priori significativity of the measurement of the diluted polarization degree.

In this work, we tested two methods to recover the intrinsic polarization degree of the MC in the 4−8 keV energy range, where the polarized reflection outshines the unpolarized plasma emission. The dilution map method, described in Sect. 10.4.1, consists in dividing pixel-by-pixel two polarization maps: the observed polarization map and a dilution map. We created the dilution map by simulating the case of clouds 100% polarized. In this way, the resulting polarization degree image maps the dilution factor over the FOV. This method allows to remove the depolarizing effect of the plasma and the emission of the Fe Kα line. For G0.11-0.11, from a polarization degree of 15 ± 6%, the dilution method recovers a value of 49±20%, consistent within the uncertainty with the input model of 55.8%. For G0.11-0.11, this method allows us to recover the line-of-sight distance of the cloud with Eq. 4.5 and 4.6 as ±19±7 pc, consistent with the -17 pc of the model.

The subtraction method, described in Sect. 10.4.2, is based on the subtraction from the observed Stokes I parameters map of the Stokes maps of the unpolarized components only. The residual parameters are employed for the calculation of the polarization degree. For G0.11-0.11, the subtraction method gives a polarization degree value of 53±13%. Again, the reconstructed line-of-sight distance of ±18±4 pc is consistent with the input model. In both cases, the ambiguity on the position can be removed through spectroscopic means by studying the shape of the reflection continuum, as explained in Churazov et al. [31]: if the cloud is closer to the observer with respect to the illuminating source, the reprocessed radiation from the farthest, directly illuminated, side of the cloud would be suppressed at low energies by photo-absorption.

Besides the uncertainty in the cloud distance and, hence in the theoretical polarization degree, there are other potential challenges for the planning of an IXPE observation of the GC. The MC exhibit a time variability in flux and morphology on a timescale of several years. For instance, Terrier et al. [150] note that the flux of the clouds Sgr B2 and G0.74-0.10 decreased by a factor factor $4-5$ over 12 years. In the case of G0.11-0.11, the brightest Fe K$\alpha$ feature shifted towards the Galactic East by ~3' in 12 years. Hence, it is fundamental that the IXPE observation is complemented by a quasi simultaneous pointing with another X-ray facility that provides the up-to-date morphology and spectrum of the clouds. Using IXPE maps only, it is difficult to pinpoint the location of the brightest Fe K$\alpha$ patches. The spectral resolution of IXPE at 6 keV is ~1 keV, thus the Fe K$\alpha$ line from the clouds is blended with the Fe$_{xxv}$-He$\alpha$ and Fe$_{xxvi}$-Lyman lines of the hard plasma. This means that in using IXPE data only, we may be not able to identify the optimal regions where the reflection of the clouds prevails over the plasma emission. The up-to-date Fe K$\alpha$ morphology of the Sgr A region can be provided either by Chandra, XMM-Newton, or eROSITA. However, only Chandra maps can provide the input in our procedures to compute the synthetic maps of the unpolarized components. This is because the resolution of Chandra is infinite from the IXPE point of view. For this reason, a Chandra observation is the best complement to the IXPE observation and would allow to treat the data exactly as we outlined in this paper. In the absence of a simultaneous Chandra coverage, it would still be feasible to apply our correction methods using Chandra archival maps and spectra for the plasma components and complement them with an up-to-date spectrum of the clouds provided by e.g. XMM-Newton or eROSITA, as the latter performs a monitoring of the GC every six months. However, in this case, a uniform morphology must be assumed for the cloud while computing the synthetic maps the unpolarized components. We already checked in Di Gesu, L. et al. [38] that simulating the clouds as a uniform source does not change the results. Without an up-to-date spectrum of the clouds, our methods cannot be applied, and the intrinsic polarization of the clouds cannot be retrieved correctly. The better the quality of the spectrum, the better the final uncertainties on the measurement of the distance along the line-of-sight of the cloud. In order to quantify this point we make the exercise of checking how much the uncertainty on both our methods would change if the knowledge of the up-to-date spectrum comes from IXPE data only. We fit a simulated IXPE spectrum of G0.11-0.11 with an absorbed APEC+APEC+PEXMON model. We computed the errors for $F_{crefl}$, $F_{tot}$, and $F_{K\alpha}$ and we used Eq. 10.3 and 10.5 to evaluate the uncertainties on the reconstructed polarization. We find that the uncertainty on the reconstructed polarization worsens by a relative factor of 20% and 46% with the dilution and subtraction methods, respectively. The uncertainty on the line-of-sight distances increases to $19^{+14}_{-8}$ pc and $18^{+9}_{-6}$ pc with the dilution and subtraction methods, respectively. Thus, we expect that the spectral quality of Chandra, XMM-Newton or eROSITA will ensure that the distance of the cloud along the line-of-sight is determined with an uncertainty on the order of a few parsec.

Finally, we note that, in case a strong variability in either flux or morphology of the clouds is detected before the IXPE pointing, the MDP of the clouds must be updated to decide the best pointing strategy. Recomputing the MDP with new input spectra and cloud location is straightforward using our procedure.

The possibility of reconstructing the 3D distribution of the gas in the CMZ depends on the constrains on the polarization degree. Even in case of non-detection, an X-ray polarimetric study of the MC will put useful constraints on their position along the line-of-sight. These values will be determined by the nominal MDP in the cloud region at the time of the IXPE observation. As an example, we made the exercise of computing $|\vec{d}_{los}|^{MDP}$ for all the clouds considered here (see Table 10.2). We note that these would be model-independent constraints because only the number of counts collected during the observation is needed to determine the MDP. The other method currently available to derive the line-of-sight [e.g., 27] rely on the measurement of the equivalent width of the Fe K$\alpha$ line, which depends on the scattering angle because of the angular dependence of the scattering continuum. This requires a careful modeling of the reflection continuum and an accurate knowledge of the iron abundances in the GC region.

We note that the uncertainty on the reconstructed value of the polarization degree is always slightly larger with the dilution method with respect to the uncertainty of the subtraction method. This is because the uncertainty on the dilution method depends on the dilution factor, $D$, that in our model is no larger than the $39\pm4\%$ estimated in the Bridge E region. Even in the case where the polarization is in principle undetected, this results in different $d_{los}$ estimates because of the different environmental dilution in each cloud region. For this reason, the subtraction methods returns more accurate results.

All in all, the capability of both our methods to recover the intrinsic polarization properties of the clouds is supported by the results described in Sect. 10.4. The comparison between the undiluted polarization map shown in Fig. 10.6 (b) and the reconstructed polarization maps with the dilution and subtraction methods shown in Fig. 10.6 (e) and 10.6 (f), respectively, visually highlights the efficiency of our methods in cleaning up the data from the contamination of the unpolarized emission. Both methods presented here to recover the intrinsic X-ray polarization degree are not limited to IXPE but could also be employed to treat the data coming from future X-ray polarimetry missions such as the enhanced X-ray Timing and Polarimetry mission [eXTP, 181], the Next Generation X-ray Polarimeter [NGXP, 141], or the X-ray Polarization Probe [XPP, 73]. These missions will have even greater sensitivity and spatial resolution with respect to IXPE. This highlights the importance of having tested general methods including detailed morphological information, allowing in the future to produce synthetic products suitable for instruments with any angular resolution.

# Chapter 11

# First IXPE flight data

After the launch of IXPE on December 9th 2021, a one-month long commissioning phase started in order to check the correct functioning of all the spacecraft systems. The GPDs were switched on and illuminated with the calibration sources of the FCW I described in Chapter 6. In Fig. 11.1 are shown the modulation curve, spectra, and image of the four on-board calibration sources of DU3 taken in flight.

From the observed modulation curves, spectra, and images, a one-on-one comparison can be done with the information acquired during the TV measurement described in Chapter 6 and shown in Fig. 6.11, 6.13, and 6.15.

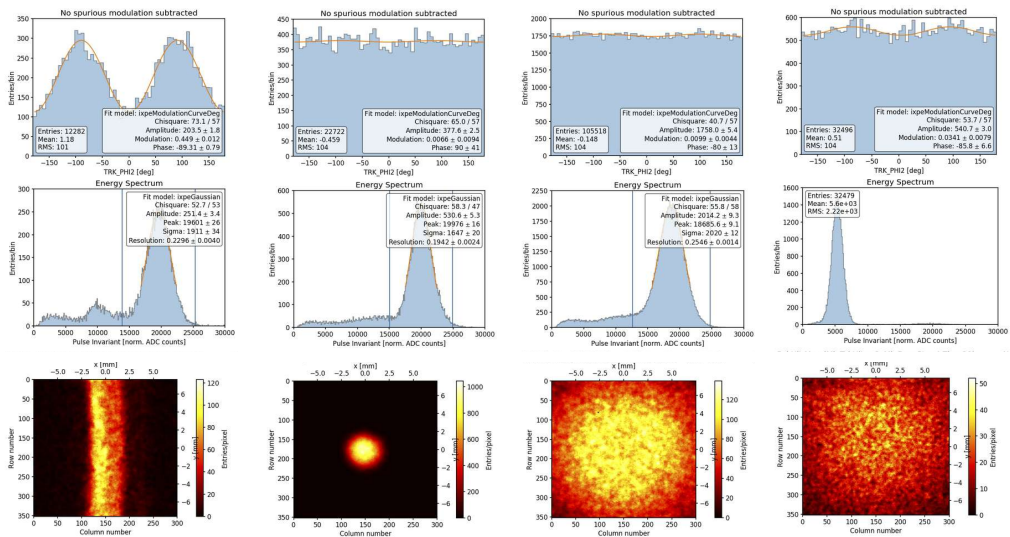

**Figure 11.1.** From top to bottom: modulation curve, spectrum, and image of the four on board calibration sources acquired during the IXPE commissioning. From left to right the plots in each column refer to CalA, CalB, CalC, and CalD of the DU3.

After correcting for the radioactive source activity at the actual measurement date, the observed counting rates of the source are compatible within 4%-20% with respect to the ones measured in Chapter 6 as summarized in Table 11.1. The differences can be explained by the uncertainties on the radioactive source emission rate.

At the time of the writing of this Chapter, the monitoring of the energy resolution,

**Table 11.1.** Comparison between the mean counting rate across the three IXPE DUs of the on board calibration sources expected at December 2021 (based on the TV measurements presented in Chapter 6), and the mean counting rate of the same sources measured in space during January 2022.

| Calibration source | Expected mean rate from TV (c/s) | Measured mean rate in space (c/s) |
|---|---|---|
| CalA @ 3 keV | 1.35 | 1.3 |
| CalA @ 5.9 keV | 16.6 | 16.3 |
| CalB | 51.5 | 44.1 |
| CalC | 118.9 | 106.6 |
| CalD | 76.3 | 60.8 |

gain, and modulation factor in time has started, as well as the check of the spurious modulation. The definition of the on-orbit calibration plan is ongoing, and will provide a fundamental contribution to the scientific production of IXPE.

At the end of the commissioning, a high celestial latitude point source, the blazar 1ES 1959+650, was pointed to check the mirror-detector alignment.

This was the very first astrophysical light of IXPE, and the quick look of this event is shown in Fig. 11.2. From this first data, I did a preliminary evaluation of the actual background. After removing from the observation the region of the detector swept by the dithered source, the remaining events can be assumed to be due to the unrejected instrumental background and the CXB, as at the latitude of 1ES 1959+650 the DGXB is negligible (see Chapter 7). I found that in the $2-8$ keV energy band, the background counting rate detected by IXPE is 6.4E-2 c/s/cm$^2$/DU. From my simulations shown in Chapter 7, the expected value was 5.9E-2 c/s/cm$^2$/DU. Hence, the observed combined CXB and unrejected instrumental background is within 10% of the simulated value. More in-depth analysis to evaluate with more precision the characteristic of the instrumental background and the application of the rejection algorithm to it are underway, and will be the subject of a future publication. At the end of the commissioning, IXPE begun its scientific operational phase. The first target was the the SNR CasA: in Fig.11.3 is shown the combined image from the three DUs at the end of the 1 Ms-long observation campaign in January 2022. I compare the real image with a simulation with *ixpeobssim*: the real image is very similar to the simulated one, and it is possible to distinguish between different structures. Although at the time of the writing of this thesis it is still not possible to show polarimetric data of this source, this picture already underscores that the imaging capabilities of IXPE are indeed as good as advertised. This bodes well for future observations of extended sources such as the Tycho SNR and the MC in the GC.

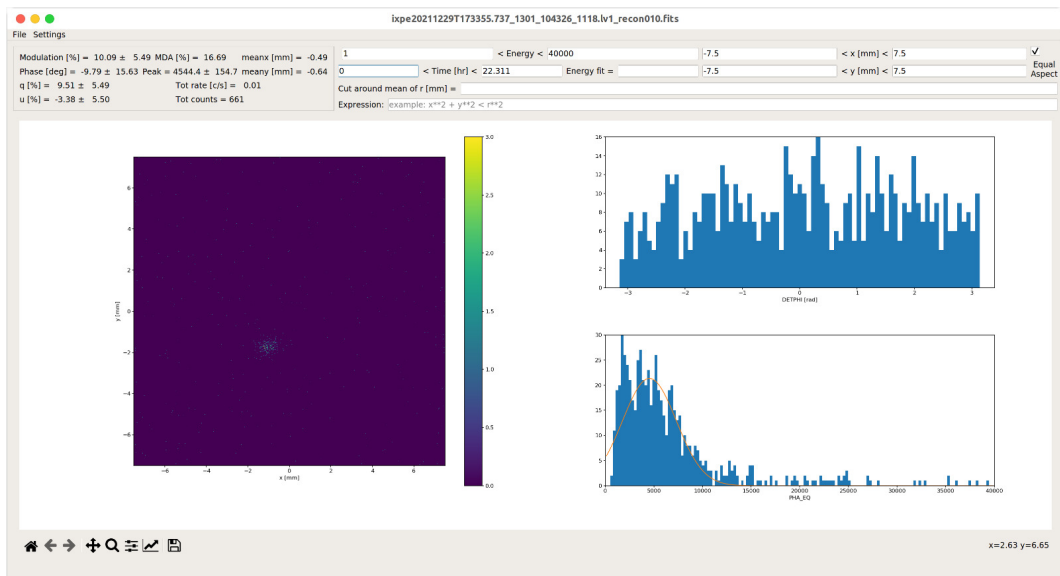

**Figure 11.2.** Quick-look of first ever celestial X-rays collected by IXPE, coming from the blazar 1ES 1959+650 during the commissioning phase (credits: John Rankin). On the left is shown the source image . On the right are shown the modulation curve of the events (top) and the spectrum in PHA (bottom).

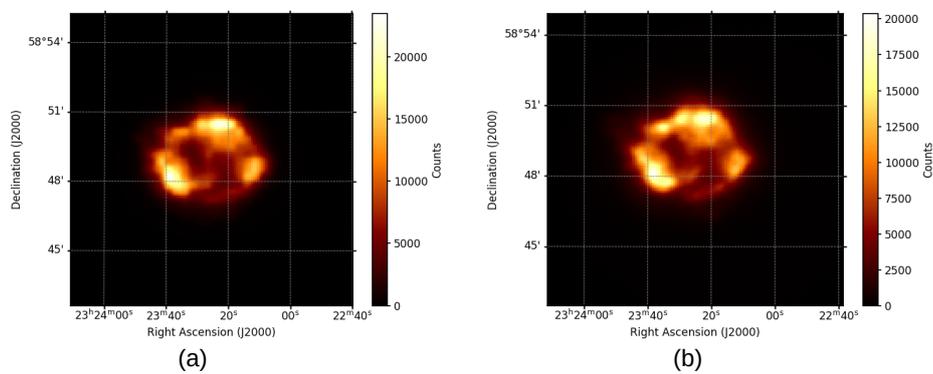

**Figure 11.3.** Comparison between a 1 Ms-long simulation of an IXPE observation of CasA (a), and the first real IXPE image of the same source observed for 1 Ms during January 2022 (b).

# Chapter 12

# Conclusions

The objectives of this thesis were the study of the calibrability in orbit of the detectors aboard the Imaging X-ray Polarimetry Explorer (IXPE) and the study of the detectability of the polarization properties of challenging sources such as Supernova remnants (SNRs) and the molecular clouds (MC) in the Galactic center (GC).

The former objective was pursued by means of laboratory measurements on the Flight Models of the polarized and unpolarized calibration sources of the IXPE Filter and Calibration Sets (FCS).

The results of this activity were presented in Chapter 6 and published in Ferrazzoli et al. [50]. The items of the Filter and Calibration Set were tested with commercials SDD and CCD to verify their functionality. The calibration sources were then tested in thermal vacuum with the flight detector units to derive their spectra, images on the detectors and polarimetric performance. The spectra and images of the sources, studied independently with CCD and GPD, were found to be consistent. Finally, I determined the in orbit calibration time needed to achieve the necessary sensitivity. The expected counting rates were comparable across the different flight models, with differences that can be ascribed to the energy resolution of each DU and on the intrinsic strength of the radioactive nuclide used. The counting rates satisfied the requirements and the modulation of the polarized sources was consistent with the one expected from Bragg diffraction. Three FCS are installed on the three IXPE DUs, while one is on a fourth DU that acts as a spare for off-line testing before and after the launch. Initially expected for the Spring of 2021, the launch of IXPE was delayed to December 9th 2021 due to the ongoing Covid-19 pandemic. We consider this a great success under very difficult circumstances. The activity of the radioactive sources have then further decayed, with respect to the values previously estimated, being now $\sim 71.1\%$ of their initial value. IXPE is expected to last at least two years, so that by the end of the mission lifetime, the activity of the sources will be reduced to $\sim 40\%$. The first measurements taken in orbit with the on-board calibration sources were shown in Chapter 11: the spectrum and image of the sources are consistent with the ones taken in Thermal vacuum. During the mission, the FCS will help validate the scientific results of IXPE by checking the detector response to point-like and extended sources. In summary, the results obtained on-ground, extrapolated to the ones expected in orbit, and then actually compared with the flight data, allows us to be confident that the FCS will be able to monitor properly the performance of the DUs during the IXPE lifetime.

The latter objective of the thesis was pursued first by evaluating all the sources of background IXPE observations of faint extended sources will be subjected to, then through realistic simulations of IXPE observations of extended sources such as the Tycho SNR and the molecular clouds in the GC.

The sources of background affecting IXPE are the instrumental background, the diffuse Galactic plane emission, and the Cosmic X-ray background. In Xie et al. [178], a work that I coauthored, we evaluated the expected instrumental background of the GPD, produced by the interaction of cosmic ray particles and photons with the spacecraft and the detector itself, using simulations based on matter-radiation interaction. We found it to be comparable to the one detected by the GPD on the PolarLight cubesat. After application of rejection algorithms based on track properties we found a residual background rate of $1.16 \times 10^{-2}$ c/s/cm$^2$/DU in the $2-8$ keV energy range. Of course this is the result of a simulation, but since late December 2021 we started collecting background data from IXPE in orbit, and we plan to apply soon the rejection method which we developed to confirm that the background can be sensitively rejected. Indeed, preliminary evaluations (see Chapter 11) suggest that our estimates of the background were correct. I found that the level of the instrumental background is negligible for point sources, but not for the faintest extended ones. The Galactic plane diffuse X-ray emission and the CXB are also sources of unpolarized background due to unresolved point sources. The latter will be uniform, the former depends on the distance of the source from the Galactic plane and its environment. To evaluate the effect of the astrophysical sources of background, in Chapter 7 I have shown simulated IXPE observations of five extended sources that will be observed during the first year: CasA, Tycho, SN1006, MSH 15-52, G0.11-0.11. I extracted the background spectrum from Chandra observations and fitted it with a template spectral model. I used the resulting fit to simulate the background counting rate for each source. I calculated the dilution of the polarization signal for each source due to the different sources of background. Both the residual instrumental and astrophysical background are small to negligible for the brightest SNR, CasA and Tycho, and for the PWN MSH 15-52. However, the residual instrumental background causes a non-negligible dilution for the faintest SNR, SN1006. Finally, being both faint and close to the plasma-rich Galactic center, the molecular Cloud G0.11-0.11 is heavily diluted.

The final results obtained in this thesis is the study of the feasibility of the observation with IXPE of two among the most interesting extended sources included in the first year observing plan: the Tycho SNR and the molecular clouds of the Sgr A* complex.

In Chapter 8 I showed the results of a simulated 1 Ms observation of Tycho with IXPE. I concluded that this exposure time is sufficient to detect in the stripes region polarization as low as $\sim$5% with at least 3$\sigma$ significance. This allows to distinguish by means of polarization between purely tangential, or radial, fields, and more complex geometries as well, such as a magnetic field aligned to the stripes. If the magnetic field is primarily radial, we should be able to probe it as long as the synchrotron polarization degree is >10%. However, if the field is not radial, but becomes tangential just before the shock, for example, or if it is not uniform over regions of the order of ~30", it would become challenging to detect anything unless the polarization degree is considerably higher. The polarization maps shown here underscore the possibility of performing spatially resolved X-ray polarimetry across the rim region of the Tycho SNR. If the synchrotron polarization fraction is high enough, it will be also possible to get significant spatially resolved results for multiple energy bins. A 1 Ms IXPE observation of the stripes region of the Tycho SNR allows to estimate at 99% confidence level the value of the magnetic field variance, and hence how much turbulent the magnetic field is, up to values of $\sigma^2 = 46.8\%$. This measurement has implication on our understanding of how the turbulence is generated in the SNR shock and on how CRs drive the magnetic field amplification. The Tycho SNR remnant will be observed for 1 Ms starting from June 19th of 2022. The work presented in this thesis will represent the basis of the Discovery paper of the IXPE observation of Tycho led by myself.

X-ray polarimetry can also be used to test an intriguing hypothesis: was SgrA*, the Supermassive Black Hole at the center of our Galaxy, a low luminosity AGN just ~300 years ago? By determining the polarization properties of the X-ray emission from the giant molecular clouds in the ~100 pc around SgrA* this hypothesis can be confirmed or discarded. If the clouds are indeed shining from scattered emission from SgrA*, they should be polarized, with the polarization direction orthogonal to the scattering plane. If the polarization vector will point perpendicularly to the cloud-Sgr A* direction, it will demonstrate that the Milky Way was, not long ago, a low luminosity AGN. Measuring the X-ray polarization property of a molecular cloud in the GC allows us to confirm (or discard) that is illuminated by a past outburst of Sgr A* (through the polarization angle) and to determine the position of the cloud along the line of sight (through the polarization degree). Assessing the history of our Galactic nucleus has implications for our understanding of the duty cycle of mass accretion onto SMBH that is believed to drive to the coevolution of SMBH and galaxies. In the paper by Di Gesu, L. et al. [38] that I coauthored, presented in Chapter 9, we have evaluated the feasibility of this experiment with IXPE and with eXTP, which is scheduled for launch in 2027. We simulated IXPE observations of the molecular clouds MC2, Bridge B2, Bridge E, G0.11-0.11, Sgr B2, Sgr C1, Sgr C2 and, Sgr C3. We considered the polarization properties predicted by the model of Marin et al. [90] and used the Monte Carlo-based simulation tool *ixpeobssim* to individually simulate IXPE images of these targets. Our source model considered the spectrum (using Chandra spectra), the polarization properties, and (when possible, using Chandra images) the spatial morphology for the molecular clouds and of the diffuse emission in the region of interest. Finally, we included in our simulations the instrumental background and the cosmic X-ray background. We determined for each cloud the minimum flux that would be detectable by IXPE in 2 Ms and found that the molecular clouds become undetectable when the total flux decreases by a factor $3-100$ (depending on the cloud) with respect to the level considered here. It cannot be excluded, however, that in the future their brightness will increase, facilitating the X-ray polarimetric observations. Moreover, we found that the polarization degree is diluted between ~99% and ~80% in the $2.0-4.0$ keV band and between ~80% and ~40% in the $4.0-8.0$ keV band, depending on the cloud considered. The morphological smearing of the sources contributes with additional dilution, whose value varies from cloud to cloud. The diluted polarization degree however at the IXPE spatial resolution does not depend on the internal morphology of the gas in the region of interest. For the flux levels we considered and the polarization degrees computed by Marin et al. [90], the most promising targets for IXPE observations are G0.11-0.11 and Sgr B2. For these two cases, we found that the $4.0-8.0$ keV polarization, even after being diluted by the surrounding plasma, is detectable by IXPE with a 2 Ms observation. The theoretical polarization degree strongly depends on the assumed position of the cloud along the line of sight. If the assumption on the distance is relaxed within the range reported in the literature, a wider range of possible polarization degrees can be derived. If this is the case, then also Bridge B2, Bridge E, and Sgr C1 might be detectable by IXPE in 2 Ms. Because its effective area is larger by a factor $\sim 4$, with the same exposure time eXTP will be able to detect the $4.0-8.0$ keV polarization degree predicted by Marin et al. [90] for G0.11-0.11, Sgr B2, and the Sgr C clouds. If a more relaxed constraint on the distance along the line of sight is assumed, then all the targets considered here may be detectable by eXTP.

In Chapter 10 I presented a follow up of the work performed in [38] that I published in Ferrazzoli, R. et al. [51]. I tackled the issues that make the observation of the molecular clouds in the GC a challenging and ambitious IXPE objective. I described data analysis techniques for the upcoming measurement of X-ray polarization from the MC in the Sgr A complex, expected in March 2022. I simulated a 2 Ms-long IXPE observation of the Sgr A region that includes four molecular clouds (MC2, Bridge B2, Bridge E, and G0.11-0.11) embedded in the diffuse plasma of the GC region. I used Chandra maps and spectra to model the spectrum and the morphology of the clouds and of the diffuse unpolarized thermal emission that has the effect of diluting the polarized signal. I also included the Cosmic X-ray Background and the instrumental background. I produced a map of the minimum detectable polarization (MDP) that can be obtained with a 2 Ms-long exposure of the Sgr A MC complex in the $4-8$ keV energy band, and I evaluated the effect on the reduction of the MDP because the clouds are observed off-axis in the FOV. I found that the MDP, with respect to the on-axis observation of each cloud, increases by a factor $\sim 1-15\%$ due to vignetting effects depending on the cloud position in the FOV and its spectral shape. I also presented two independent techniques to recover the intrinsic polarization degree of the MC: the dilution method consists in dividing pixel-by-pixel two polarization maps: the observed polarization map and a dilution map; the subtraction method is based on the subtraction from the observed Stokes I parameters map of the Stokes maps of the unpolarized components only. I found that these two techniques can recover the polarization degree and, hence, the line-of-sight distance $\tilde{d}_{\mathrm{los}}$ of a cloud whose polarization is detected at a 99% confidence level. For instance, for G0.11-0.11 I found that with a 2 Ms-long IXPE observation we can constrain the distance along the line-of-sight with respect to the Galactic plane to $\pm 19 \pm 7$ pc and $\pm 18 \pm 4$ pc with the dilution and subtraction method, respectively. Because the brightness of these clouds changes with time, we will make use of quasi-simultaneous Chandra observations that are quasi-simultaneous with the IXPE pointing of the Galactic center to assess the illumination status of the clouds, and to derive the morphological and spectral information needed to apply the polarization recovery methods. I found that with this observing strategy, the uncertainty on the measurement of the line-of-sight distance of a detected cloud will be of the order of a few parsec. The same approach can be applied to future X-ray polarimetric missions such as eXTP, NGXP, and XPP. As for the case of Tycho, the works presented in this thesis have been invaluable for the preparation of this ambitious observation.

In conclusion, IXPE is be the first mission capable of performing spatially resolved X-ray polarimetry. Thanks to the synergy of instrumental activities and astrophysical simulations, the works presented in this thesis gave a fundamental contribution to the preparation of the mission and the future interpretation of scientific data. Polarimetry was once the missing piece of the X-ray astronomy puzzle. This is no longer the case thanks to the launch of IXPE, that opened a new era of discovery opportunities.

This thesis argued that even the observation of the most challenging targets will be possible, thanks to the careful evaluation of the sources of background, data analysis based on the Stokes parameters formalism, and extensive on-board calibrations.

Next generation X-ray polarimetry missions will continue to benefit from the studies here presented.

# Glossary of acronyms

**ADC:** Analog to Digital Converter
**AGN:** Active Galactic Nucleus
**ASI:** Agenzia Spaziale Italiana
**ASIC:** Application Specific Integrated Circuit
**BH:** Black Hole
**CALDB:** CALibrations Data Base
**CCD:** Charge Couple Device
**CIAO:** Chandra Interactive Analysis of Observations
**CMOS:** Complementary Metal-Oxide Semiconductor
**CMZ:** Central Molecular Zone
**CR:** Cosmic Ray
**CXB:** Cosmic X-ray Background
**DME:** DiMethyl-Ether
**DSA:** Diffusive Shock Acceleration
**DSU:** Detector Service Unit
**DU:** Detector Unit
**EEF:** Encircled Energy Function
**ESA:** European Space Agency
**eXTP:** enhanced X-ray Timing and Polarimetry mission
**FCS:** Filter and Calibration Set
**FITS:** Flexible Image Transport System
**FCW:** Filter and Calibration Wheel
**FM:** Flight Model
**FOV:** Field Of View
**FWHM:** Full Width Half Maximum
**GC:** Galactic Center
**GEM:** Gas Electon Multiplier
**GPD:** Gas Pixel Detector
**GPDE:** Galactic Plane Diffuse Emission
**GPS:** Global Positioning System
**HEASARC:** High-Energy Astrophysics Science Archive Research Center
**HPD:** Half Power Diameter
**INAF:** Istituto Nazionale di Astrofisica
**INFN:** Istituto Nazionale di Fisica Nucleare
**IRF:** Instrument Response Function
**IXO:** International X-ray Observatory
**IXPE:** Imaging X-ray Polarimetry Explorer

**MC:** Molecular Cloud
**MDA:** Minimum Detectable Amplitude
**MDP:** Minimum Detectable Polarization
**MMA:** Mirror Module Assembly
**MSFC:** Marshall Space Flight Center
**NASA:** National Aeronautic and Space Administration
**NGXP:** Next Generation X-ray Polarimeter
**NS:** Neutron Star
**OAC:** Osservatorio Astronomico di Cagliari
**OSO-8:** Orbital Solar Observatory 8
**PCUBE:** Polarization CUBE
**PHA:** Pulse Height Amplitude
**PI:** Principal Investigator/Pulse Invariant **PMAPCUBE:** Polarization MAP CUBE
**PWN:** Pulsar Wind Nebula
**PSF:** Point Spread Function
**RQ:** Radio Quiet
**SAA:** South Atlantic Anomaly
**SASWG:** Science Analysis and Simulations Working Group
**SDD:** Silicon Drift Detector
**SMBH:** SuperMassive Black Hole
**SMEX:** SMall EXploration mission
**SNR:** Supernova Remnant
**SOC:** Science Operations Center **SRG:** Spectrum-X-Gamma
**STWG:** Scientific Topical Working Group
**SXRP:** Stellar X-Ray Polarimeter
**TPC:** Time Projection Chamber
**TV:** Thermal Vacuum
**XEUS:** X-ray Evolving Universe Spectroscopy
**XIPE:** X-ray Imaging Polarimetry Explorer
**XPP**: X-ray Polarimetry Probe

# List of Publications

List of scientific publications produced during the PhD.

**(2022)**

- F. Muleri, R. Piazzolla, A. Di Marco, S. Fabiani, F. La Monaca, C. Lefevre, A. Morbidini, J. Rankin, P. Soffitta, A. Tobia, F. Xie, F. Amici, P. Attinà, M. Bachetti, D. Brienza, M. Centrone, E. Costa, E. Del Monte, S. Di Cosimo, G. Di Persio, Y. Evangelista, **R. Ferrazzoli**, P. Loffredo, M. Perri, M. Pilia, A. Ratheesh, A. Rubini, F. Santoli, E. Scalise, A. Trois,
  **"The IXPE Instrument Calibration Equipment"**,
  Astroparticle Physics, Volume 136, March 2022, 102658.

**(2021)**

- M. Weisskopf, A. Ratheesh, A. Rubini, A. Marscher, A. Manfreda, A. Marrocchesi, A. Brez, A. Di Marco, A. Paggi, A. Profeti, A. Nuti, A. Trois, A. Morbidini, A. F. Tennant, A. L. Walden, A. Sciortino, A. Antonelli, A. Tobia, B. Negri, B. Garelick, B. Forsyth, B. D. Ramsey, B. Weddendorf, C. Lefevre, C. Sgro', C. Alexander, C. O. Speegle, C. Oppedisano, C. Pentz, C. Boree, C. Schroeder, C. Caporale, C. Cardelli, C. Peterson, D. Zachery Allen, D. Brienza, D. Osborne, D. Dolan, D. Mauger, D. Welch, D. Zanetti, E. Gurnee, E. Mangraviti, E. D'Alba, E. Cavazzuti, E. Scalise, E. Costa, E. Kelly, E. Del Monte, F. Borotto, F. D' Amico, F. La Monaca, F. Muleri, F. Amici, F. Marin, F. Mosti, F. Xie, F. Massaro, F. Santoli, F. Zanetti, G. Matt, G. Di Persio, G. Spandre, G. Hibbard, G. Magazzu, H. Kyle Bygott, H. L. Marshall, H. Nasimi, I. Mitsuishi, I. Donnarumma, I. Nitschke, J. Ranganathan, J. Andersen, J. Sanchez, J. Bladt, J. J. Kolodziejczak, J. McCracken, J. Wedmore, J. Rankin, J. Footdale, J. Poutanen, K. Sosdian, K. Ferrant, K. Kilaru, K. L. Dietz, L. Di Gesu, L. Reedy, L. Lucchesi, L. Orsini, L. Baldini, L. Cavalli, L. Latronico, M. Ferrie, M. Tardiola, M. Castronuovo, M. Ceccanti, M. Marengo, M. Vimercati, M. C. Weisskopf, M. Minuti, M. Bachetti , M. Perri, M. Barbanera, M. Pilia, M. Centrone, M. Pesce-Rollins, M. Head, M. Dovciak, M. McEachen, M. Negro, M. Pinchera, M. Foster, M. Onizuka, N. Bucciantini, N.' Di Lalla, N. E. Thomas, N. Root, P. Lorenzi, P. Mereu, P. Sarra, P. Soffitta, P. Loffredo, P. Slane, P. Attina', R. Bonino, R. Piazzolla, R. M. Baggett, **R. Ferrazzoli** et al.,
  **"The Imaging X-Ray Polarimetry Explorer (IXPE): Pre-Launch"**,
  Accepted for publication on J. of Astronomical Telescopes, Instruments, and Systems.

43rd COSPAR Scientific Assembly. Held 28 January - 4 February, 2021. Abstract E1.2-0018-21 (oral), id.1469.

**Observation proposals:**